\documentclass{aastex701}

\usepackage{amsmath}
\usepackage{subcaption}
\usepackage{placeins}
\usepackage{multirow}

\begin{document}

\title{Discovery of Trailing Spiral Dust Alignment in the Dust Polarization of HL~Tau at $\boldsymbol{\lambda = 2.1}$~mm}

\author[0009-0004-2356-9483]{Yi-Wei Jerry Xu}
\affiliation{Department of Astronomy, University of Illinois, 1002 West Green St, Urbana, IL 61801, USA}
\email{yiweix3@illinois.edu}

\author[0000-0002-4540-6587]{Leslie W. Looney}
\affiliation{Department of Astronomy, University of Illinois, 1002 West Green St, Urbana, IL 61801, USA}
\email{lwl@illinois.edu}

\author[0000-0003-3017-4418]{Ian W. Stephens} 
\affiliation{Department of Earth, Environment, and Physics, Worcester State University, Worcester, MA 01602, USA}
\email{istephens@worcester.edu}

\author[0000-0001-5811-0454]{Manuel Fern\'{a}ndez-L\'{o}pez}
\affiliation{Institut de Ci\`encies de l'Espai (ICE), CSIC, Campus UAB, Carrer de Can Magrans s/n, 08193 Bellaterra, Barcelona, Spain}
\affiliation{Instituto Argentino de Radioastronom\'{i}a (CCT-La Plata, CONICET; UNLP; CICPBA), C.C. No. 5, 1894, Villa Elisa, Buenos Aires, Argentina}
\affiliation{Facultad de Ciencias Astron\'{o}micas y Geof\'{i}sicas, Universidad Nacional de La Plata, Paseo del Bosque S/N, B1900FWA La Plata, Argentina}
\email{manferna@gmail.com}

\author[0000-0001-7233-4171]{Zhe-Yu Daniel Lin}
\affil{National Radio Astronomy Observatory, 520 Edgemont Rd., Charlottesville, VA 22903 USA} 
\altaffiliation{Jansky Fellow of the National Radio Astronomy Observatory}
\email{zlin@carnegiescience.edu}

\author[0000-0002-7402-6487]{Zhi-Yun Li}
\affiliation{University of Virginia, 530 McCormick Rd., Charlottesville, Virginia 22904, USA}
\email{zl4h@virginia.edu}

\author[0000-0002-8537-6669]{Haifeng Yang}
\affiliation{Institute for Astronomy, School of Physics, Zhejiang University, Hangzhou, 310027 Zhejiang, China}
\email{hfyang@zju.edu.cn}

\author[0000-0001-7387-3898]{Austen Fourkas}
\affiliation{Department of Astronomy, University of Illinois, 1002 West Green St, Urbana, IL 61801, USA}
\email{fourkas2@illinois.edu}

\author[0000-0003-2118-4999]{Rachel Harrison}
\affiliation{Department of Physics and Astronomy, Rice University, 6100 Main Street -- MS 108, Houston, TX 77005, USA}
\email{rh128@rice.edu}

\author[0000-0003-1451-6836]{Ryo Tazaki}
\affiliation{Department of Earth Science and Astronomy, The University of Tokyo, 3-8-1 Komaba, Meguro, Tokyo 153-8902, Japan}
\email{rtazaki@g.ecc.u-tokyo.ac.jp}

\author[0000-0002-1886-0880]{Satoshi Okuzumi}
\affiliation{Department of Earth and Planetary Sciences, Institute of Science Tokyo, 2-12-1 Ookayama, Meguro, Tokyo 152-8551, Japan}
\email{okuzumi@eps.sci.isct.ac.jp}

\author[0000-0003-2862-5363]{Carlos Carrasco-Gonz\'{a}lez}
\affiliation{Instituto de Radioastronomía y Astrofísica, Universidad Nacional Autónoma de México, Antigua Carretera a Pátzcuaro 8701, Ex-Hda. San José de la Huerta, 58089 Morelia, Michoacán, Mexico}
\email{c.carrasco@irya.unam.mx}

\begin{abstract}

Polarimetric observations of protoplanetary disks at (sub)millimeter wavelengths constrain dust grain properties, dynamics, and alignment mechanisms. We present new high-resolution ALMA Band~4 (2.1~mm) polarization observations of HL~Tau with spatial resolutions of $\sim0\farcs08$ (11.78~au) and $\sim0\farcs19$ (27.99~au). Linearly decomposing the polarization into contributions from dust scattering and thermal emission by aligned grains, we analyze the new Band~4 data and published high-resolution Band~7 (870 $\rm{\mu m}$) observations. Thermal emission from aligned grains dominates the polarization across the disk at Band~4, in contrast to the scattering-dominated Band~7, and both mechanisms vary radially with the disk substructures, consistent with expected optical depth variations. At Band~4, we detect a coherent angular deviation from purely azimuthal grain alignment, in which the streamlines of the alignment field yield trailing spirals with respect to disk rotation. A single global offset angle quantifies this deviation as $2.83^\circ\pm0.33^\circ$ and $4.10^\circ\pm0.09^\circ$ (statistical uncertainties) from two robust weightings, and the detection reaches a significance of $\approx7\sigma$ under conservative uncertainty estimates. This trailing orientation is consistent with aerodynamic (``badminton birdie-like'') alignment driven by gas-dust drift near pressure bumps. However, under an axisymmetric disk velocity structure, this interpretation requires Stokes numbers two to three orders of magnitude larger than expected for the grains dominating the millimeter emission. At Band~7, the polarization angle is weakly sensitive to the alignment direction and exhibits a near/far-side antisymmetric structure that the model cannot reproduce. The Band~7 data therefore neither support nor contradict the Band~4 result.

\end{abstract}



\section{INTRODUCTION} \label{sec:Introduction}
Observations of dust continuum polarization at (sub)millimeter wavelengths have provided unique insight into the properties of dust grains in protoplanetary disks. The morphology and intensity of polarization encodes information about the grain sizes \citep[e.g.,][]{Kataoka_2015, Ohashi_2019}, shapes and internal structures \citep[e.g.,][]{Tazaki_2019_dust_structure, Brunngraber_2021, Zhang_2023, Vargas_2026}, alignment orientations with respect to local radiation, magnetic, or velocity fields \citep[e.g.,][]{Lazarian_2007_RAT, Lazarian_B-field, Tazaki_2017_RAT, Kataoka_2019_helicity}, and the dynamical coupling between dust and gas \citep[e.g.,][]{Kataoka_2019_helicity, badminton_birdie-like}. These grain properties and dynamics are fundamentally important in understanding planet formation, as the growth of interstellar grains into aggregates and eventually planetesimals is the critical first step in the core-accretion pathway \citep{Dominik_2007, Testi_2014_planetf}, while the aerodynamic coupling between dust and gas governs key processes such as vertical settling, radial drift, and the streaming instability that can concentrate solids into gravitationally bound clumps \citep{Youdin_2005, Rucska_2023}. Polarization provides constraints on grain size and optical depth that are independent of, and complementary to, those derived from total intensity observations alone \citep{Kataoka_2015, Bacciotti_2018, Haifeng_scattering_2016, Haifeng_2017_s_od}. 

There are two main mechanisms that can produce observable polarization patterns in protoplanetary disks. First, dust grains can polarize thermal photons emitted by other grains through scattering \citep{Kataoka_2015}, regardless of whether the initial photon is polarized or not. This polarization is a generic feature of light scattering of particles and is most efficient when the grain size is comparable to the observing wavelength divided by $2\pi$ $\left(a_{\text{max}}\sim\lambda / 2\pi \right)$, making this scattering-induced polarization a good probe of grain size \citep{Kataoka_2015, Kataoka_2016_HLTau_grainsize}. For a geometrically thin, inclined disk, the observing geometry causes the observed scattered radiation to be preferentially polarized in the direction parallel to the disk minor axis \citep{Haifeng_scattering_2016}. 

Second, elongated dust grains can emit intrinsically polarized thermal radiation. The observed polarization morphology across the disk depends on how these grains are aligned. Several mechanisms have been proposed to explain grain alignment in protoplanetary disks. Radiative alignment torques \citep[RATs,][]{Lazarian_2007_RAT} can align grains with either the magnetic field or the local radiation field, depending on the magnetic field strength \citep{Tazaki_2017_RAT, Haifeng_Band3_2019}. The expected polarization morphology from magnetically aligned grains can be modified by wrong internal alignment \citep{Hoang_2009_wrong_alignment, Thang_2024_MRAT}, or by polarization reversal in the Mie regime \citep{Guillet_2020_Mie}. Relative motion between gas and dust can also align grains mechanically \citep{Gold_mechanism, Lazarian_2007_MAT, Reissl_2023_MAD, badminton_birdie-like}. Most recently, \cite{badminton_birdie-like} demonstrated that gas-dust drift can align grains through aerodynamic torques under subsonic drift conditions when the grain's center of mass is offset from its geometric center along the long axis, producing a ``badminton birdie-like" alignment effect that can align prolate grains with their long axes parallel to the gas flow direction.

Several protoplanetary disks exhibit coexistence of polarization from scattering and from thermal emission of aligned grains, and the two contributions vary with observing wavelength \citep[e.g.,][]{Harrison_2019, Harrison_2024}. Among these sources, HL~Tau has been one of the best examples for studying the wavelength-dependent polarization transition. HL~Tau is located at a distance of $\sim147.3$ pc \citep{Gaia_2018_distance}, exhibiting prominent concentric ring and gap substructures \citep{ALMA_Partner_2015}, and currently has the most complete wavelength coverage in spatially resolved polarization for a protoplanetary disk \citep[e.g.,][]{Ian_2017_HLTau, Ian_Nature_2023, Daniel_Multiband_2024}. Observations of HL~Tau have shown polarization morphologies that exhibit a clear transition from aligning in the azimuthal direction at $3.1 \rm{mm}$ ALMA Band~3 \citep{Kataoka_2017} to aligning with the disk minor-axis at 870 $\rm{\mu m}$ ALMA Band~7 \citep{Ian_2017_HLTau}. Building upon modeling framework developed by \cite{Daniel_model_2022}, \cite{Daniel_Multiband_2024} also carried out a multi-wavelength study on HL~Tau to decompose the polarization mechanisms, and characterized wavelength dependence of the scattering and aligned-grain thermal emission contributions based on observations by ALMA at Bands 3-7 and by the VLA at \textit{Q}-band. \cite{Ian_Nature_2023} obtained high-resolution ($\sim0\farcs033$, $4.86$ au) Band~7 polarization observations of HL~Tau that resolved the polarization within individual rings and gaps for the first time, revealing distinct polarization signatures induced from dust grain alignment and scattering between ring and gap substructures. 

In this paper, we present new high-resolution ALMA Band~4 ($2.1$ mm) polarization observations of HL~Tau, imaged with robust weightings of $-0.5$ ($\sim0\farcs08$, 11.78 au) and 0.5 ($\sim0\farcs19$, 27.99 au), which resolve the polarization within prominent ring and gap substructures. At this wavelength, the polarization is expected to be dominated by thermal emission from aligned grains rather than scattering, which is complementary to the scattering-dominated Band~7 observations of \cite{Ian_Nature_2023}. By applying the linear decomposition modeling framework of \cite{Daniel_model_2022} to both bands, our work provides a multi-wavelength comparison that helps break degeneracies inherent to single-wavelength analysis, such as those between grain size, optical depth, and the relative contributions of scattering and thermal emission.  

This paper is organized as follows: Section~\ref{sec:observations} describes the observations and data reduction processes; Section~\ref{sec:B4_obs_result} presents the observational results of HL Tau at Band~4, including total intensity maps and polarization morphologies; Section~\ref{sec:Linear_Polarization_Modeling} describes linear polarization modeling and the fitting procedures; Section~\ref{sec:B4_Modeling_Results} presents the Band~4 modeling results, while Section~\ref{sec:B7_Modeling_Results} presents the Band~7 modeling results with the same framework for comparison. In Section~\ref{sec:discussion}, we provide a detailed examination of interpretations from the modeling results, including the existence and origin of trailing spiral streamlines in grain alignment, variation of polarization mechanisms with substructures, as well as their wavelength-dependence. In Section~\ref{sec:conclusions}, we conclude our findings. 

\section{OBSERVATIONS} \label{sec:observations}
This paper presents results from the ALMA projects 2019.1.00134.S and 2023.1.00515.S (principal investigator: Ian W. Stephens), taken in three different array configurations (C4, C6, and C8) and combined as discussed below. Both projects observed HL Tau in full Stokes polarization mode at ALMA Band~4 (2.07 mm). The different array configurations covered baselines between 20 k$\lambda$ and 4000 k$\lambda$. HL Tau was observed under project 2019.1.00134.S in July 2021 (C6 configuration) and under project 2023.1.00515.S in October 2023 (C4 configuration) and December 2023 (C8 configuration). The July 2021 data were originally presented in \cite{Daniel_Multiband_2024}. The total on-source integration time was 7.3 hours, which, including overheads, amounted to 13 hours of observations. The number of available antennas varied between 39 and 44. The correlator setup included four 1.875 GHz spectral windows devoted to continuum observations tuned to central frequencies 138.0 GHz, 139.9 GHz, 150.0 GHz and 152.0 GHz. Weather conditions were generally good for observations at the targeted frequency at Chajnantor, with a precipitable water vapor varying between 0.2 mm and 3.8 mm and phase fluctuations smaller than 15$^\circ$.

The calibrated data were delivered by the ALMA staff. Bandpass, flux, phase, and polarization calibrators were quasars J0510+1800, J0431+1731, and J0440-4333 respectively for most of the tracks. J0423-0120 was used as bandpass and flux calibrator in one of the October 2023 tracks, and as polarization calibrator during the July 2021 observations. With the combined data from the three configurations, we performed three iterations of phase-only self-calibration with decreasing integration times down to times ranging between 5 and 10 seconds, depending on the observed track. After that, we performed an additional amplitude and phase self-calibration step which, in general, made significant improvements in signal-to-noise ratio and image quality. Before combining the three epochs, we aligned the disk peak in all datasets to the reference position from the 2023.1.00515.S long-baseline data (C8 configuration) using the CASA tasks \texttt{fixplanets} and \texttt{phaseshift}. By using the CASA task \texttt{gencal}, we re-scaled their fluxes matching the amplitudes of the visibilities that overlap in uv-distance. We deconvolved and restored the visibility measurements in CASA version 6.6.5.31 \citep{The_CASA_Team_2022} using the \texttt{tclean} task with two different robust settings: $+0.5$ and $-0.5$. The former selection provides a good compromise between angular resolution and sensitivity, whereas the latter results in an enhanced resolution at the expense of a lower sensitivity. The robust = +0.5 images have a synthesized beam of $0\farcs19\times0\farcs17$ (position angle of $-12^\circ$), reaching an rms noise level of 10.1 $\mu \rm{Jy}$ $\rm{beam}^{-1}$ (Stokes $I$), and 3.9 $\mu \rm{Jy}$ $\rm{beam}^{-1}$ (Stokes $Q$ and $U$). The robust $=-0.5$ images have a synthesized beam of $0\farcs08\times0\farcs07$ (position angle of $-20^\circ$), and an rms noise level of 8.8 $\mu \rm{Jy}$ $\rm{beam}^{-1}$ (Stokes $I$) and 8.2 $\mu \rm{Jy}$ $\rm{beam}^{-1}$ (Stokes $Q$ and $U$). We note that throughout the paper, beam sizes refer to the full width at half maximum (FWHM) of the synthesized beam. Following \cite{Hull_2015_debias}, debiased images of polarized intensity and polarization fraction were constructed from these products, along with an image of the polarization direction.

In addition to the new Band~4 data, this paper also analyzes archival ALMA Band~7 (870~$\mu$m) polarization observations of HL~Tau, which have been previously presented in \cite{Ian_Nature_2023} and analyzed in \cite{Yang_2025}. These Band~7 data are a combination of the ALMA projects 2016.1.00115.S and 2019.1.01051.S, which observed HL~Tau in full Stokes polarization mode at 870~$\mu$m. In this work, we use images with robust weighting = 0.5 and also prepared new images with robust weighting = 2.0. The synthesized beam sizes are $\sim0\farcs035$ and $\sim0\farcs066$, respectively. The rms noise levels of the Stokes $I$/$Q$/$U$ images are 39.7/8.4/8.4$\mu\rm{Jy}~\rm{beam}^{-1}$ and 167/13.3/15.4$\mu \rm{Jy}~\rm{beam}^{-1}$ for robust = 0.5 and robust = 2.0, respectively. 

\section{OBSERVATIONAL RESULTS} \label{sec:B4_obs_result}
\subsection{Band~4 total intensity and polarization morphology} \label{sec:B4_I_pfrac_pi}
\begin{figure}[h]
    \centering
    \includegraphics[width=1.0\linewidth]{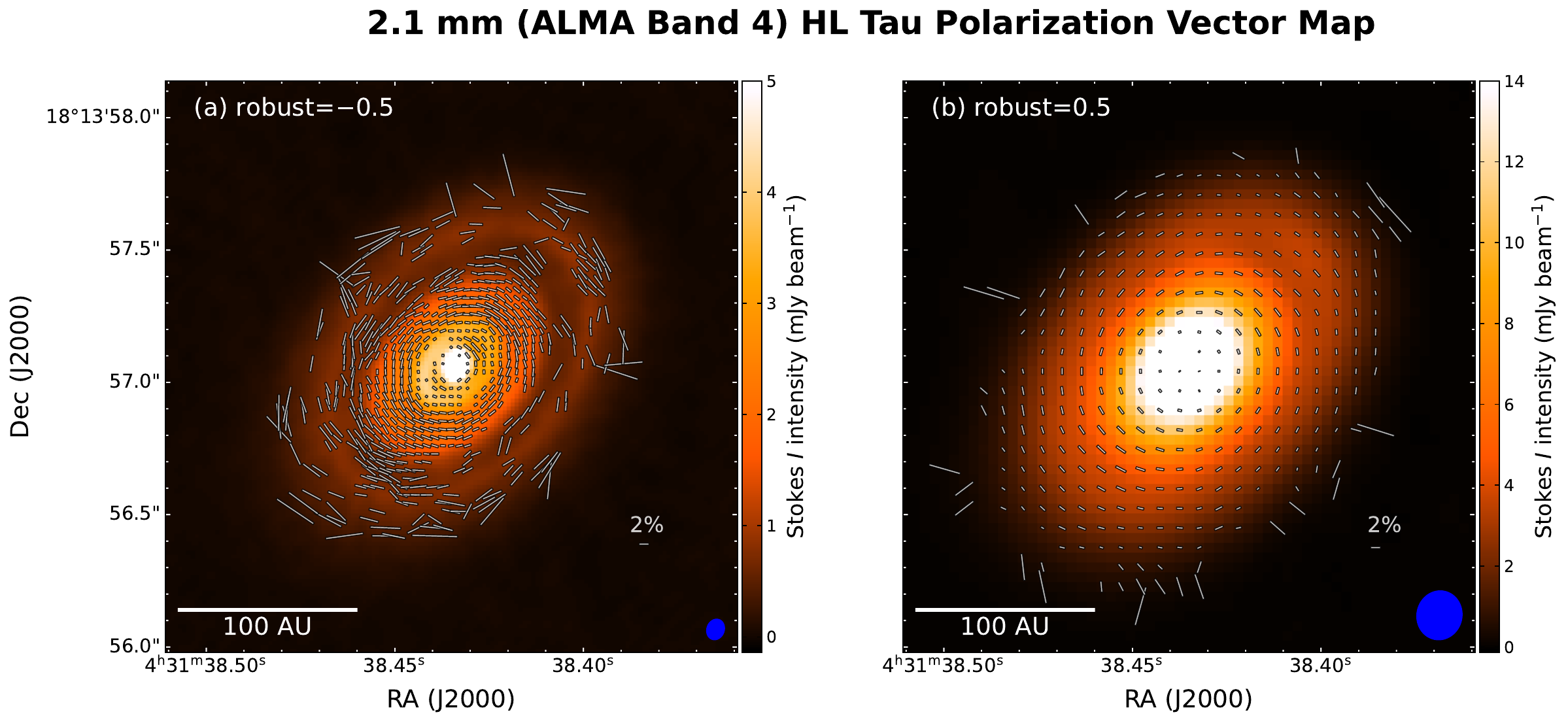}
    \caption{The 2.1-mm (ALMA Band~4) polarization vectors overlaid on total-intensity (Stokes $I$) images of HL~Tau with robust weightings of (a) $-0.5$ and (b) 0.5. The length of the polarization vectors is proportional to the polarization fraction ($p$), with a scale bar of $2\%$ shown at the bottom right. Vectors are masked out at locations where either Stokes $I$ or polarized intensity $P$ is below the $2\sigma$ level. The blue ellipse in the bottom right shows the synthesized beam.}
    \label{fig:B4pvector}
\end{figure}

\begin{figure}[h]
    \centering
    \includegraphics[width=1.0\linewidth]{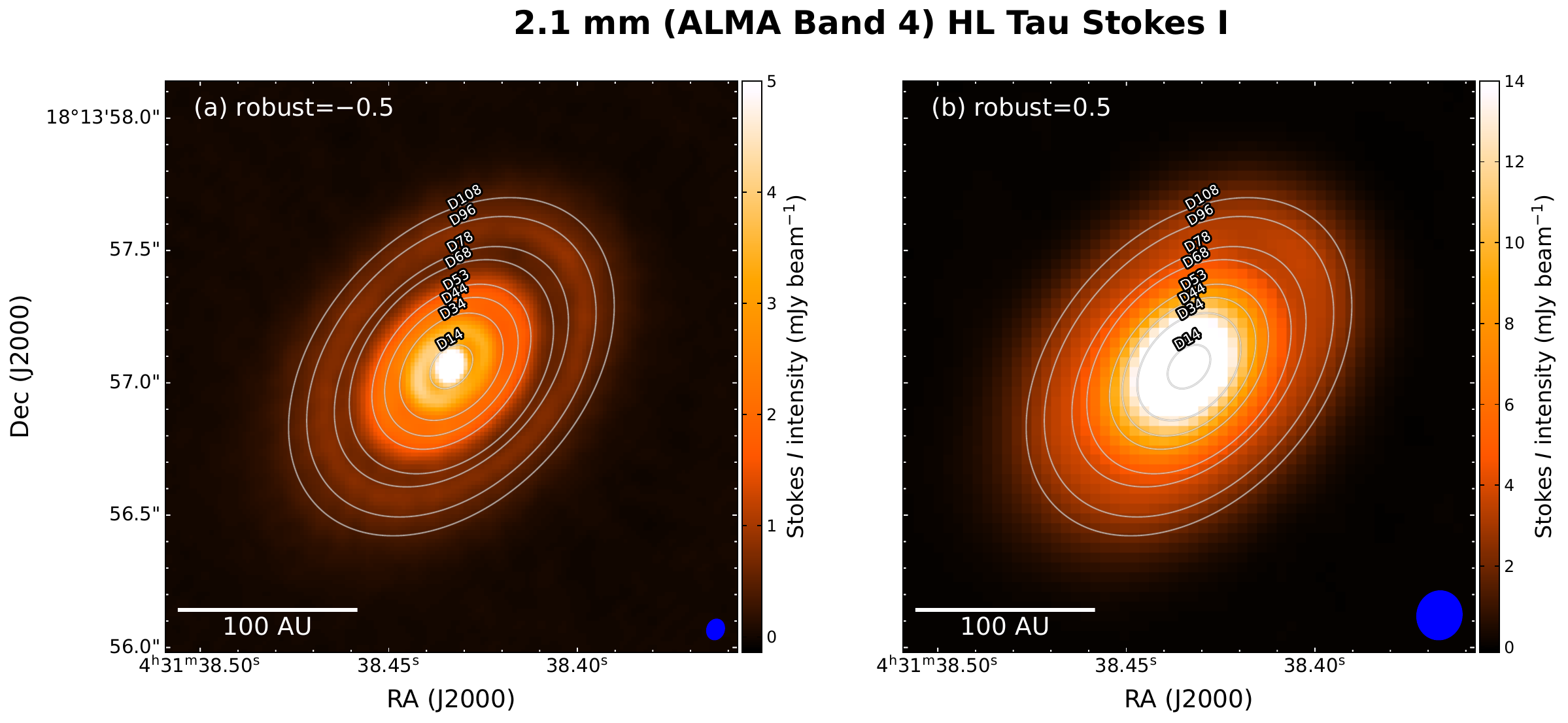}
    \caption{The 2.1-mm (ALMA Band~4) total-intensity (Stokes $I$) image of HL~Tau, imaged with (a) robust $=-0.5$ and (b) robust = 0.5, with  ellipses overlaid representing the locations of gaps previously identified by \cite{ALMA_Partner_2015} and \cite{Ian_Nature_2023}. Each gap is labeled D (dark lane) followed by the semi-major axis of the ellipse in units of au. The blue ellipse in the bottom right shows the synthesized beam.}
    \label{fig:B4StokesI}
\end{figure}

Figure~\ref{fig:B4pvector} shows the polarization direction morphology of HL~Tau at Band~4 overlaid on its Stokes $I$ image. Line segments represent polarization directions (hereafter referred to as ``polarization vectors"). The robust $=-0.5$ image provides a higher resolution for resolving ring-gap structures, while the robust = 0.5 image provides higher S/N with lower spatial resolution. The polarization vectors are predominantly azimuthally oriented in an elliptical pattern that follows the projected disk geometry in both robust maps. Figure~\ref{fig:B4StokesI} presents Stokes $I$ maps at both robust values with ellipses overlaid at the locations of the gaps. These gaps were previously identified by \cite{ALMA_Partner_2015} using combined Band~6 and Band~7 observations (the first seven gaps) and \cite{Ian_Nature_2023} using Band~7 data (the outermost eighth gap). 

Not all gaps are resolved with our Band~4 observations. At both robust weightings, the relatively faint gaps at 44 and 53 au (D44 and D53) appear blended with the neighboring rings into a single bright belt, while the gaps at 68 and 78 au (D68 and D78) merge into a single dark belt. Along with the vector map in Figure~\ref{fig:B4pvector}, we note that polarization vector orientations are generally consistent across the ring and gap regions. This differs from the Band~7 behavior observed by \cite{Ian_Nature_2023}, where polarization vectors are predominantly aligned with the minor axis in the rings but exhibit significant azimuthal components in the gaps.

\begin{figure}
    \includegraphics[width=1.0\linewidth]{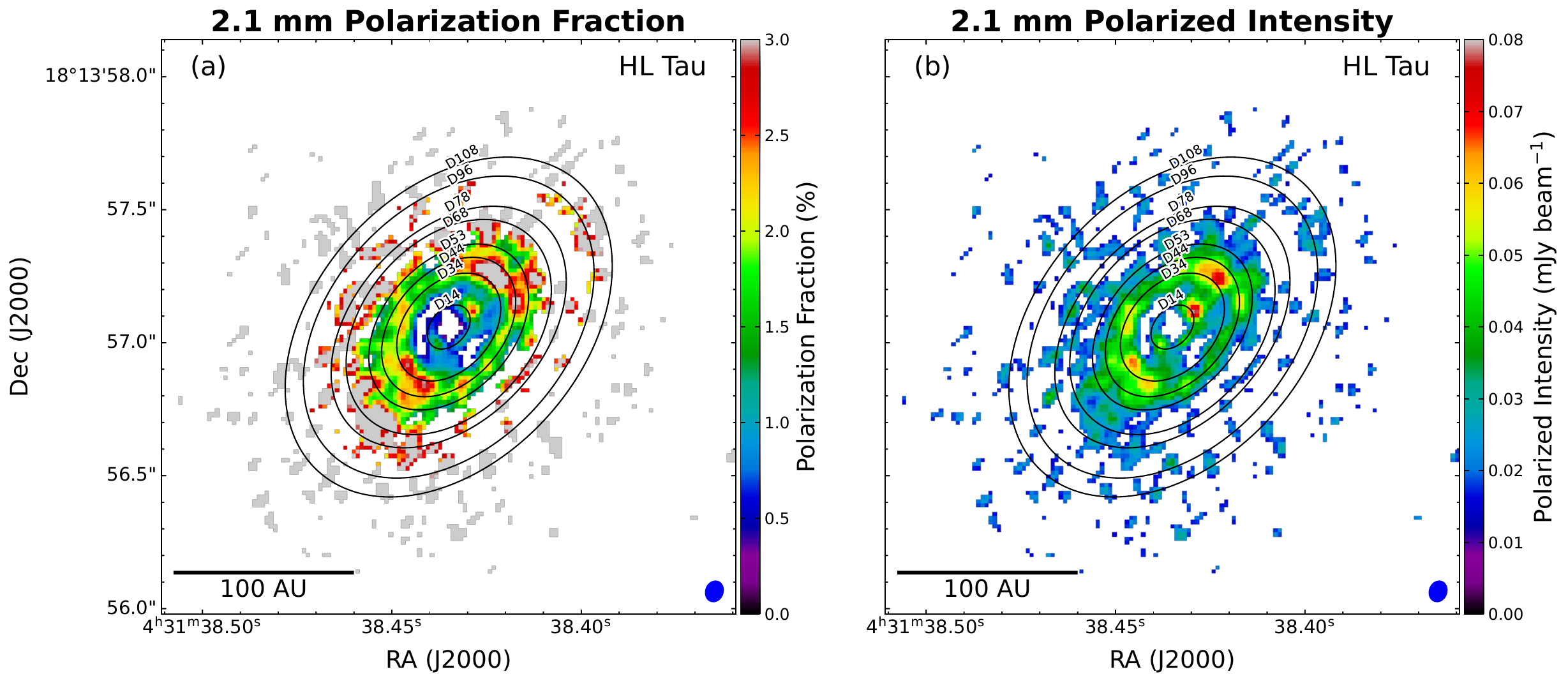}
    \caption{The polarization fraction map (a) and polarized intensity map (b) of HL~Tau, imaged with robust $=-0.5$. Pixels are masked out at locations where either Stokes $I$ or polarized intensity $P$ is below the $2\sigma$ level. Ellipses representing the gaps in the disk are overlaid as indicated in Figure~\ref{fig:B4StokesI}. The blue ellipse in the bottom right shows the synthesized beam.}
    \label{fig:B4rm0.5_pfrac&pintensity}
\end{figure}

\begin{figure}
    \includegraphics[width=1.0\linewidth]{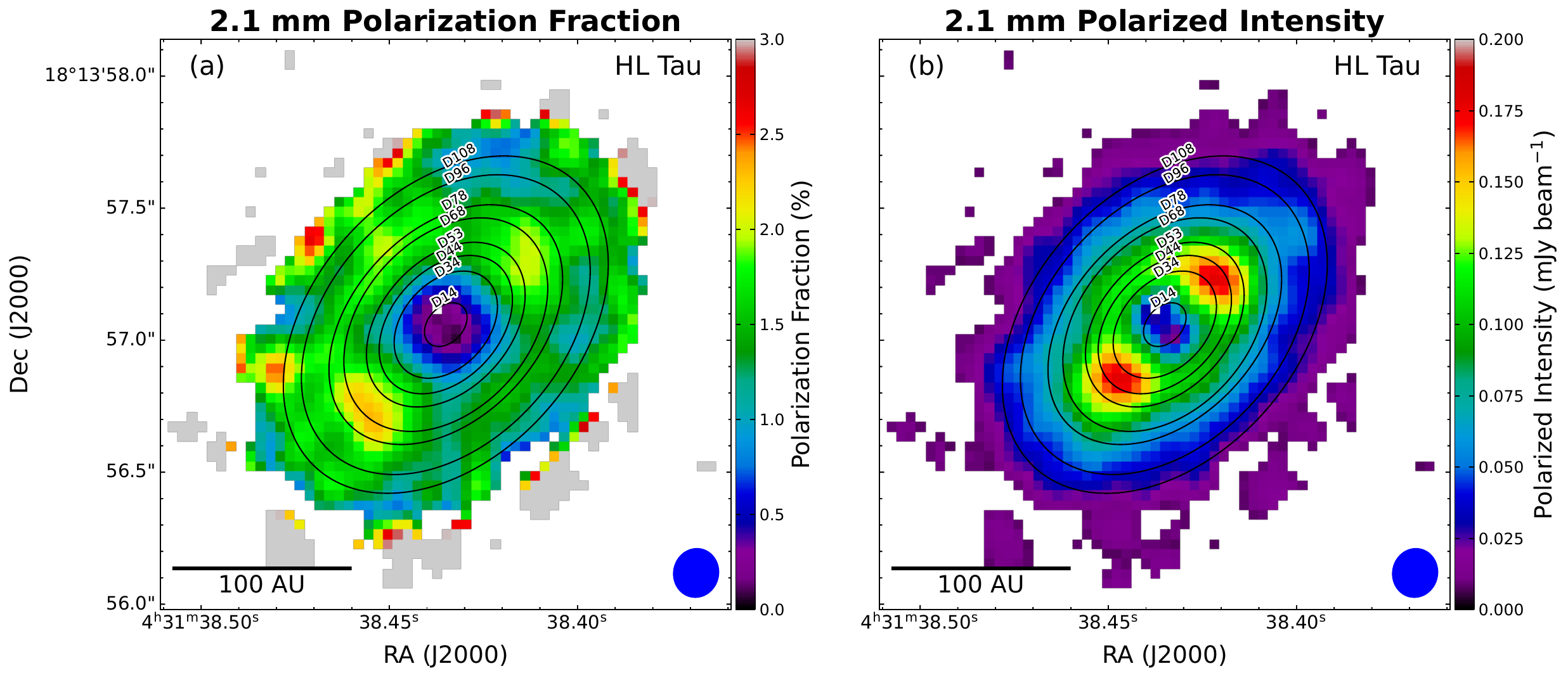}
    \caption{The polarization fraction map (a) and polarized intensity map (b) of HL~Tau, imaged with robust = 0.5. The plotting convention is the same as Figure~\ref{fig:B4rm0.5_pfrac&pintensity}.}
    \label{fig:B4r0.5_pfrac&pintensity}
\end{figure}

Figures~\ref{fig:B4rm0.5_pfrac&pintensity} and~\ref{fig:B4r0.5_pfrac&pintensity} show the polarization fraction and polarized intensity maps at robust weightings $-0.5$ and 0.5, respectively. In the robust $=-0.5$ polarization fraction map, we observe very slight enhancements in the innermost gap D14 along the major axis, and significant enhancements around gaps D34 and D44 across all azimuths, but mainly along the major axis. Similar enhancements are observed in the robust $=-0.5$ polarization intensity map, at the same location. These enhancements show an asymmetry, where the signal along the northwest is stronger than that along the southeast. We also see strong enhancements around D68 and D78 in polarization fraction, but they are hard to characterize under robust $=-0.5$ due to poor S/N. In the robust = 0.5 polarization fraction map, the enhancements around D14 and D34 are smoothed by the larger synthesized beam. Some enhancement is visible along the major axis in the bright belt that covers D44 and D53, and a more pronounced enhancement appears in the subsequent region corresponding to the dark belt that covers D68 and D78 in the Stokes $I$ emission. The polarization fraction shows asymmetries along both axes, with stronger emission on the southeast side along the major axis and the northeast side along the minor axis. In the robust = 0.5 polarized intensity map, the primary peak appears in the region covering D34 to D53 along the major axis, without clear distinction between rings and gaps. This robust = 0.5 data has resolution similar to the previous Band~4 data in \cite{Daniel_Multiband_2024}.
In both polarization fraction and polarized intensity maps, values along the major axis are significantly higher than those along the minor axis, which is consistent with the morphology observed in Band~7 \citep{Ian_Nature_2023}. 

\FloatBarrier
\section{LINEAR POLARIZATION MODELING} \label{sec:Linear_Polarization_Modeling}
We model the observed polarization using a linear decomposition approach following \cite{Daniel_model_2022}. The model assumes that the polarization is a simple addition of the scattering-induced polarization and polarization from aligned grains. This simplification is motivated by proper solutions to the plane-parallel slab of scattering of aligned grains. Plane-parallel slab solutions have been used to model HL~Tau \citep{Ian_Nature_2023} and are reasonable given the geometrically thin dust layer of the disk \citep{Pinte_2016_HLTau_od}. However, rather than solving the full radiative transfer at each radius, we utilize the linear decomposition model, which allows empirical constraints on the scattering and thermal emission contributions as a function of radius. 

With the assumption of an axisymmetric disk and geometrically thin dust midplane, from the azimuthal variation of polarization, the model decomposes polarization into contributions from two mechanisms: dust scattering ($s$) and thermal emission from aligned grains ($t$). As concluded by \cite{Ian_Nature_2023}, we adopt effectively prolate grains (see \cite{Haifeng_Band3_2019} for discussion), i.e., grains spinning about their long axes, with their long axes toroidally aligned as our grain alignment model. We also explore additional mechanisms including circular alignment pattern as predicted from radiative alignment \citep{Haifeng_Band3_2019} and a toroidal alignment pattern with an angular offset from the purely azimuthal direction (hereafter, deviated toroidal alignment).

\subsection{Model Formulations}
We describe the formulation of each model below. Section~\ref{sec:st_model_formulation} introduces the coordinate system and geometric definitions adopted by all models, along with the toroidal alignment model formulation. The deviated toroidal alignment model is presented in Section~\ref{sec:spiral_alignment_model}, and the circular alignment model is presented in Section~\ref{sec:circular_alignment_model}.

\subsubsection{Toroidal Alignment Model formulation} \label{sec:st_model_formulation}
\begin{figure}[h]
    \centering
    \begin{subfigure}{0.465\linewidth}
        \includegraphics[width=1.0\linewidth]{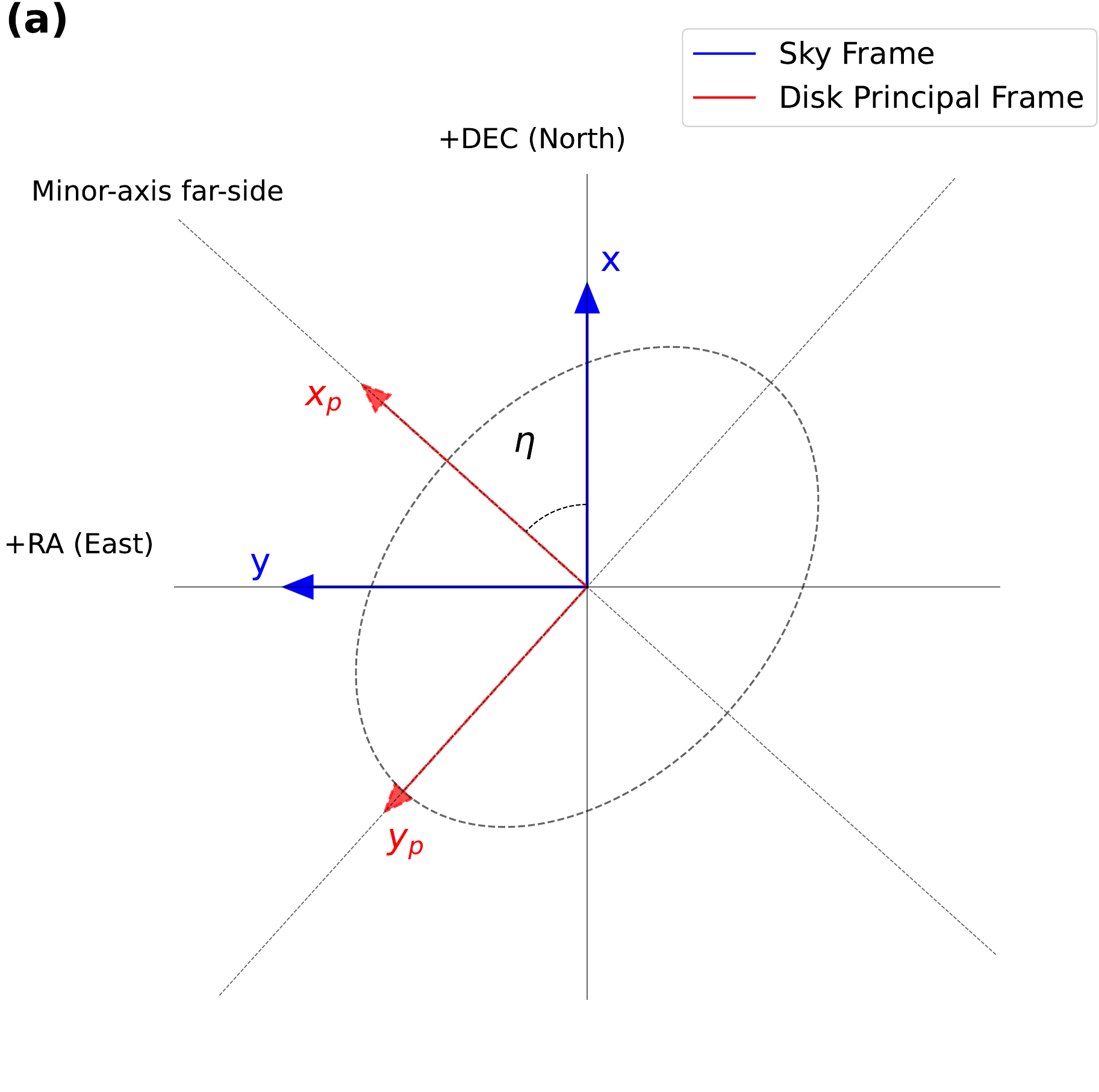}
        \phantomcaption
        \label{fig:principal_frame}
    \end{subfigure}
    \begin{subfigure}{0.525\linewidth}
        \includegraphics[width=1.0\linewidth]{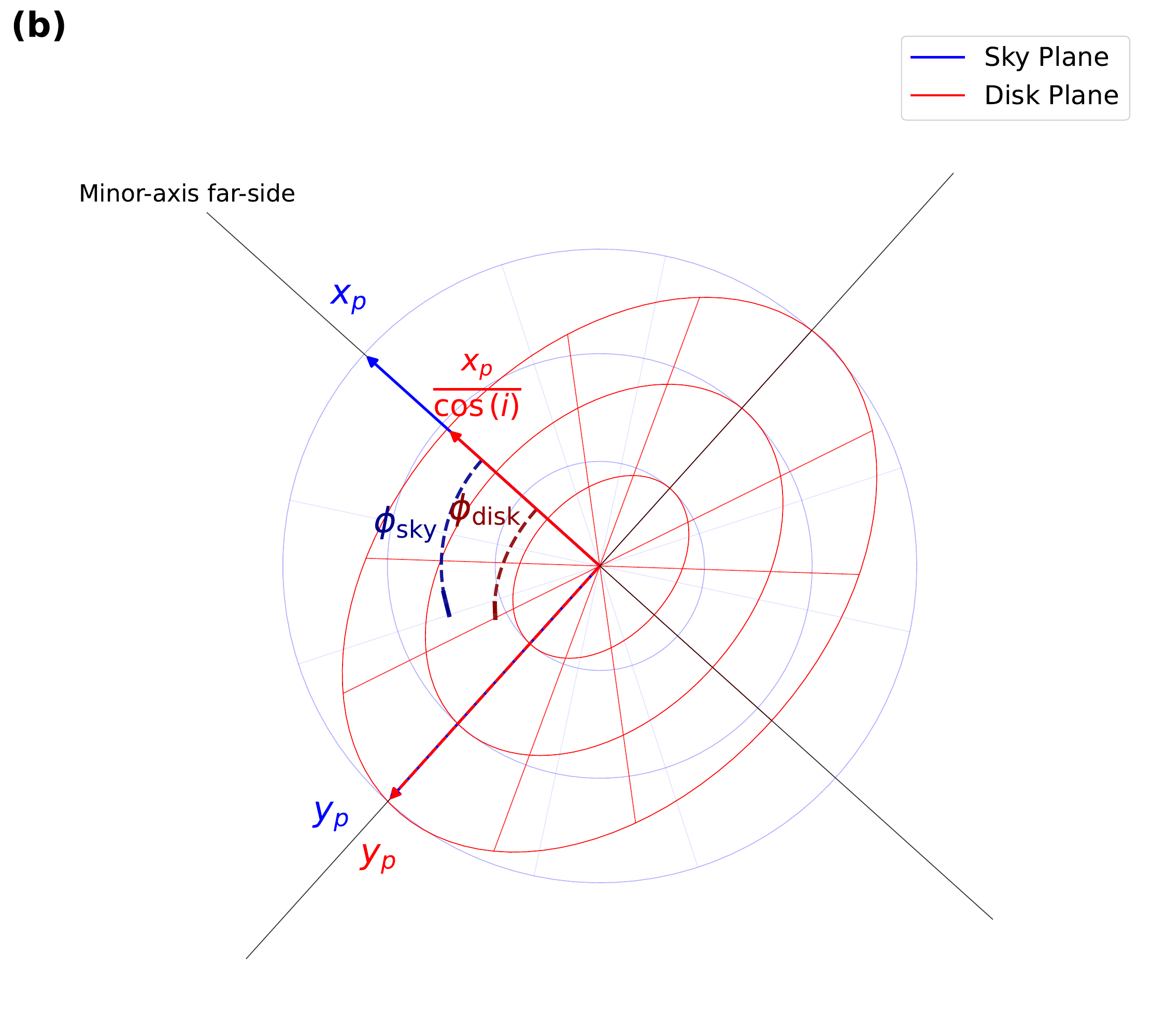}
        \phantomcaption
        \label{fig:disk_plane}
    \end{subfigure}
    \vspace{-40pt}
    \caption{Coordinate system definitions. Panel (a) shows the definition of the disk principal frame ($x_p, y_p$), which involves a pure 2D rotation from the sky frame ($y,-x$) by the position angle of the minor-axis far-side $\eta$. Panel (b) shows the definition of the disk plane, which involves a 3D rotation from the sky plane along the disk major axis by the inclination $i$. The azimuthal angle on the sky plane $\phi_{\rm{sky}}$ is uniformly distributed across the image, while the azimuthal angle on the disk plane $\phi_{\rm{disk}}$ is not uniform when projected on the image.}    
    \label{fig:geometry}
\end{figure}

We adopt the same coordinate system as \cite{Daniel_Multiband_2024} and work in the principal frame of the disk, which has coordinates $x_p$ and $y_p$, such that the $x_p$ axis is along the disk minor axis with positive direction pointing to the far-side (East-of-North in the observed sky frame) of the disk; and the $y_p$ axis is along the disk major axis, with positive direction $90^\circ$ (counterclockwise) from $+x_p$. In the original sky frame, the coordinates $x$ and $y$ correspond to $+\Delta\mathrm{DEC}$ and $+\Delta\mathrm{RA}$. Let $\eta$ be the position angle of the minor axis corresponding to the far-side of the disk in the sky frame. Then, the principal frame coordinates can be obtained through a rotation of the sky frame (Figure~\ref{fig:principal_frame}) by:
\begin{equation}
\begin{pmatrix}
x_p \\ y_p
\end{pmatrix}
=
\begin{pmatrix}
\cos{\eta} & \sin{\eta} \\
-\sin{\eta} & \cos{\eta}
\end{pmatrix}
\begin{pmatrix}
\Delta \mathrm{Dec} \\
\Delta \mathrm{RA}
\end{pmatrix}
=
\begin{pmatrix}
    \cos{\eta} & \sin{\eta} \\
    -\sin{\eta} & \cos{\eta}
\end{pmatrix}
\begin{pmatrix}
    x \\ y
\end{pmatrix}
\end{equation}
The Stokes parameters are defined based on the International Astronomical Union convention \citep{IAU_convention_1996}. We define Stokes $Q_p$ and $U_p$ in the disk principal frame coordinates $x_p$ and $y_p$. Let $\phi$ be the angle in the image plane measured counterclockwise from the positive $x_p$-axis in the principal frame. Positive $Q_p$ corresponds to polarization oriented along the $x_p$-axis ($\phi = 0^\circ$ or $180^\circ$), while positive $U_p$ corresponds to polarization oriented along the bisector of the positive $x_p$ and $y_p$ axes ($\phi = 45^\circ$ or $225^\circ$). We can obtain $Q_p$ and $U_p$ from the originally observed Stokes $Q$ and $U$ in the sky frame by:
\begin{equation}
    \begin{pmatrix}
        Q_p \\ U_p
    \end{pmatrix}
    =
    \begin{pmatrix}
        \cos{2\eta} & \sin{2\eta} \\
        -\sin{2\eta} & \cos{2\eta}
    \end{pmatrix}
    \begin{pmatrix}
        Q \\ U
    \end{pmatrix}
\end{equation}

We adopt the geometric parameters of the disk from \cite{ALMA_Partner_2015}. The disk inclination is $i = 46.7^\circ$ ($i$ = 0 corresponds to face-on), and the position angle of the disk major axis is $138.02^\circ$ measured east of north. \cite{ALMA_Partner_2015} and \cite{Yen_2017_outflow} detected blueshifted outflow to the northeast of HL~Tau, and redshifted outflow to the southwest, both of which are perpendicular to the disk. Thus, the northeast of the disk is the far-side, and we determine the position angle of the disk minor axis on the far-side to be $\eta = 48.02^\circ$.

We construct two different azimuthal angles in the principal frame. The first is the sky-plane azimuthal angle: 
\begin{equation} \label{eq:phi_sky}
    \phi_{\rm{sky}}=\operatorname{atan2}(y_p, x_p)
\end{equation}
which represents the azimuthal angle measured directly in the sky plane. $\phi_{\rm{sky}}$ is measured counterclockwise from the positive $x_p$-axis, which corresponds to the disk minor axis on the far-side. $\phi_{\rm{sky}}$ is uniformly distributed over the observed image. $\operatorname{atan2}(y, x)$ denotes the two-argument arctangent, i.e., the polar angle of the point $(x, y)$, with values in $(-180^\circ, 180^\circ]$. Throughout the paper, azimuthal angles are quoted in the equivalent range $[0^\circ, 360^\circ)$.

The second is the disk-plane azimuthal angle:
\begin{equation} \label{eq:phi_ell}
    \phi_{\rm{disk}}=\operatorname{atan2}\left(y_p,\frac{x_p}{\cos{(i)}}\right)
\end{equation}
which represents the true azimuthal angle in the physical disk plane, accounting for the projection effect along the line of sight due to inclination. $\phi_{\rm{disk}}$ is also defined to increase counterclockwise from the positive $x_p$-axis, but it is not uniformly distributed in the observed image because of the projection (Figure~\ref{fig:disk_plane}). In the face-on disk case ($i=0^\circ$), $\phi_{\rm{sky}} = \phi_{\rm{disk}}$.

With these definitions, we are able to construct our model to linearly decompose the contributions from the two polarization mechanisms. Specifically, we model the normalized Stokes parameters in the principal frame $q_p = \frac{Q_p}{I}$ and $u_p=\frac{U_p}{I}$, with two components, $s$ and $t$, corresponding to the polarization fraction contributed by scattering and thermal emission from aligned grains, respectively. For the scattering component $s$, the inclination-induced polarization direction is parallel to the disk minor axis \citep{Haifeng_scattering_2016}. Since the positive $Q_p$ axis in the principal frame is defined along the minor axis, the scattering contribution to $q_p$ is a constant, while its contribution to $u_p$ is zero:
\begin{align}
    q_s &= s \\
    u_s &= 0
\end{align}
For the thermal emission component $t$ from toroidally aligned prolate grains, the alignment configuration causes grains on the disk major axis to produce polarization parallel to the minor axis, while grains on the minor axis produce polarization parallel to the major axis that partially cancels the scattering contribution. At intermediate azimuthal positions, the thermal component contributes to both $q_p$ and $u_p$. We adopt the formulation of the elliptical polarization pattern produced by thermal emission from toroidally aligned prolate grains, as derived in \cite{Daniel_model_2022,Daniel_Multiband_2024}:
\begin{align}
    q_t &= t \cdot G_q \\
    u_t &= t \cdot G_u
\end{align}
where the geometric functions $G_q$ and $G_u$ are defined as
\begin{align}
    G_q &= \cos^2{(i)} \sin^2{(\phi_{\rm{disk}})} - \cos^2{(\phi_{\rm{disk}}}) \\
    G_u &= -\cos{(i)} \sin{(2\phi_{\rm{disk}})}
\end{align}
Adding two components together, we are able to construct:
\begin{align} \label{toroidal_model}
    q_p &= s + t[\cos^2{(i)} \sin^2({\phi_{\rm{disk}})} - \cos^2{(\phi_{\rm{disk}})}] \\
    u_p &= -t[\cos{(i)} \sin{(2\phi_{\rm{disk}})}]
\end{align}

\subsubsection{Deviated Toroidal Alignment Model Formulation} \label{sec:spiral_alignment_model}
The aforementioned toroidal alignment model assumes prolate grains are perfectly aligned in the azimuthal direction. However, observations have revealed polarization patterns in some disks that deviate from the azimuthal direction by a few degrees (e.g., AS~209, \citealt{Mori_2019_AS209}; GG~Tau~A, \citealt{Tang_2023_GGTau}), see Section~\ref{sec:spiral_mechanism_discussion} for detailed discussion on the physical origin of such deviation. To account for this effect, following the framework in \cite{Daniel_model_2022}, we extended the toroidal alignment model by introducing an angular offset $\Delta\phi$ to the disk-plane azimuthal angle $\phi_{\rm{disk}}$ in the geometric functions for the thermal component:
\begin{align}
    G_q &= \cos^2{(i)} \sin^2{(\phi_{\rm{disk}}+\Delta\phi)} - \cos^2{(\phi_{\rm{disk}}+\Delta\phi)} \\
    G_u &= -\cos{(i)} \sin{(2(\phi_{\rm{disk}}+\Delta\phi))}
\end{align}
A positive $\Delta\phi$ corresponds to a counterclockwise rotation of the grain alignment direction from purely azimuthal direction in the disk plane. The physical interpretation of the sign of $\Delta\phi$ is discussed in Section~\ref{sec:spiral_mechanism_discussion}.
Polarization due to scattering is still assumed to be a constant component contributing to Stokes $Q_p$. Thus, the deviated toroidal alignment model has:
\begin{align} 
    q_p &= s + t[\cos^2{(i)} \sin^2{(\phi_{\rm{disk}}+\Delta\phi)} - \cos^2{(\phi_{\rm{disk}}+\Delta\phi)}] \label{eq:spiral_alignment_qp}\\
    u_p &= -t[\cos{(i)} \sin{(2(\phi_{\rm{disk}}+\Delta\phi))}] \label{eq:spiral_alignment_up}
\end{align}
The angular offset $\Delta\phi$ can be treated as a global parameter shared by all annuli, or as a per-annulus parameter (same as $s$ and $t$) that allows radial variation of the deviation angle. 

\subsubsection{Circular Alignment Model Formulation} \label{sec:circular_alignment_model}
In the case of radiative alignment, the magnetic field is weak and radiation dominates the grain precession, such that the short-axis of the dust grain would be aligned with the radiation direction \citep{Lazarian_2007_RAT, Tazaki_2017_RAT}. For a disk where the radiation flux is predominantly radial, the grains are effectively oblate, with their shortest (alignment) axes oriented radially in the disk plane \citep{Haifeng_Band3_2019}. The thermal emission from such grains produces a circular polarization vector morphology on the sky plane, distinct from the elliptical pattern produced by toroidally aligned prolate grains.

To investigate the potential existence of radiatively aligned oblate grains and model the polarization in this scenario, we follow the formalism of \cite{Haifeng_Band3_2019}. The polarization fraction of an effectively oblate grain depends on the angle $i_d$ between its shortest (alignment) axis and the line of sight:
\begin{equation} \label{eq:pid}
    p(i_d) \approx p_0 \sin^2{(i_d)}
\end{equation}
where $p_0$ is the intrinsic polarization fraction. For radiatively aligned grains in an inclined disk, the alignment axis at each location is the radial direction in the disk plane, whose inclination to the line of sight varies with azimuth:
\begin{equation} \label{eq:cosid}
    \cos{(i_d)} = \sin{(i)}\cos{(\phi_{\rm{disk}})}
\end{equation}
where $i$ is the disk inclination and $\phi_{\rm{disk}}$ is the disk-plane azimuthal angle. On the major axis ($\phi_{\rm{disk}}=90^\circ$ and $270^\circ$), the radial direction is perpendicular to the line of sight ($i_d = 90^\circ$), yielding the maximum polarization fraction. On the minor axis ($\phi_{\rm{disk}}=0^\circ$ and $180^\circ$), the radial direction is tilted toward the observer, and polarization fraction is minimized. This azimuthal dependence is opposite to that of toroidally aligned prolate grains.

We define the polarization component from circularly aligned oblate grains as $c$, similar to $t$ in the toroidal alignment model. Both $c$ and $t$ describe the amplitude of polarized thermal emission from aligned grains, and we use different symbols solely to distinguish between the circular and elliptical polarization patterns they produce in different models. The scattering-induced polarization $s$ and the disk geometry follow the same definitions as previous models.

Substituting Eq.~\ref{eq:cosid} into Eq.~\ref{eq:pid}, we obtain the azimuthal variation of the polarization fraction contributed by the thermal emission as a function of disk-plane azimuthal angle:
\begin{equation} \label{eq:circular_p_azimuthal_var}
    p(\phi_{\rm{disk}}) = c \cdot (1-\sin^2{(i)}\cos^2{(\phi_{\rm{disk}})})
\end{equation}
The polarization direction from radiatively aligned oblate grains is perpendicular to the projection of the radial alignment axis on the sky plane, and traces a circular pattern. 
Decomposing this into fractional Stokes parameters in the principal frame, the contribution to $q_p$ and $u_p$ from a polarization fraction of amplitude $p$ at sky-plane angle $\phi_{\rm{sky}}$ is:
\begin{align}
    q_c &= p \cdot (-\cos{(2\phi_{\rm{sky}})}) \\
    u_c &= p \cdot (-\sin{(2\phi_{\rm{sky}})})
\end{align}
Combining the azimuthal variation in Eq.~\ref{eq:circular_p_azimuthal_var}, we define the geometric functions $C_q$ and $C_u$ as:
\begin{align}
    C_q &=  -(1-\sin^2{(i)}\cos^2{(\phi_{\rm{disk}})}) \cdot \cos{(2\phi_{\rm{sky}})}\\
    C_u &=  -(1-\sin^2{(i)}\cos^2{(\phi_{\rm{disk}})}) \cdot \sin{(2\phi_{\rm{sky}})}
\end{align}
such that $q_c = c \cdot C_q$ and $u_c = c \cdot C_u$. Still, scattering-induced polarization $s$ is added as a constant component to $q_p$. Thus, the full circular alignment model is:
\begin{align} 
    q_p &= s - c \cdot[(1-\sin^2{(i)}\cos^2{(\phi_{\rm{disk}})}) \cdot \cos{(2\phi_{\rm{sky}})}]\\
    u_p &= -c \cdot [(1-\sin^2{(i)}\cos^2{(\phi_{\rm{disk}})}) \cdot \sin{(2\phi_{\rm{sky}})}]
\end{align}

\subsection{Fitting Procedure} \label{sec:fit_procedure}
To fit the models to the observed HL~Tau polarization, we divide the disk into several elliptical annular regions, where the radial width of each annulus is determined based on the geometric mean of the FWHM major and minor axes of the synthesized beam for the corresponding data. We use half a beam as the annulus spacing (Nyquist sampling) to retain sufficient radial sampling. Since different robust weightings produce different beam sizes, the annulus widths differ accordingly. The outer boundary and width of a representative annulus are illustrated by the green ellipses shown in the result figures in Section~\ref{sec:B4_Modeling_Results}. For each annulus, we assign a set of parameters, such as $(s,t)$ for the toroidal alignment model, representing the amplitude of the scattering and thermal emission contributions to the polarization fraction, respectively. We assume that $s$ and $t$ are constant within each annulus over a given radius range. This assumption is motivated by the expectation that the polarization fraction amplitudes from both scattering and thermal emission are governed by the local optical depth and grain properties, which are expected to vary primarily with radius rather than azimuth in an approximately axisymmetric disk such as HL~Tau.

We use \texttt{emcee}, a Markov Chain Monte Carlo (MCMC) based sampling package \citep{emcee_2013} to constrain the model parameters for each elliptical annulus. We construct an equilateral triangular sampling grid (distributed on the sky-plane) with a spacing of half the beam FWHM size between neighboring points, satisfying the Nyquist sampling. 

We note that the annulus boundaries are determined solely by the beam size rather than designed to coincide with the known ring and gap locations. When a substructure falls near the boundary between two adjacent annuli, neither annulus captures its full contribution, and the fitted $s$ and $t$ values in either annulus may not be accurately representative of the polarization properties at that substructure. These effects can introduce slight radial offsets between the fitted parameter peaks and the nominal substructure locations.

For each trial set of parameters (e.g., $(s,t)$ for the toroidal alignment model, or $(s,t,\Delta\phi)$ for the deviated toroidal alignment model), we construct the fractional Stokes parameter models $q_{p,\rm mod}$ and $u_{p,\rm mod}$ following the formulation described in Section~\ref{sec:st_model_formulation}. These are then multiplied by a Stokes $I$ model, $I_{\rm{mod}}$ to obtain the model of Stokes parameters $Q_{p,\rm mod}$ and $U_{p,\rm mod}$, which are subsequently convolved with the same synthesized beam as the corresponding observations. The convolved models are then sampled at the grid points and compared directly with the observed Stokes $Q_p$ and $U_p$ data to compute the likelihood. 
To strictly account for the beam convolution step, this procedure requires constructing an infinitely resolved model of Stokes $I$ in principle, which would then be multiplied by $q_{p,\rm mod}$ and $u_{p,\rm mod}$ before convolution. However, the complexity of ring and gap structures of HL~Tau makes such a treatment difficult and beyond the scope of this paper. As a practical approach, we directly adopt the Stokes $I$ image of the corresponding robust as a proxy for $I_{\rm mod}$.

The convolved model Stokes parameters are then sampled at the grid points and compared directly with the observed Stokes $Q_p$ and $U_p$ data through a Gaussian log-likelihood function:
\begin{equation} \label{eq:log_likelihood}
    \ln \mathcal{L} = -\frac{1}{2} \left[ \sum_i \left(\frac{Q_{p,i} - Q_{p,\text{mod},i}}{\sigma_Q}\right)^2 + \sum_i \left(\frac{U_{p,i} - U_{p,\text{mod},i}}{\sigma_U}\right)^2 \right]
\end{equation}
where the sum is over all sampling grid points within each annulus, and $\sigma_Q$ and $\sigma_U$ are the rms noise levels of the Stokes $Q_p$ and $U_p$ data, respectively. 

The best-fit parameters for each annulus are determined from the posterior median of the MCMC samples, with uncertainties estimated from the standard deviation of the posterior distribution. These parameters represent the intrinsic (pre-convolution) polarization amplitudes at each radius. 

\section{BAND 4 MODELING RESULTS} \label{sec:B4_Modeling_Results}
For each model, we assess (1) data–model agreement and residuals, (2) the radial profiles of the best-fit parameters, and (3) the overall goodness of fit via the reduced $\chi^2$. All models are fit to the data processed with both robust $=-0.5$ and robust=0.5. 
The robust $=-0.5$ images provide the highest angular resolution and are used to examine the radial variations of the inferred polarization mechanisms across ring and gap substructures. The robust=0.5 images offer a higher signal-to-noise ratio, which better reveals larger-scale polarization structures and disk-wide trends. 

We present the reduced $\chi^2$ values for each model in Table~\ref{tab:chi2_band4} for reference. We note that in robust $=-0.5$ $U_p$, the per-annulus $\Delta\phi$ model yields a marginally higher (worse) reduced $\chi^2$ than the global $\Delta\phi$ model despite having more free parameters. This is because the additional parameters are poorly constrained at this signal-to-noise level and provide negligible improvement in the total $\chi^2$ while reducing the degrees of freedom.

\begin{table}[h]
\centering
\caption{Reduced $\chi^2$ values for different models and robust weightings for Band~4 observations. For each entry, values are shown for $Q_p$, $U_p$, and combined fit.}
\label{tab:chi2_band4}
\begin{tabular}{lccccc}

\hline\hline
\rule{0pt}{5.5ex} Robust & & Toroidal & \shortstack{Deviated Toroidal\\(global $\Delta\phi$)} & \shortstack{Deviated Toroidal\\(per-annulus $\Delta\phi$)} & Circular \\[1.0ex]
\hline
\multirow{3}{*}{$-0.5$} & $Q_p$ & 1.27 & 1.25 & 1.25 & 1.69 \\
                        & $U_p$ & 1.03 & 1.01 & 1.02 & 1.06 \\
                        & Combined & 1.13 & 1.11 & 1.10 & 1.35 \\
\hline
\multirow{3}{*}{$0.5$}  & $Q_p$ & 9.76 & 5.87 & 5.54 & 27.85 \\ 
                        & $U_p$ & 6.63 & 4.80 & 4.78 & 6.60 \\
                        & Combined & 7.85 & 5.10 & 4.83 & 16.11 \\
\hline
\end{tabular}
\end{table}

\subsection{Toroidal Alignment Model Results} \label{sec:B4_st_result}
\begin{figure}[h]
    \centering
    \begin{subfigure}{\linewidth} 
        \centering
        \includegraphics[width=1.0\linewidth]{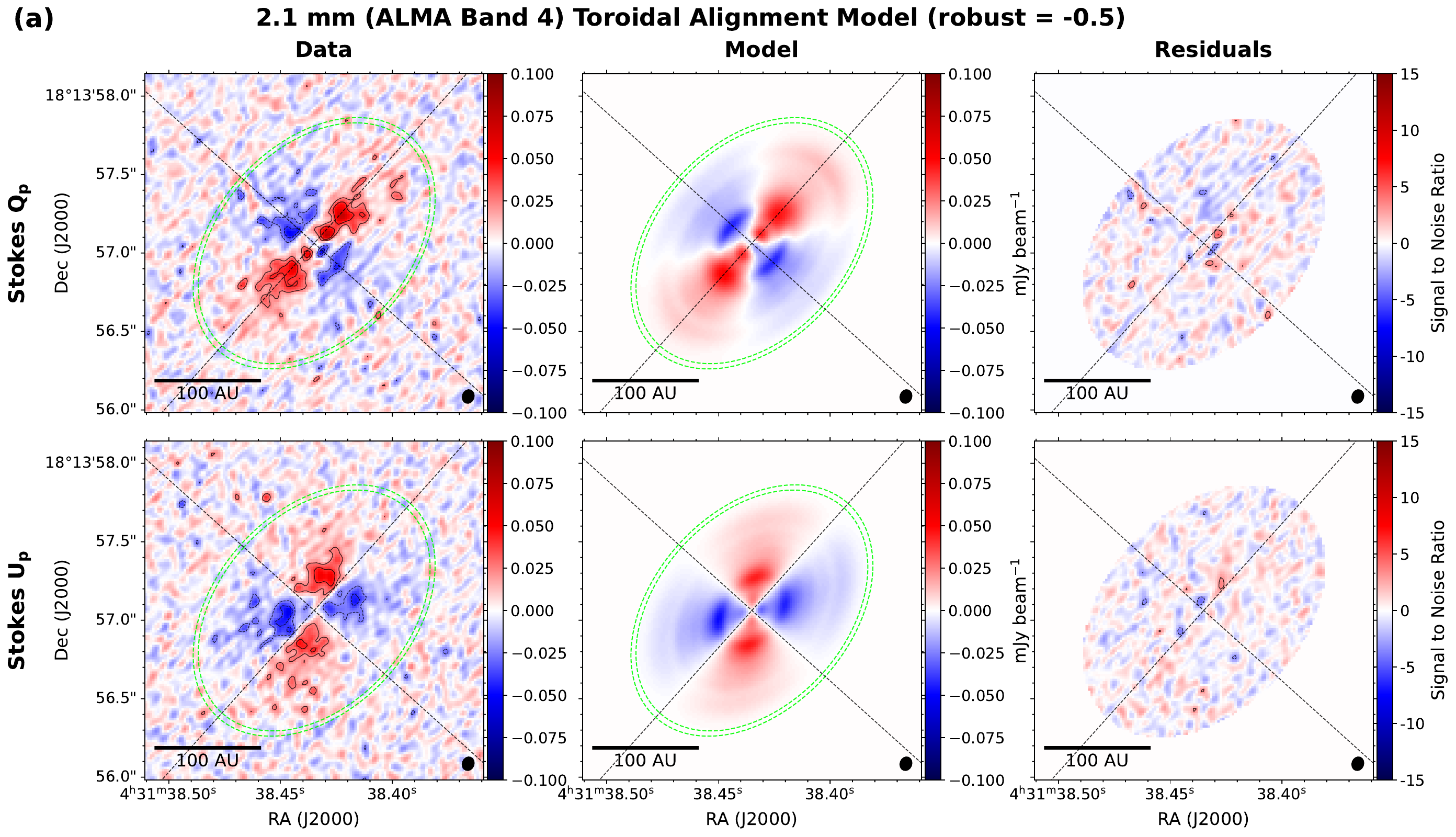}
        \phantomcaption
        \label{fig:B4rm0.5st_result}
    \end{subfigure}

    \begin{subfigure}{\linewidth} 
        \centering
        \includegraphics[width=1.0\linewidth]{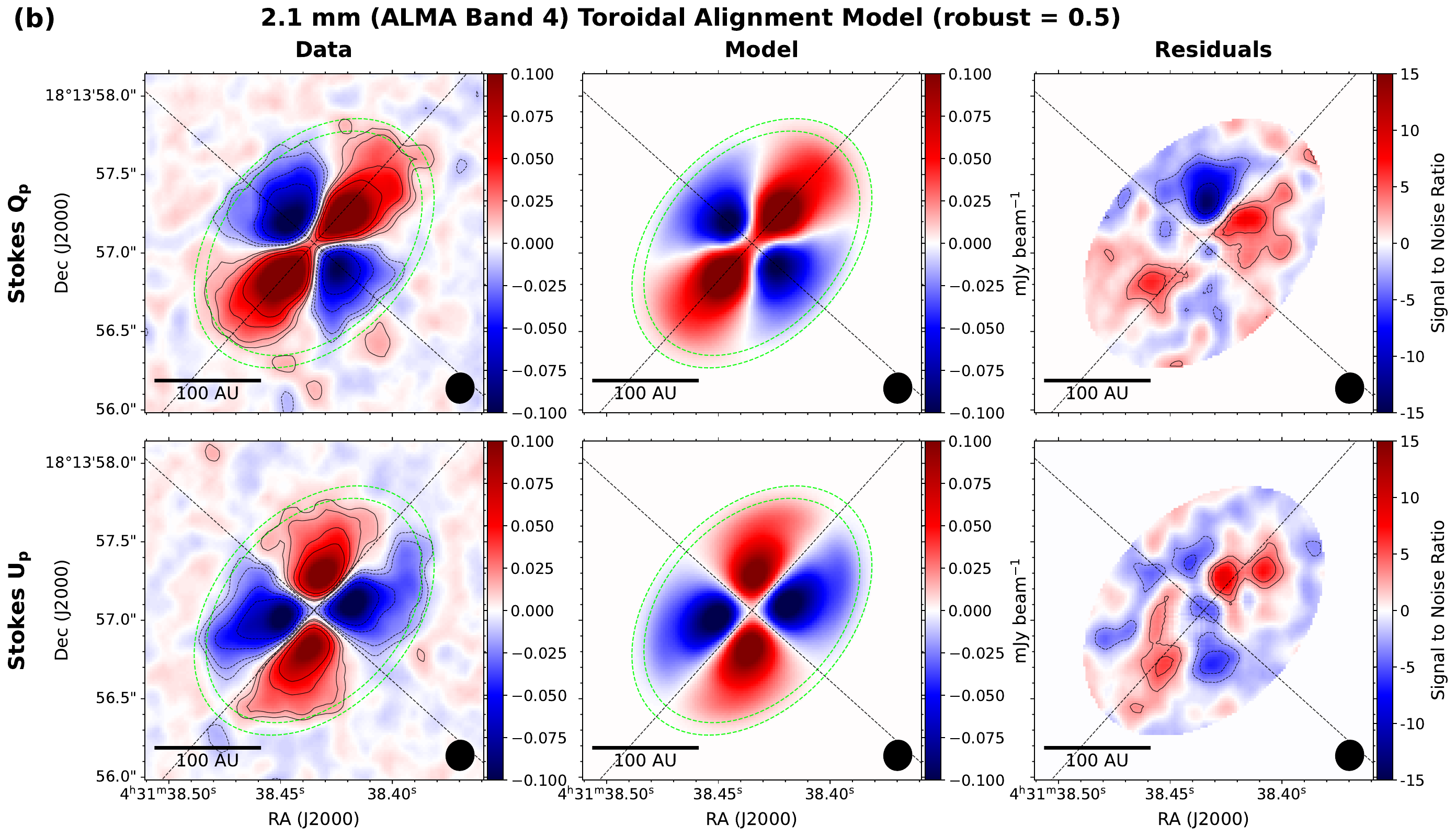}
        \phantomcaption
        \label{fig:B4r0.5st_result}
    \end{subfigure}

    \caption{Results of the toroidal alignment model fitting for robust $=-0.5$ (a) and robust=0.5 (b). For each robust weighting, the top and bottom rows correspond to Stokes $Q_p$ and $U_p$, respectively. From left to right, each column shows the observed data, the model, and the corresponding residuals. The green elliptical annulus indicates the outer boundary of the fitted region, as well as the individual annulus width adopted in the fitting as explained in Section~\ref{sec:fit_procedure}. The black dashed lines mark the disk major and minor axes. In the residuals panels, the color scale shows the signal-to-noise ratio, where $\sigma$ is the background rms of the corresponding Stokes $Q_p$ and $U_p$ data for each robust weighting. Contours in the data and residuals panels are plotted at $[\pm 3, \pm 5, \pm 10, \pm 15] \sigma$.}
    \label{fig:B4st_results}
\end{figure}

\begin{figure}[h]
    \centering
    \begin{subfigure}{0.495\linewidth}
        \includegraphics[width=1.0\linewidth]{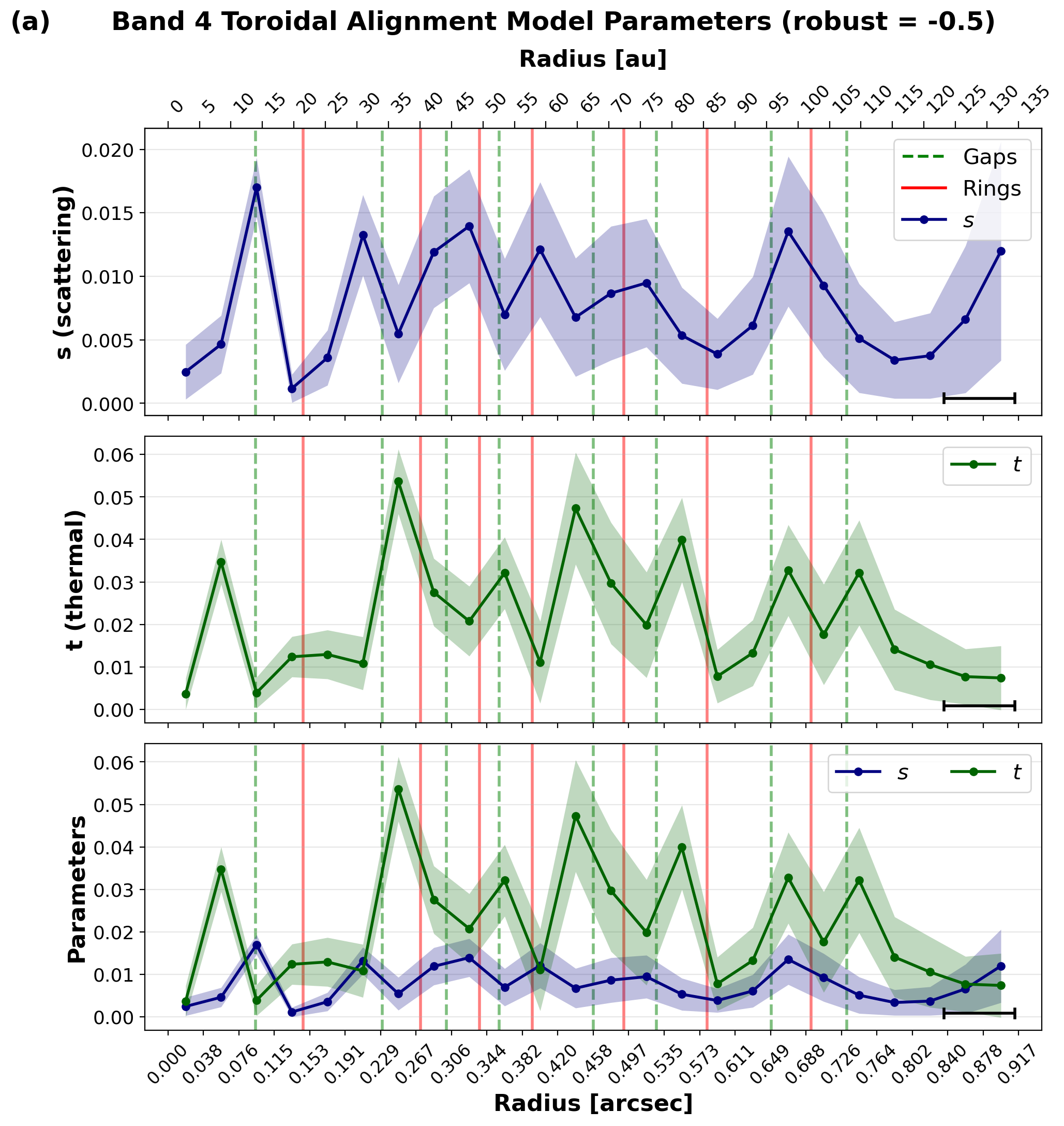}
        \phantomcaption
        \label{fig:B4rm0.5_st_st_profile}
    \end{subfigure}
    \begin{subfigure}{0.49\linewidth}
        \includegraphics[width=1.0\linewidth]{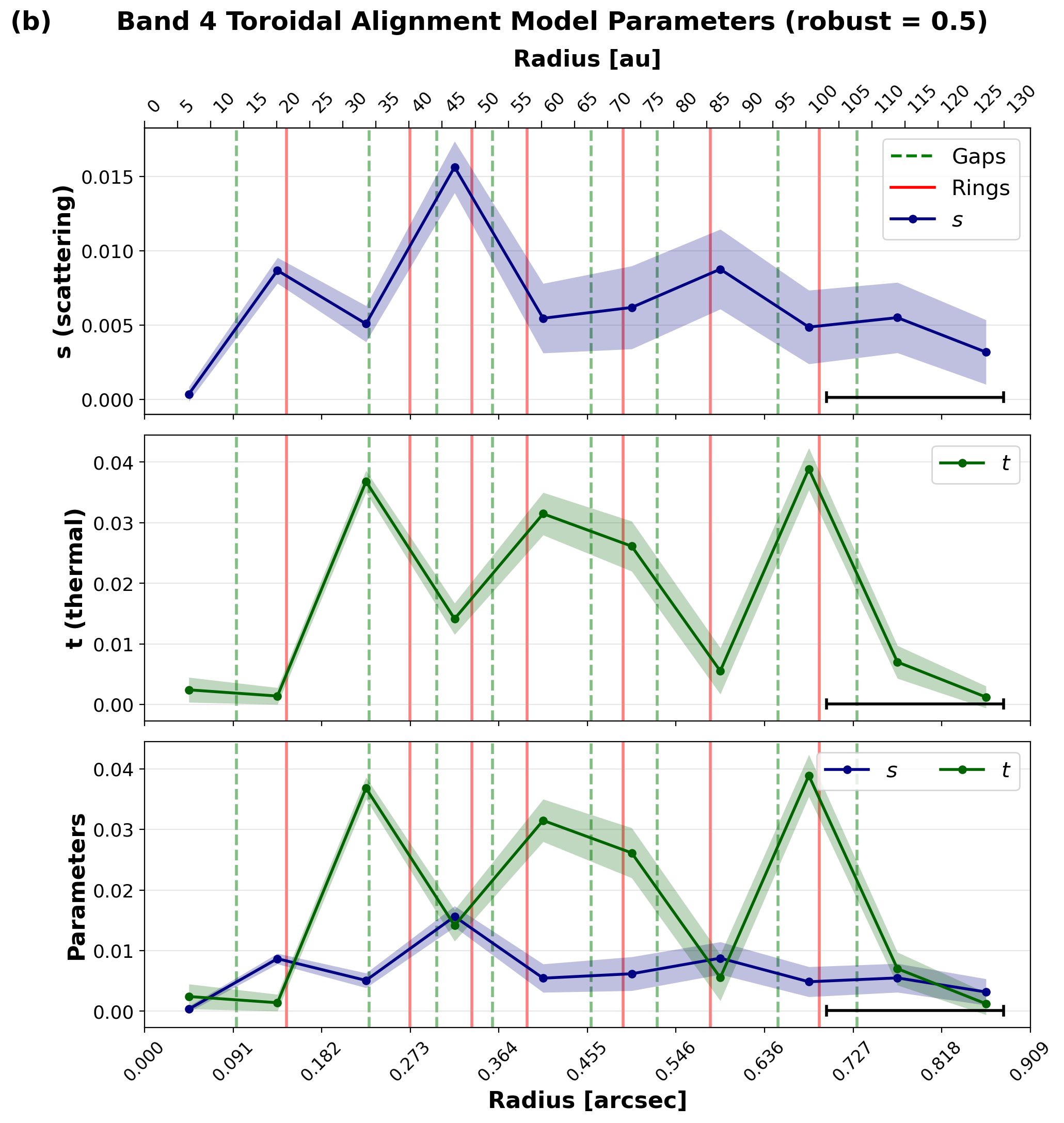}
        \phantomcaption
        \label{fig:B4r0.5_st_st_profile}
    \end{subfigure}
    
    \caption{Radial profiles of the parameters of the toroidal alignment model fitting for robust $=-0.5$ (a) and robust = 0.5 (b). The vertical red (solid) line and green (dashed) lines indicate the locations of rings and gaps, respectively, identified by \cite{ALMA_Partner_2015} and \cite{Ian_Nature_2023}. The ticks on the horizontal axis denote the radial boundaries of the individual elliptical annuli used in the fitting, measured along the disk major axis. The horizontal bar in the lower right corner of each panel indicates the angular resolution of the corresponding data, i.e., the geometric mean of the FWHM major and minor axes of the synthesized beam (Section~\ref{sec:observations}). The bottom row shows both parameters plotted on the same scale for direct comparison. Shaded regions represent 1$\sigma$ (posterior standard deviation).}
    \label{fig:B4_st_params}
\end{figure}

We present the fitting results for the toroidal alignment model, including the data-model comparison and residual (data minus model) maps (Figure~\ref{fig:B4st_results}), followed by the best-fit parameters ($s$ and $t$) radial profiles for both robust weightings (Figure~\ref{fig:B4_st_params}). 

The toroidal alignment model achieves an overall good fit to the Band~4 robust $=-0.5$ data as shown in Figure~\ref{fig:B4rm0.5st_result}, with reduced $\chi^2 \sim 1.13$ of the combined fit. The residuals in both Stokes $Q_p$ and $U_p$ are largely consistent with noise, with no significant coherent structures above the $3\sigma$ level. However, this should be interpreted in the context of the high angular resolution but relatively low signal-to-noise ratio of the robust $=-0.5$ data, which limits sensitivity to extended, low-contrast residual structures.

In contrast, the robust = 0.5 data reveal significant structured residuals, as shown in Figure~\ref{fig:B4r0.5st_result}. In Stokes $Q_p$, a prominent region of negative residuals appears on the northern side of the disk, reaching peak amplitudes of $\sim10\sigma$, while the southern side also exhibits negative residuals, but with much weaker amplitudes that largely remain below the $3\sigma$ level, showing a strong north--south asymmetry. On the eastern and western sides, extended regions of positive residuals are present at the $3\sigma$ to $5\sigma$ level and also show some asymmetry. These asymmetries in Stokes $Q_p$ are consistent with asymmetries previously identified in the polarization fraction, along both axes of the disk (see Section~\ref{sec:B4_I_pfrac_pi} and Figure~\ref{fig:B4r0.5_pfrac&pintensity}). Similar near-far side asymmetries in Stokes $Q_p$ have also been discussed in \cite{Ian_Nature_2023} and \cite{Daniel_Multiband_2024}. We note that our model assumes an azimuthally constant scattering component, which does not account for potential near-far side asymmetries in the scattering contribution.

The Stokes $U_p$ residuals for the robust = 0.5 fits show more complex spatial structures. Positive residuals are located along the major axis, while negative residuals appear along an axis that is slightly rotated clockwise with respect to the minor axis. Additional residuals are scattered throughout the disk away from the principal axes, showing radial variations and amplitudes typically around $3\sigma$ to $5\sigma$. Despite this complexity, we notice that the overall distribution of Stokes U residuals exhibits an approximate centrosymmetric ($180^\circ$) pattern. A similar centrosymmetric arrangement is also present in Stokes $Q_p$ residuals if we only focus on the sign distribution and ignore the asymmetric amplitudes between opposite quadrants, where positive and negative residuals occupy alternating quadrants defined by the disk major and minor axes.

The radial profiles of the $s$ and $t$ contributions to the polarization fraction exhibit radial variations that correlate with the disk substructures in robust $=-0.5$ results (Figure~\ref{fig:B4rm0.5_st_st_profile}). Such variations are expected, because both amplitudes depend on the local optical depth and grain properties (Section~\ref{sec:fit_procedure}), which differ between rings and gaps. We discuss them in Section~\ref{sec:P_mechanism_discussion}. In contrast, robust = 0.5 data have a significantly larger synthesized beam, and accordingly adopt wider annuli in the modeling. As a result, radial structures that are resolved in the robust $=-0.5$ modeling are smoothed out in the robust = 0.5 modeling, and the peak values of polarization fraction are reduced (Figure~\ref{fig:B4r0.5_st_st_profile}). The description below focuses on the higher angular resolution robust $=-0.5$ results. 

As shown in Figure~\ref{fig:B4rm0.5_st_st_profile}, the thermal emission component $t$ starts from a low value in the innermost disk, rises to a high peak before the first prominent gap D14, then drops to a distinct trough at D14. It rises again to form a broad bump at the bright ring around 20 au, followed by a high peak slightly exterior to the next major gap D34, reaching $\sim 5.4\%$. Beyond the ring near 40 au, $t$ decreases and forms a trough spanning the bright belt that contains D44, then rises to another peak at D53. The profile then exhibits a deep trough at the ring around 58 au, before reaching another high peak of $\sim 4.7\%$ interior to the prominent gap D68. Through the subsequent alternation of rings and gaps, $t$ exhibits repeated troughs and peaks accordingly, and decreases steadily in the outer disk beyond D108 where there is a lack of identified substructures.

The scattering component $s$ exhibits a different radial behavior. It reaches a sharp peak up to $\sim1.7\%$ at the innermost prominent gap D14. Across the intermediate disk regions interior to the bright ring near $\sim$ 86 au, $s$ displays a complementary profile to $t$, with peak corresponding to troughs in $t$ and vice versa. Beyond the 86 au ring, $s$ drops to a broad trough, then rises to another peak exterior to D96, followed by a decline and a subsequent rise in the outer disk where there is a lack of identified substructures.

\subsection{Deviated Toroidal Alignment Model Results} \label{sec:B4_st_dphi_result}
The deviated toroidal alignment model extends the toroidal alignment model by introducing an extra angular offset $\Delta \phi$ at each annulus, which allows the grain alignment direction to deviate from the purely azimuthal direction. This results in a rotation of the thermal-emission polarization pattern associated with aligned grains at each radius. As described in Section~\ref{sec:spiral_alignment_model}, the angular offset $\Delta\phi$ can be implemented as either a global parameter or a per-annulus parameter. We first present results with a global $\Delta\phi$, which provides a single, well-constrained quantification of the overall deviation from purely azimuthal direction across the disk. We then present results with a per-annulus $\Delta\phi$, which allows investigation of radial variations of the deviation angle.

\subsubsection{Global $\Delta\phi$ Model Results} \label{sec:B4_st_dphi_global_result}
\begin{figure}[h]
    \centering
    \begin{subfigure}{\linewidth}
        \centering
        \includegraphics[width=1.0\linewidth]{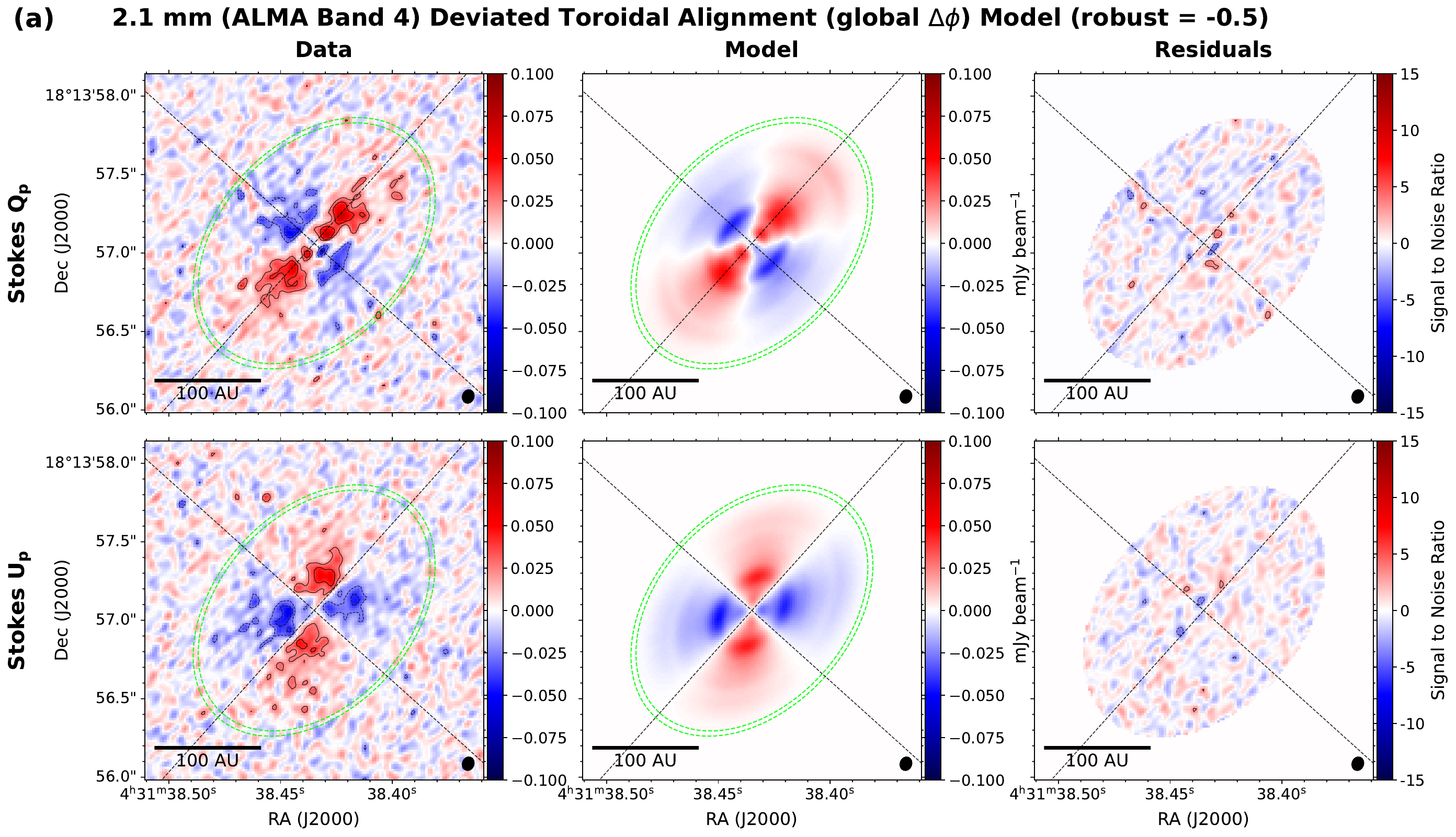}
        \phantomcaption
        \label{fig:B4rm0.5st_dphi_const_result}
    \end{subfigure}

    \begin{subfigure}{\linewidth}
        \centering
        \includegraphics[width=1.0\linewidth]{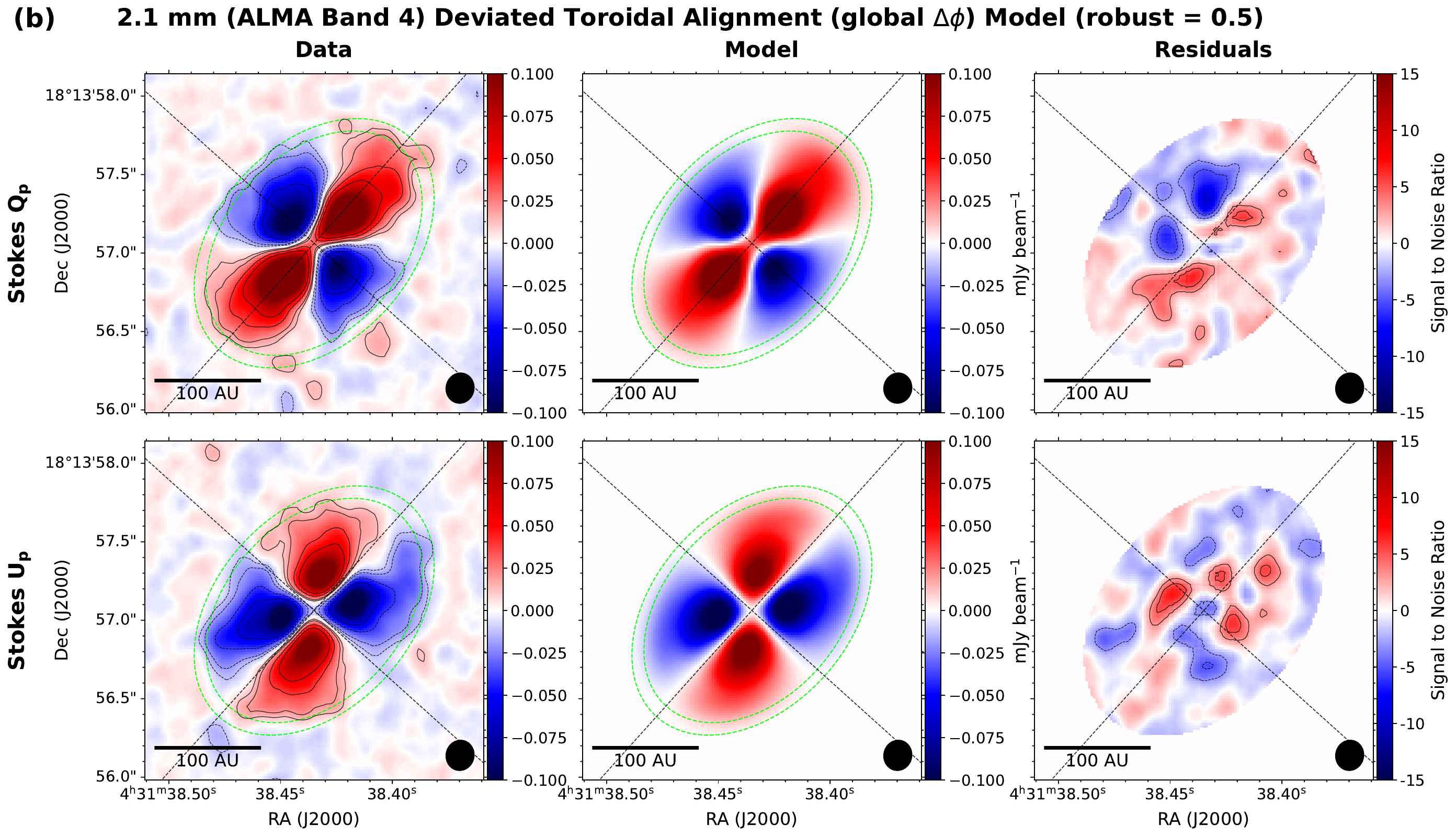}
        \phantomcaption
        \label{fig:B4r0.5st_dphi_const_result}
    \end{subfigure}
    \caption{Results of the deviated toroidal alignment model with global $\Delta\phi$ fitting for robust $=-0.5$ (a) and robust=0.5 (b). The layout and plotting conventions are the same as Figure~\ref{fig:B4st_results}.}
    \label{fig:B4st_dphi_const_results}
\end{figure}

For the robust $=-0.5$ data (Figure~\ref{fig:B4rm0.5st_dphi_const_result}), the deviated toroidal alignment model with a global $\Delta\phi = 2.83^\circ \pm 0.33^\circ$ achieves a reduced $\chi^2 \sim 1.11$ for the combined fit, which is almost identical to that of the toroidal alignment model ($\sim1.13$). The residuals show no significant differences compared to the toroidal alignment model, with both Stokes $Q_p$ and $U_p$ residuals remaining largely consistent with noise below the 3$\sigma$ level.

For the robust = 0.5 data (Figure~\ref{fig:B4r0.5st_dphi_const_result}), the global $\Delta\phi$ model shows substantial improvements over the toroidal model. The model yields $\Delta\phi = 4.10^\circ \pm 0.09^\circ$ for the combined fit, with a reduced $\chi^2$ of $\sim5.10$, significantly improved from $\sim7.85$ for the purely toroidal model. 
The uncertainties of the fitted parameters reported here and throughout are statistical only, i.e., posterior standard deviations. Systematic effects are assessed in Section~\ref{sec:B4_dphi_robustness}. 
In Stokes $Q_p$, the centrosymmetric residual pattern in the toroidal alignment model as identified in Section~\ref{sec:B4_st_result} is no longer present, while the north--south asymmetry persists (possibly due to a near-far side asymmetry in the scattering component not accounted for in this work), with structured strong negative residuals on the northern side and positive residuals $\lesssim 3\sigma$ on the southern side. In Stokes $U_p$, the residuals still exhibit a subtle centrosymmetric tendency, but they are more diffuse and random compared to those seen in the toroidal alignment model. 

Compared to the toroidal alignment model, the global $\Delta\phi$ model introduces visible differences in the model geometry, which can be seen in the locations of zero-crossings and extrema relative to the disk principal axes (black dashed lines), most apparently in the robust = 0.5 model maps, and in the azimuthal profiles of Figure~\ref{fig:B4r0.5_azimuthal_profile_68au}. We discuss these features in Section~\ref{sec:B4_st_dphi_discussion}.

\begin{figure}[h]
    \centering
    \begin{subfigure}{0.49\linewidth}
        \includegraphics[width=1.0\linewidth]{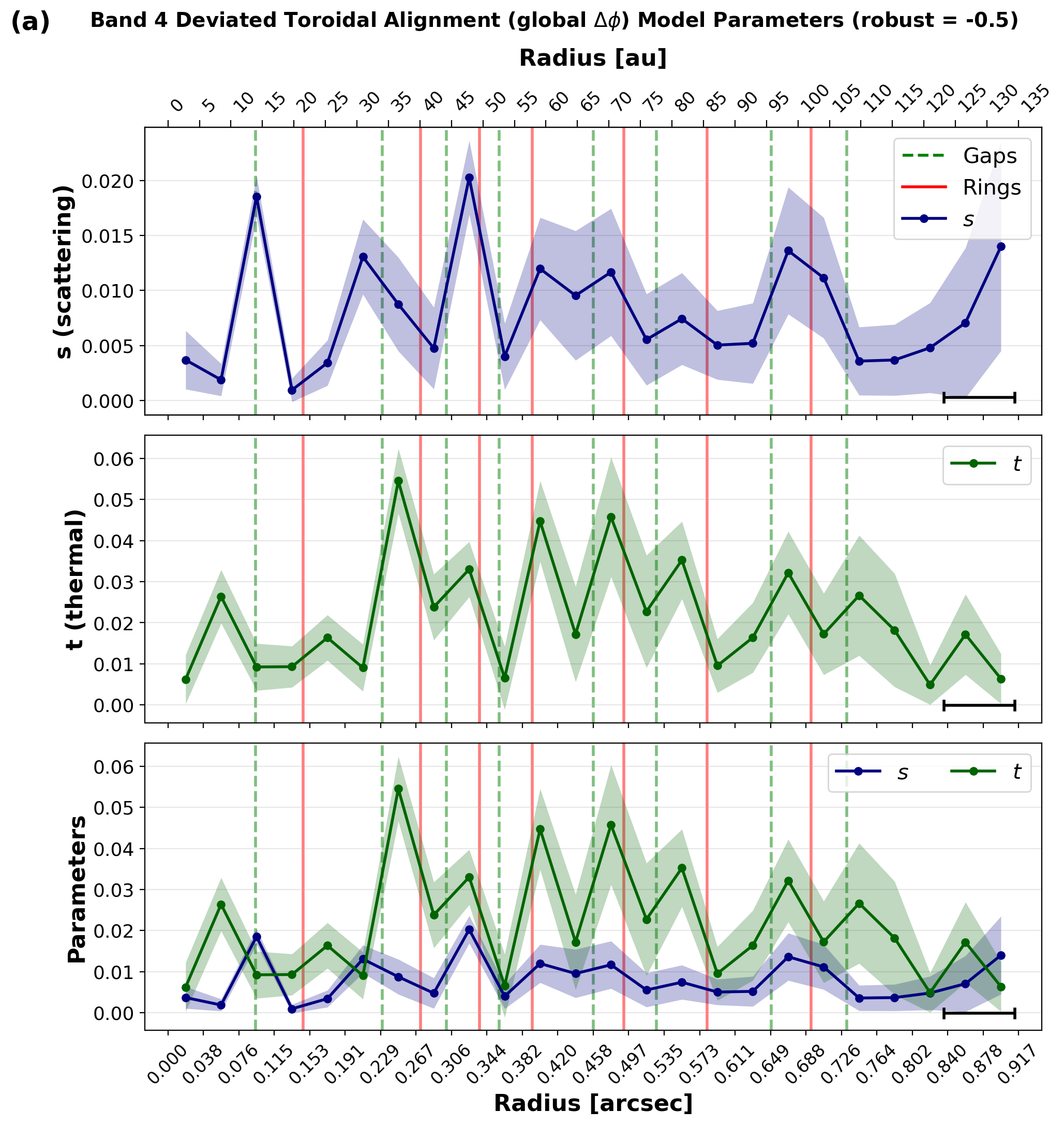}
        \phantomcaption
        \label{fig:B4rm0.5_st_dphi_const_st_profile}
    \end{subfigure}
    \begin{subfigure}{0.492\linewidth}
        \includegraphics[width=1.0\linewidth]{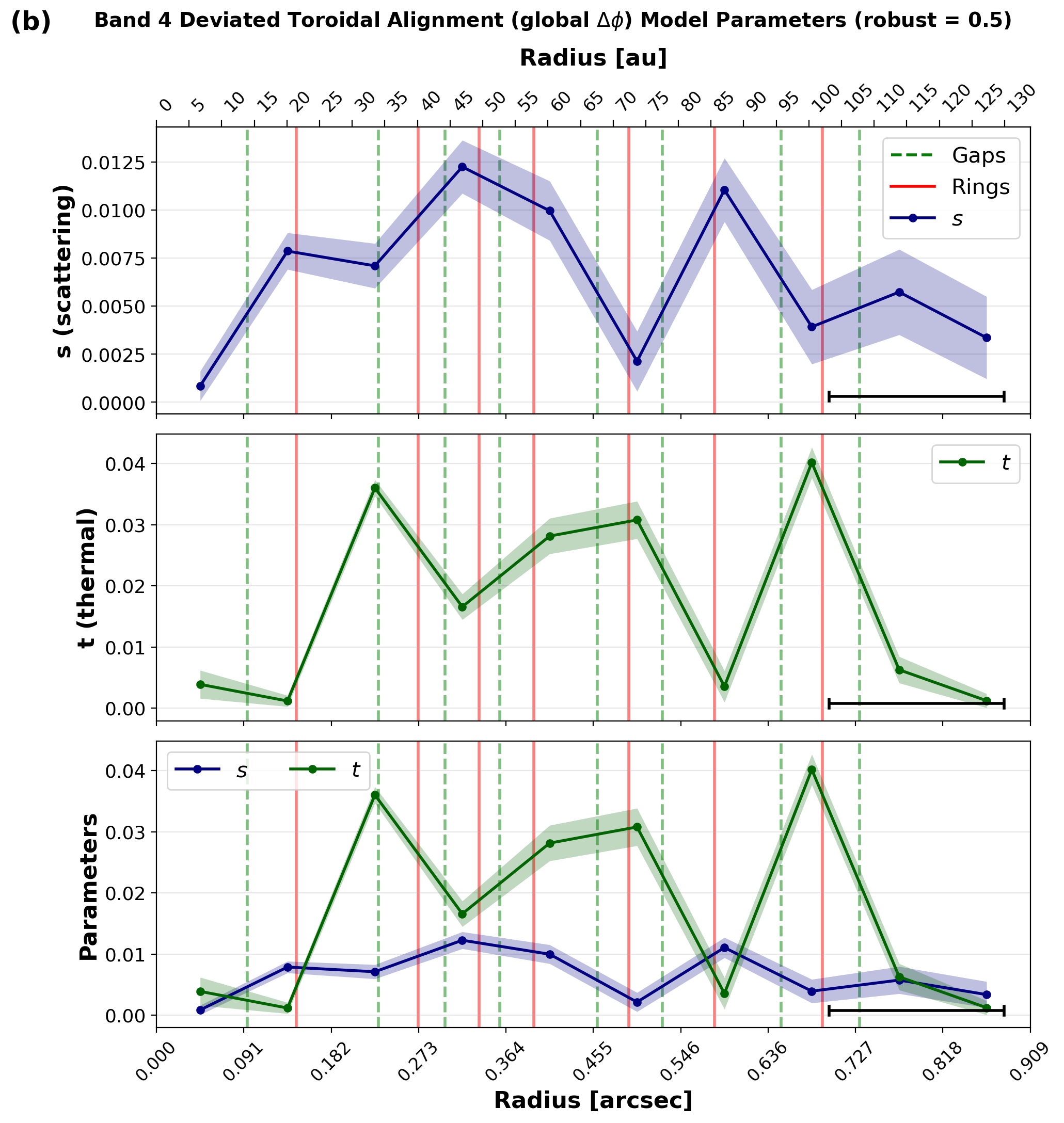}
        \phantomcaption
        \label{fig:B4r0.5_st_dphi_const_st_profile}
    \end{subfigure}
    
    \caption{Radial profiles of the parameters of the deviated toroidal alignment model (global $\Delta\phi$) fitting for robust $=-0.5$ (a) and robust = 0.5 (b). The plotting convention is the same as Figure~\ref{fig:B4_st_params}.}    
    \label{fig:B4_st_dphi_const_params}
\end{figure}

In contrast to the substantial improvement seen in the residuals, the radial profiles of the polarization mechanism amplitudes $s$ and $t$ in the deviated toroidal alignment model with global $\Delta\phi$ (Figure~\ref{fig:B4_st_dphi_const_params}) remain broadly consistent with those obtained from the toroidal alignment model at both robust weightings. No distinctly new radial features are introduced in either component across the disk, and the overall amplitudes and relative behaviors of $s$ and $t$ still follow those described in Section~\ref{sec:B4_st_result}. 

\subsubsection{Per-annulus $\Delta\phi$ Model Results} \label{sec:B4_st_dphi_per-annulus_result}
\begin{figure}[h]
    \centering
    \begin{subfigure}{\linewidth}
        \centering
        \includegraphics[width=1.0\linewidth]{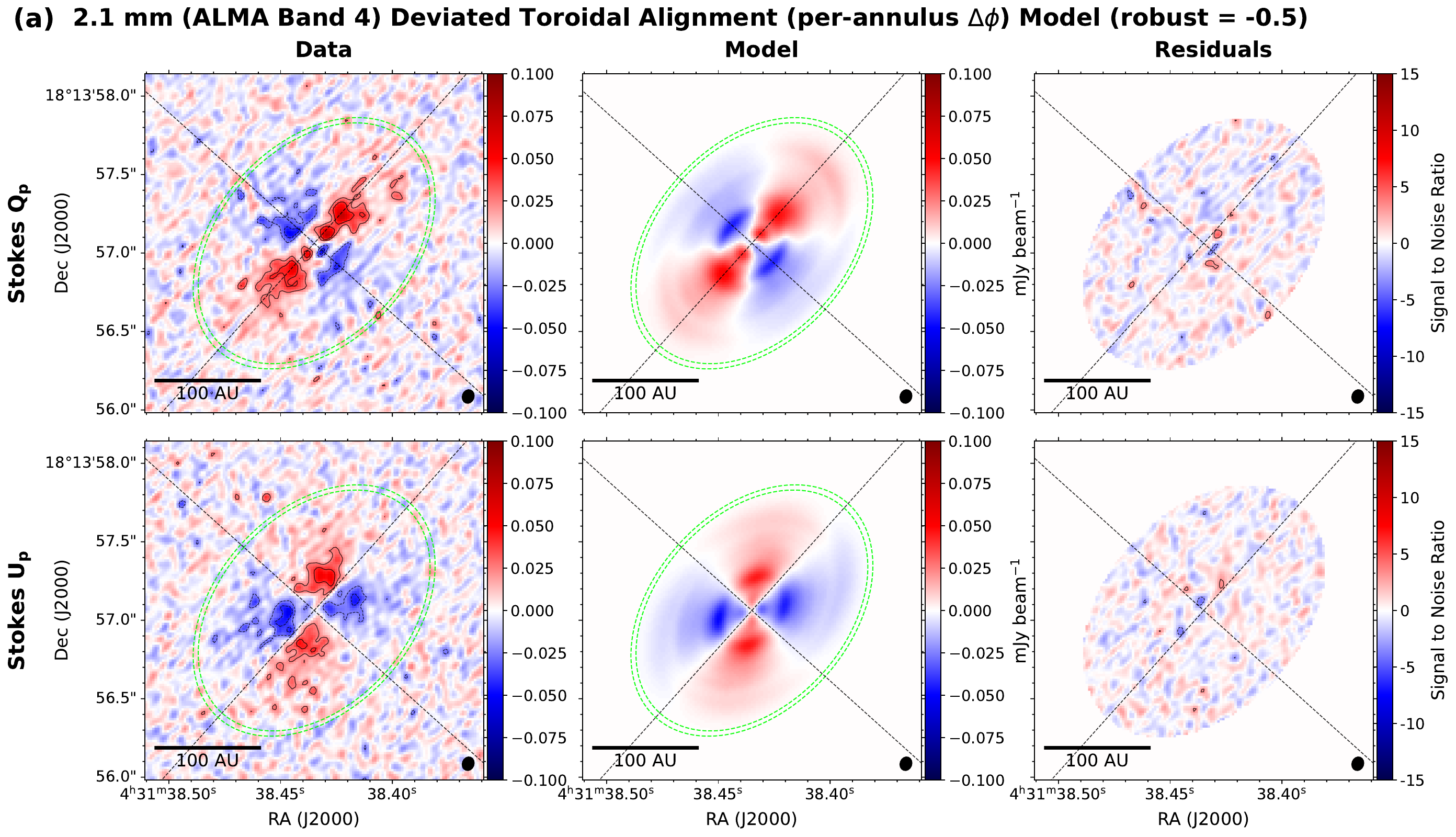}
        \phantomcaption
        \label{fig:B4rm0.5st_dphi_result}
    \end{subfigure}

    \begin{subfigure}{\linewidth}
        \centering
        \includegraphics[width=1.0\linewidth]{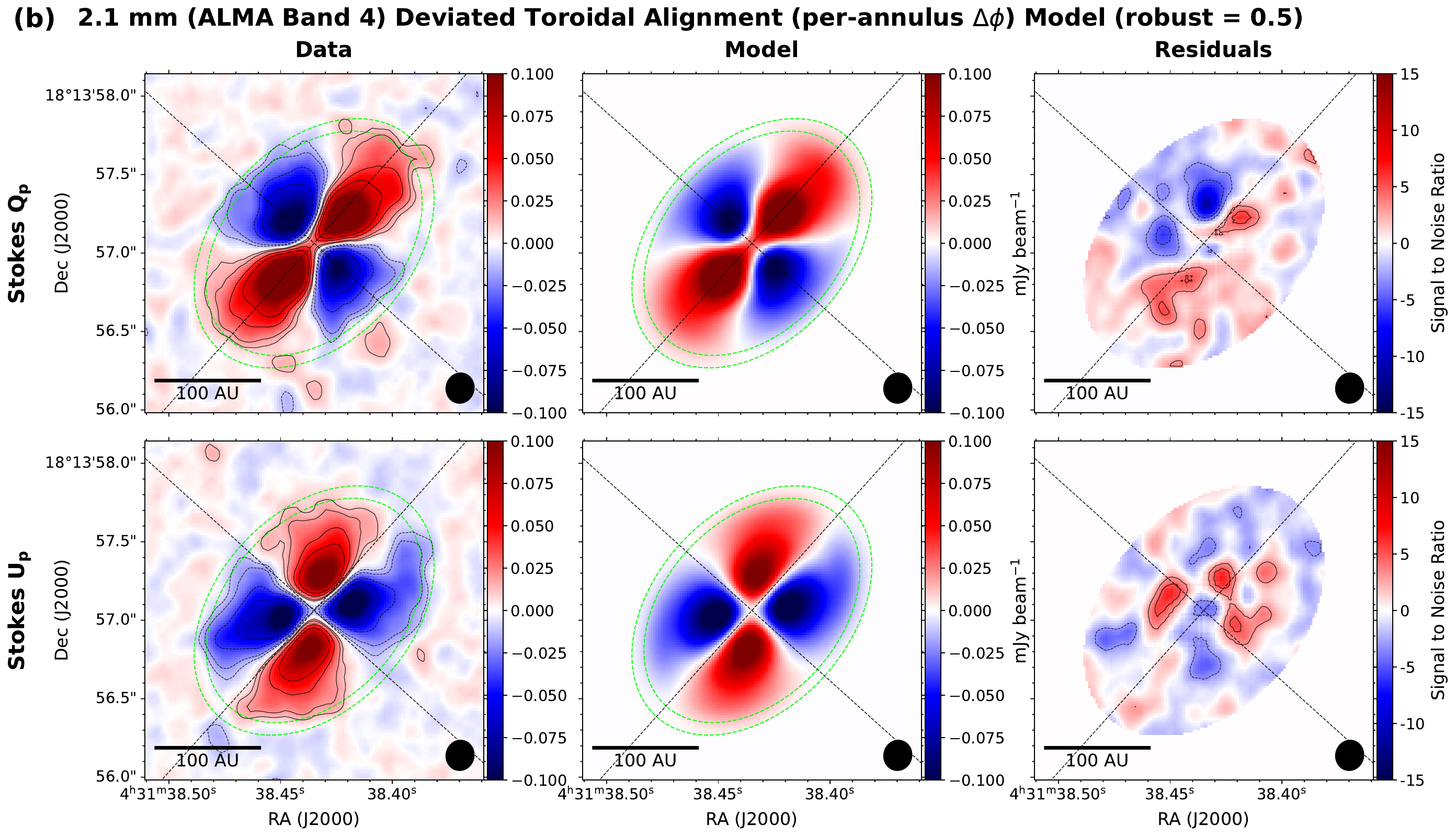}
        \phantomcaption
        \label{fig:B4r0.5st_dphi_result}
    \end{subfigure}
    \caption{Results of the deviated toroidal alignment model with per-annulus $\Delta\phi$ fitting for robust $=-0.5$ (a) and robust=0.5 (b). The layout and plotting conventions are the same as Figure~\ref{fig:B4st_results}.}
    \label{fig:B4st_dphi_results}
\end{figure}

The deviated toroidal alignment model with per-annulus $\Delta\phi$ allows the deviation angle to vary independently at each radius. For the robust $=-0.5$ data (Figure~\ref{fig:B4rm0.5st_dphi_result}), this model achieves a reduced $\chi^2 \sim 1.10$ for the combined fit, nearly indistinguishable from the global $\Delta\phi$ model ($\sim1.11$). For the robust = 0.5 data (Figure~\ref{fig:B4r0.5st_dphi_result}), the per-annulus $\Delta\phi$ model yields reduced $\chi^2\sim4.83$ for the combined fit, which is a modest improvement from $\sim5.10$ (global $\Delta\phi$). 

Compared to the global $\Delta\phi$ model, the radial variation in the deviation angle introduced by per-annulus $\Delta\phi$ model can be particularly seen in the locations of zero-crossings in the robust = 0.5 model maps (see Section~\ref{sec:B4_st_dphi_discussion} for more detailed discussion). The residuals and the radial profile of $s$ and $t$ remain broadly consistent with those obtained from the global $\Delta\phi$ model at both robust weightings.

\begin{figure}[h]
    \centering
    \begin{subfigure}{0.49\linewidth}
        \includegraphics[width=1.0\linewidth]{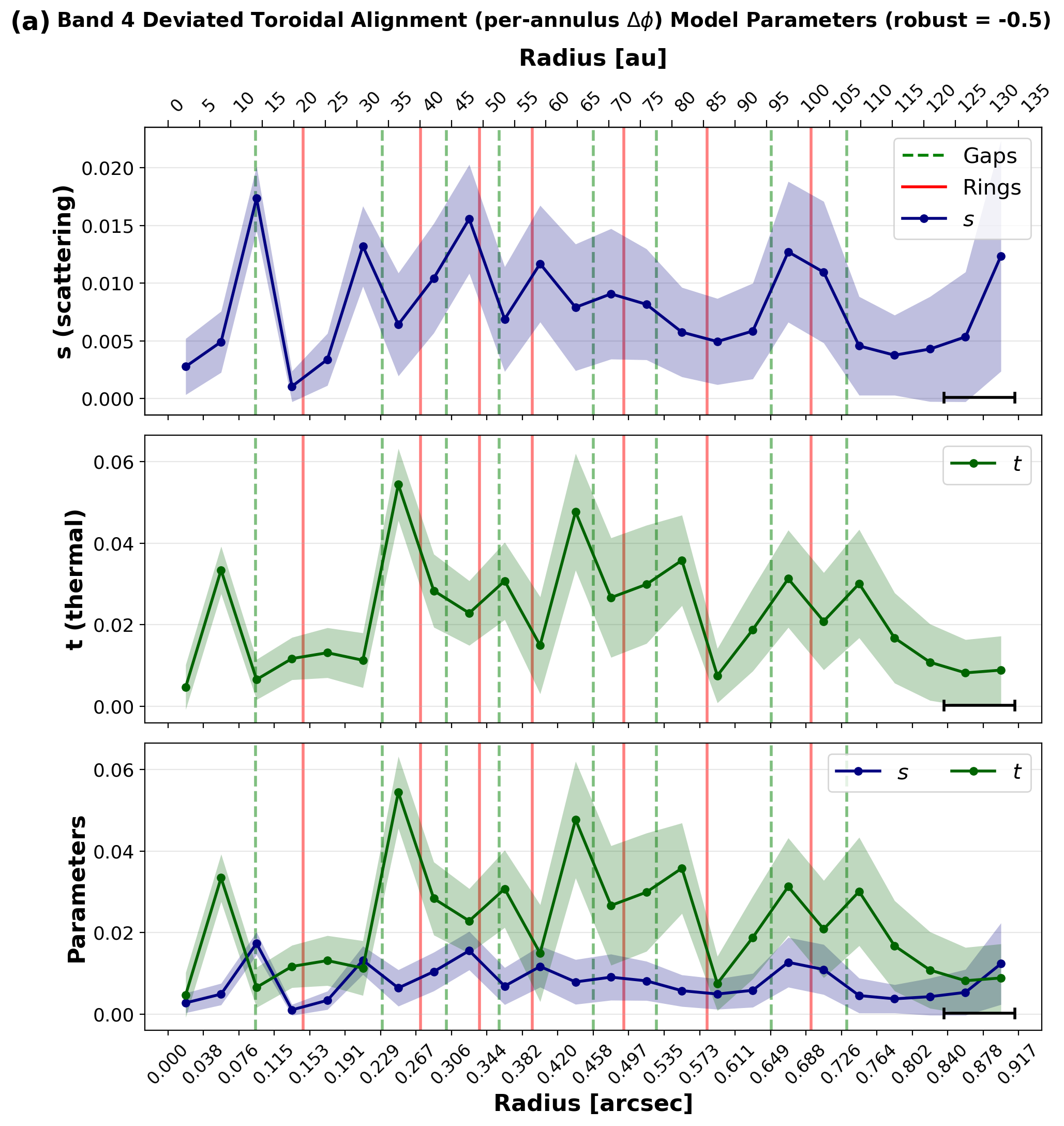}
        \phantomcaption
        \label{fig:B4rm0.5_st_dphi_st_profile}
    \end{subfigure}
    \begin{subfigure}{0.492\linewidth}
        \includegraphics[width=1.0\linewidth]{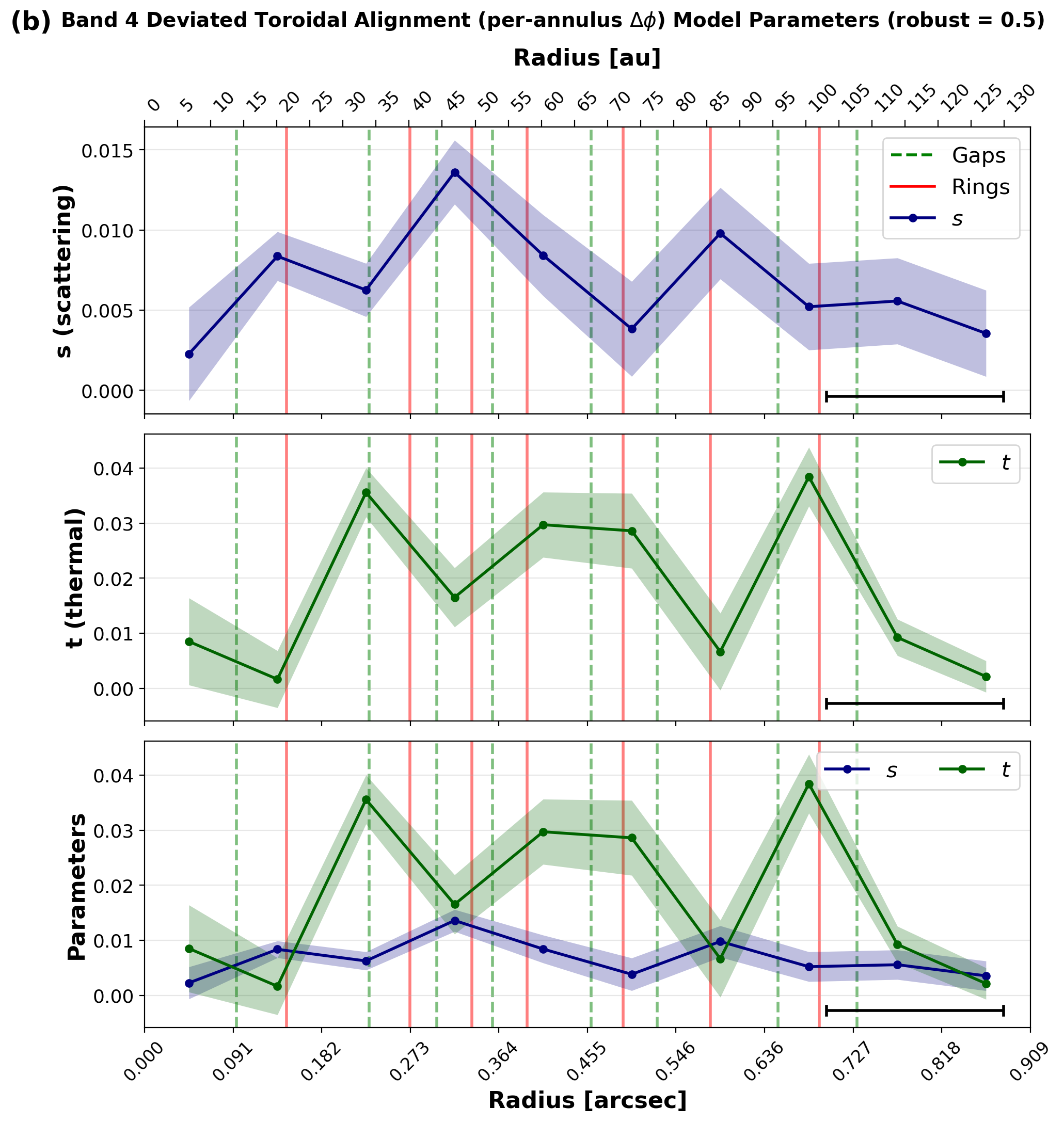}
        \phantomcaption
        \label{fig:B4r0.5_st_dphi_st_profile}
    \end{subfigure}
    
    \caption{Radial profiles of the parameters of the deviated toroidal alignment model with per-annulus $\Delta\phi$ fitting for robust $=-0.5$ (a) and robust = 0.5 (b). The plotting convention is the same as Figure~\ref{fig:B4_st_params}.}    
    \label{fig:B4_st_dphi_params}
\end{figure}

\begin{figure}[h] 
    \centering
    \begin{subfigure}{0.50\linewidth}
        \includegraphics[width=1.0\linewidth]{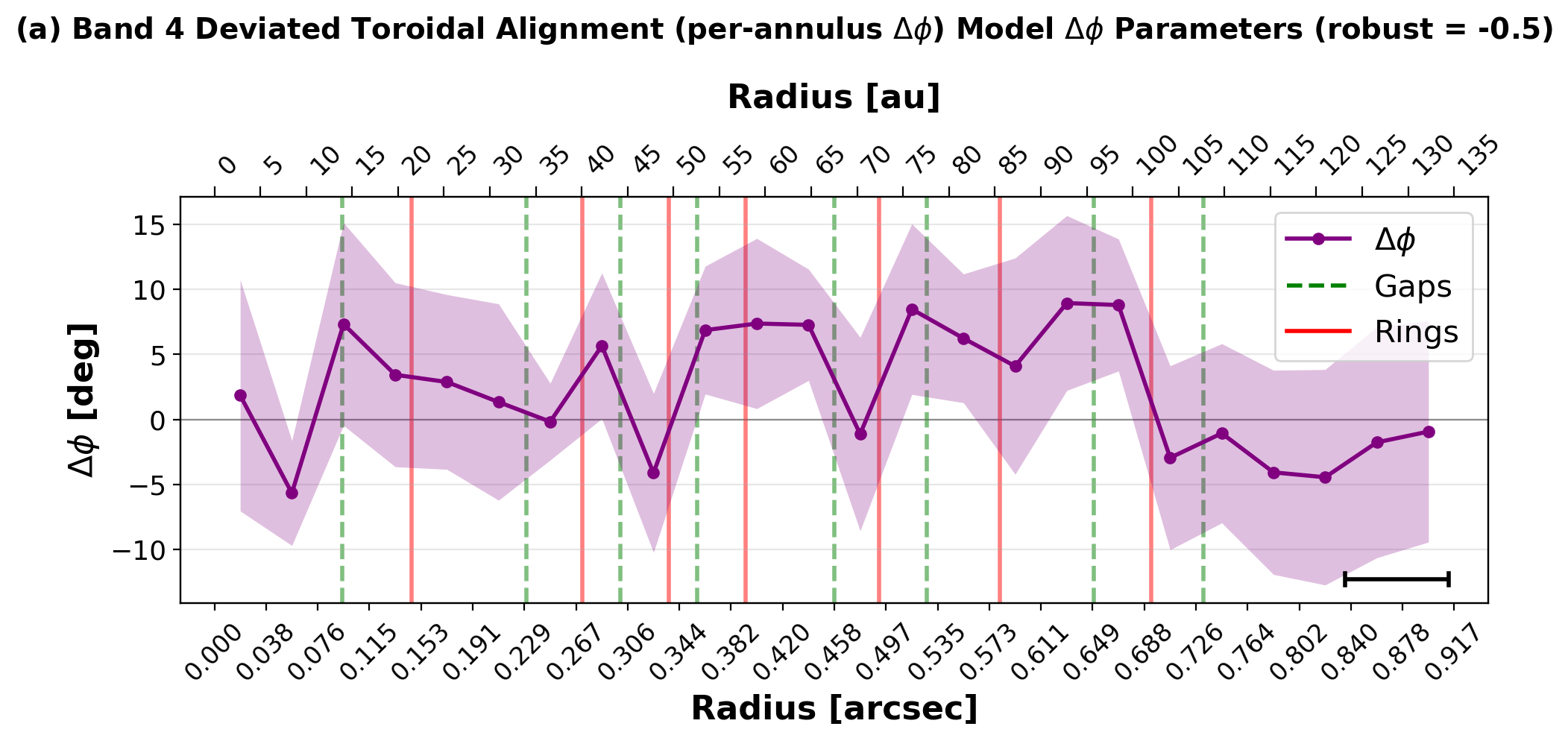}
        \phantomcaption
        \label{fig:B4rm0.5_dphi_profile}
    \end{subfigure}
    \begin{subfigure}{0.491\linewidth}
        \includegraphics[width=1.0\linewidth]{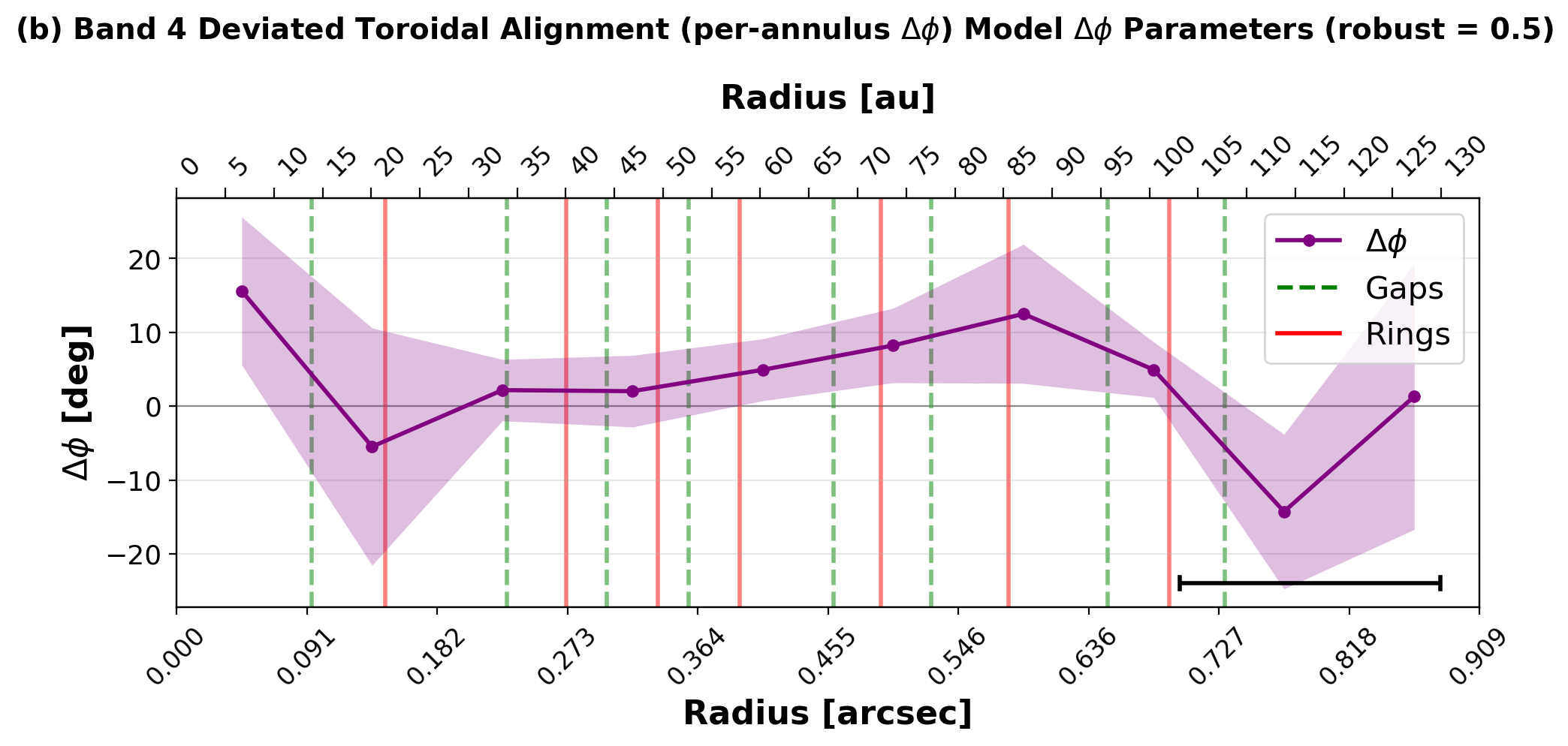}
        \phantomcaption
        \label{fig:B4r0.5_dphi_profile}
    \end{subfigure}
    \caption{Radial profiles of the angular deviation $\Delta \phi$ for the deviated toroidal alignment model with per-annulus $\Delta\phi$ fitting for robust $=-0.5$ (a) and robust = 0.5 (b). The vertical red and green dashed lines indicate the locations of rings and gaps identified by \cite{ALMA_Partner_2015} and \cite{Ian_Nature_2023}. The plotting convention is the same as Figure~\ref{fig:B4_st_params}.}    
    \label{fig:B4_dphi_profile}
\end{figure}

Figure~\ref{fig:B4_dphi_profile} shows the raw $\Delta\phi$ profiles fitted for both robust weighting results. In the robust $=-0.5$ result, $\Delta\phi$ values exhibit large uncertainties (typically $\gtrsim \pm 5^\circ$) and significant fluctuations between adjacent annuli. However, over the intermediate disk radii, where ring and gap substructures are present, $\Delta\phi$ is predominantly positive despite the large scatter. The robust = 0.5 result shows a smoother radial profile where the predominantly positive trend persists, with a broad peak around 87 au up to $\sim12.5^\circ\pm9.4^\circ$. In both robust weightings, $\Delta\phi$ drops to consistently negative values in the outer disk beyond the last identified substructures. The large uncertainties and rapid fluctuations between adjacent annuli are due to inter-annulus degeneracy introduced by beam convolution, which will be discussed in detail in Section~\ref{sec:B4_st_dphi_discussion}. 

\subsubsection{Robustness and Statistical Significance of the Angular Offset} \label{sec:B4_dphi_robustness}

To assess the robustness and statistical significance of the global $\Delta\phi$ detection, we perform a jackknife test that refits the model separately to four half-disks. Dividing the disk along the projected major axis gives the northeast (far-side) and southwest (near-side) halves, and dividing it along the projected minor axis gives the northwest and southeast halves. Because the minor axis lies at a position angle of $48.02^\circ$, each half-disk is centered on an intercardinal direction. As shown in Table~\ref{tab:B4_jackknife}, all four half-disk refits return positive $\Delta\phi$ values spread between $2.6^\circ$ and $6.9^\circ$. This spread exceeds the statistical uncertainties ($\sim0.2^\circ$), with the northeast and northwest halves returning larger offsets than the southwest and southeast halves, paralleling the north--south amplitude asymmetry in the toroidal model residual maps (discussed in Section~\ref{sec:B4_st_dphi_discussion}). Over a half-disk domain, the unmodeled asymmetric structure no longer averages out, but is partially absorbed into the fitted $\Delta\phi$, so that half-disk values are biased in opposite directions on the two sides. This spread should therefore be considered the imprint of the unmodeled structure rather than a true azimuthal dependence of the alignment deviation. In the full-disk fit, these opposing biases largely average out, and the global value lies between the values produced from each complementary pair of halves, close to their means. We therefore regard the global $\Delta\phi=4.10^\circ\pm0.09^\circ$ as an azimuthal average whose sign and overall magnitude are robust against the unmodeled asymmetric structure, with the half-disk spread bounding its possible systematic influence. 

\begin{table}[h]
\centering
\caption{Global $\Delta\phi$ from half-disk jackknife refits of the Band~4
robust = 0.5 data. Uncertainties are 1$\sigma$ (posterior standard deviation).}
\label{tab:B4_jackknife}
\begin{tabular}{lc}
\hline\hline
Region & $\Delta\phi$ [deg] \\
\hline
Full disk & $4.10 \pm 0.09$ \\
Northeast half (far side) & $5.09 \pm 0.17$ \\
Southwest half (near side) & $3.57 \pm 0.16$ \\
Northwest half & $6.94 \pm 0.15$ \\
Southeast half & $2.56 \pm 0.16$ \\
\hline
\end{tabular}
\end{table}

We also quantify the preference between the three models with the Bayesian information criterion \citep[$\mathrm{BIC}$;][]{Schwarz_1978_BIC}, defined as $\mathrm{BIC} = \chi^2 + k \ln n$, where $\chi^2 \equiv -2\ln\mathcal{L}_{\rm max}$, with $\ln \mathcal{L}_{\rm max}$ the log likelihood (Equation~\ref{eq:log_likelihood}) maximized over the model parameters, $k$ the number of free parameters, and $n$ the number of independent data points. $\mathrm{BIC}$ compares the maximum likelihood of two models while penalizing additional free parameters. A lower $\mathrm{BIC}$ indicates the preferred model, with a difference greater than 10 regarded as very strong evidence \citep{Kass_1995}. Results are shown in Table~\ref{tab:B4_significance}.

Since the best-fit residuals are substantially larger than the noise (reduced $\chi^2 \approx 5$ for robust = 0.5), the significance assigned to an extra parameter depends on whether these residuals are attributed to unmodeled structure or to underestimated, correlated noise. We therefore assess the significance of each model preference with two different uncertainty estimates. The raw uncertainty estimates use the rms noise levels $\sigma_Q$ and $\sigma_U$ as uncertainties and all $n = 492$ Nyquist-sampled values (Stokes $Q_p$ and $U_p$ at 246 positions) directly, which treats the residuals as unmodeled structure. For the conservative uncertainty estimates, we instead attribute all of the residuals, whatever their origin, to noise. We inflate the per-point uncertainties until the best-fit model reaches a reduced $\chi^2$ of unity and, because the noise is correlated over the synthesized beam, count only one independent measurement per beam area, which gives $n = 95$ (the two Stokes parameters over $\approx 47$ beam areas). Together, these two steps divide all $\chi^2$ values by a factor of 26.1 (Table~\ref{tab:B4_significance}), so that the same improvement of the fit produces a much smaller decrease in $\chi^2$. The raw and conservative estimates therefore represent the optimistic and conservative ends of the range of significances. The rescaling also provides the correctly normalized uncertainties that are required to convert a decrease in $\chi^2$ into a Gaussian significance.

Even with the conservative uncertainty estimates, the preference for the global $\Delta\phi$ model over the purely toroidal model remains decisive. Adding the single parameter $\Delta\phi$ decreases the conservative $\chi^2$ by 50. Under the null hypothesis that $\Delta\phi = 0$, this decrease follows a $\chi^2$ distribution with one degree of freedom \citep{Wilks_1938}, giving $p = 1.5\times10^{-12}$, or, expressed as an equivalent Gaussian significance, $\sqrt{50} \approx 7.1\sigma$. The conservative $\mathrm{BIC}$ decreases by 45.5, far above the very-strong-evidence threshold defined above. By contrast, the per-annulus $\Delta\phi$ model is not statistically preferred. Its nine additional parameters decrease the conservative $\chi^2$ by only 6.5, and its conservative $\mathrm{BIC}$ is higher than that of the global model by 34.5.

We therefore conclude that the per-annulus $\Delta\phi$ model is not preferred over the global $\Delta\phi$ model, which has fewer free parameters. We treat the per-annulus $\Delta\phi$ profile as exploratory rather than as an established measurement. The global positive angular offset itself is highly robust. It is preferred decisively even under the conservative uncertainty estimates, and is recovered independently with the same sign in every half-disk jackknife refit.

\begin{table}[h]
\centering
\caption{Bayesian information criterion comparison for the Band~4 robust = 0.5 models under the raw and conservative uncertainty estimates (parameters and uncertainty estimates are defined in Section~\ref{sec:B4_dphi_robustness}). The conservative columns apply the uncertainty inflation and per-beam counting described in the text, which jointly divide $\chi^2$ by a factor of 26.1 and use $n = 95$ in the BIC penalty. Values in parentheses (diff.) give the difference with respect to the model in the preceding row. With a lower BIC indicating the preferred model, a negative BIC difference favors the model of that row over the one above it.}
\label{tab:B4_significance}
\begin{tabular}{lccccc}
\hline\hline
Model & $k$ (diff.) & $\chi^2_{\rm raw}$ (diff.) &
$\mathrm{BIC}_{\rm raw}$ (diff.) & $\chi^2_{\rm cons}$ (diff.) &
$\mathrm{BIC}_{\rm cons}$ (diff.) \\
\hline
Toroidal & 20 & 3671.9 & 3795.9 & 140.6 & 231.6 \\
Deviated toroidal, global $\Delta\phi$ & 21 ($+1$) & 2365.7 ($-1306.2$) &
2495.9 ($-1300.0$) & 90.6 ($-50.0$) & 186.1 ($-45.5$) \\
Deviated toroidal, per-annulus $\Delta\phi$ & 30 ($+9$) & 2196.4 ($-169.3$) &
2382.4 ($-113.5$) & 84.1 ($-6.5$) & 220.6 ($+34.5$) \\
\hline
\end{tabular}
\end{table}

\subsection{Circular Alignment Model Results} \label{sec:B4_sc_result}
\begin{figure}[h]
    \centering
    \includegraphics[width=1.0\linewidth]{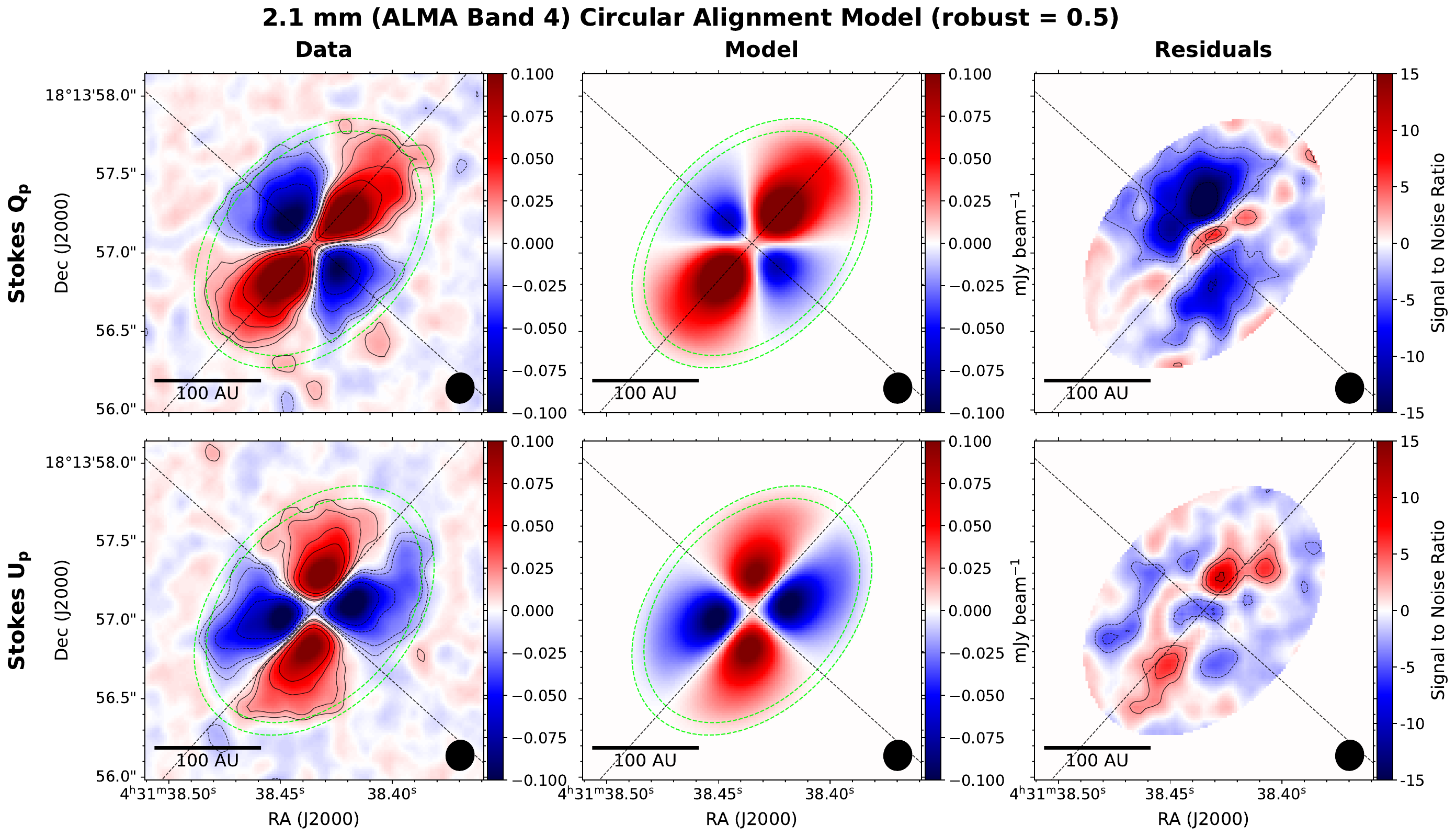}
    \caption{Results of the circular alignment model fitting for robust=0.5. The layout and plotting conventions are the same as Figure~\ref{fig:B4st_results}.}
    \label{fig:B4r0.5_sc_results}
\end{figure}

We briefly present the fitting results for the circular alignment model which is expected for radiative alignment. Since this model is not central to our analysis, and the fit is substantially worse than that of the toroidal alignment model in both the residuals and the reduced $\chi^2$ (Table~\ref{tab:chi2_band4}), we restrict the presentation to the robust = 0.5 fitting result (Figure~\ref{fig:B4r0.5_sc_results}). The relative orientation of the Stokes $Q_p$ zero-crossing lines and the shapes and locations of the extrema in both $Q_p$ and $U_p$ differ significantly from the data. Such discrepancies cannot be removed by any global rotation, as they reflect the intrinsic difference between a circular azimuthal morphology and the elliptical azimuthal morphology originating from disk inclination (see \cite{Haifeng_Band3_2019} for detailed discussion). Consistent with the Band~3 analyses of \cite{Haifeng_Band3_2019} and \cite{Mori_2021}, we conclude that radiative alignment is unlikely to be the grain alignment mechanism observed in HL~Tau at Band~4.

\section{BAND 7 MODELING RESULTS} \label{sec:B7_Modeling_Results}
We also apply the same modeling framework to the Band~7 polarization data for comparison with the Band~4 analysis. We fit Band~7 data imaged with robust = 0.5 (with beam size $\sim0\farcs035$) and robust = 2.0 (with beam size $\sim0\farcs066$). For Band~7 we adopt an annulus width of one beam, rather than the half-beam width used for Band~4, as the smaller Band 7 beam sizes would otherwise result in annuli width approaching the pixel scale. We focus on two features: the radial variation of the inferred polarization mechanisms with ring and gap substructures, and the presence of deviations from purely azimuthal alignment. The overall fitting quality and specific data features are summarized at the end of this section, with a more detailed discussion available in Section~\ref{sec:B7_limitations}.

\begin{table}[h]
\centering
\caption{Reduced $\chi^2$ values for different models and robust weightings for Band~7 observations. For each entry, values are shown for $Q_p$, $U_p$, and combined fit.}
\label{tab:chi2_band7}

\begin{tabular}{lcccc}
\hline\hline
\rule{0pt}{5.5ex} Robust & & Toroidal & \shortstack{Deviated Toroidal\\(global $\Delta\phi$)} & \shortstack{Deviated Toroidal\\(per-annulus $\Delta\phi$)} \\[1.0ex]
\hline
\multirow{3}{*}{$0.5$} & $Q_p$ & 1.61 & 1.62 & 1.63\\
                        & $U_p$ & 1.93 & 1.91 & 1.92 \\
                        & Combined & 1.75 & 1.73 & 1.73\\
\hline
\multirow{3}{*}{$2.0$}  & $Q_p$ & 2.20 & 2.27 & 2.36\\ 
                        & $U_p$ & 6.02 & 5.88 & 6.02\\
                        & Combined & 3.98 & 3.94 & 3.98\\
\hline
\end{tabular}
\end{table}

\begin{figure}
    \centering
    \includegraphics[width=1.0\linewidth]{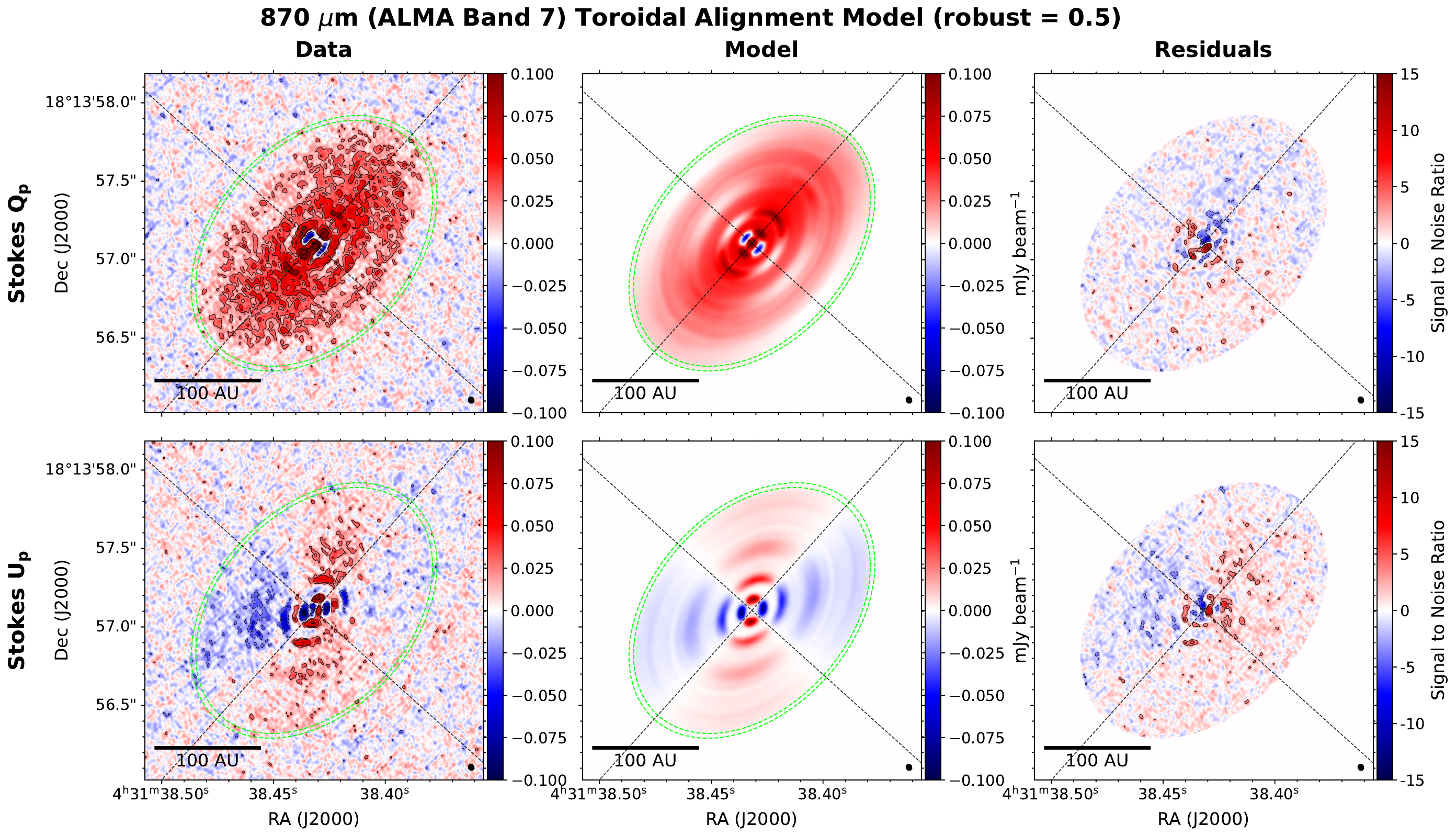}
    \caption{Results of the toroidal alignment model fitting for Band~7 robust=0.5. The layout and plotting conventions are the same as Figure~\ref{fig:B4st_results}.}
    \label{fig:B7_r0.5_st_result}
\end{figure}

\begin{figure}
    \centering
    \includegraphics[width=0.5\linewidth]{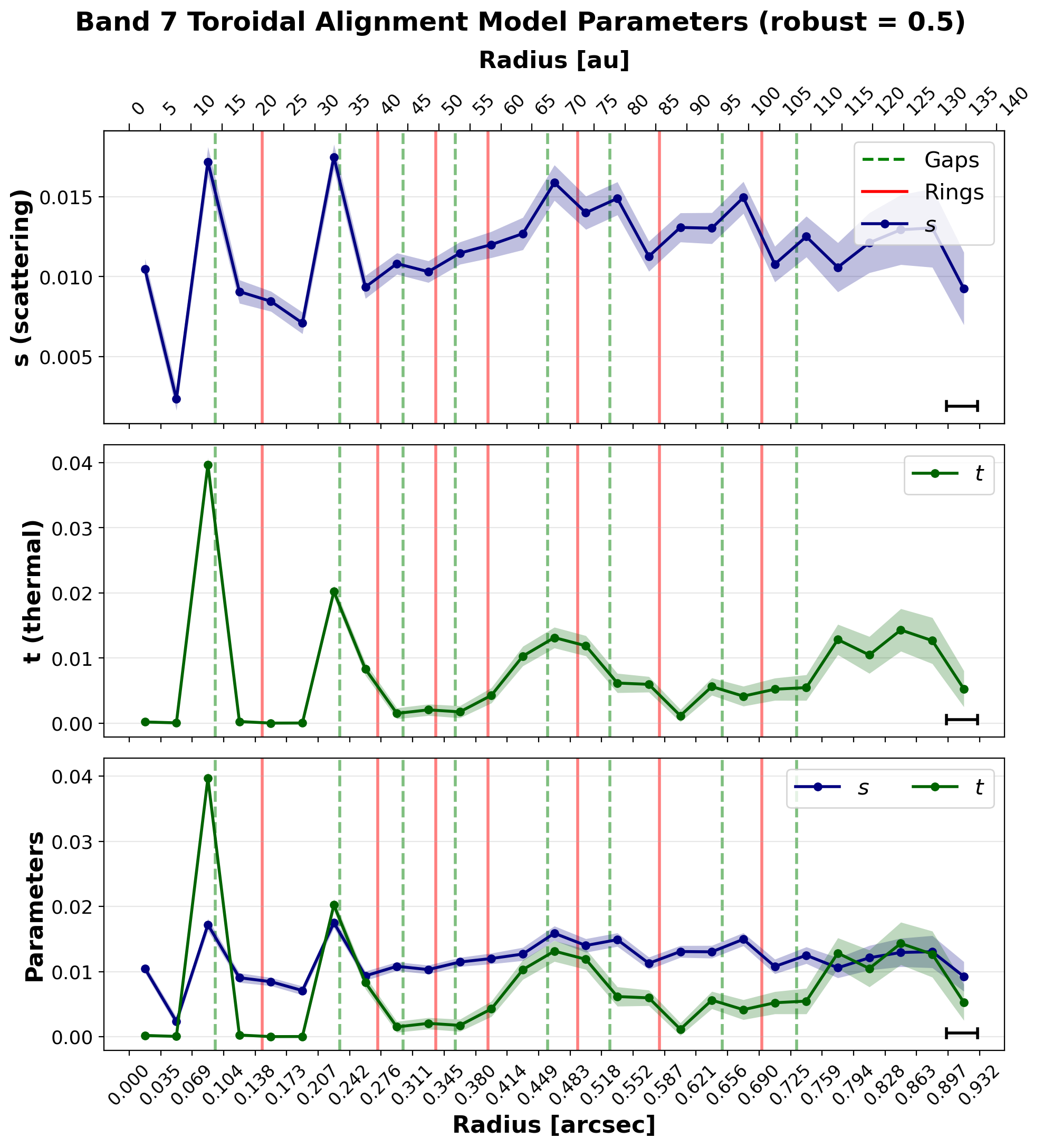}
    \caption{Radial profile of the parameters of the toroidal alignment model fitting for Band~7 robust=0.5. The layout and plotting convention are the same as Figure~\ref{fig:B4_st_params}.}
    \label{fig:B7_r0.5_st_profile}
\end{figure}

To examine the radial variation of polarization mechanisms in Band~7, we focus on the higher angular resolution data imaged with robust = 0.5. We perform model fitting with both toroidal and deviated toroidal alignment models, and find that they yield similar radial profiles of the parameters $s$ and $t$. Therefore, we present only the toroidal alignment model results here for clarity (Figures~\ref{fig:B7_r0.5_st_result} \& \ref{fig:B7_r0.5_st_profile}).

As shown in Figure~\ref{fig:B7_r0.5_st_profile}, the thermal emission component $t$ exhibits pronounced and sharp radial variations in Band~7. It reaches a high and narrow peak at the innermost prominent gap D14 up to $\sim4\%$, which is the global maximum of $t$ in Band~7. Other peaks are observed at the major gaps D34 (sharp peak) and D68 (broad peak). In contrast, deep troughs in $t$ are found at the prominent ring near 22 au, across the bright band containing D44 and D53, and around the ring region beyond $\sim$ 85 au. In the outer disk where no clear substructures are identified, $t$ shows a shallow and broad increase with no clear local maximum.

The scattering component $s$ displays a more intricate but lower-amplitude response to disk substructures. Two sharp peaks are present at D14 and D34, with troughs at the ring near 22 au in between. Then, $s$ increases smoothly across the bright band that includes D44 and D53. In the subsequent dark band, at D68 gap, the ring near 73 au, and the D78 gap, subtle rise-fall-rise variations are measured. Then, $s$ reaches a trough at the bright ring near 85 au, and shows another rise-fall-rise variation corresponding to the D96 gap, 103 au ring, and D108 gap. 

To investigate the presence of deviations from purely azimuthal alignment in Band~7, we further analyze the data imaged with robust~=~2.0, which provides higher sensitivity to polarization morphology on larger scales. We apply the toroidal alignment model and the deviated toroidal alignment model (with both global and per-annulus $\Delta\phi$) to this dataset, and compare the resulting fits and the $\Delta \phi$ parameter to assess the presence of alignment deviation features. The corresponding fitting results are shown in Figure~\ref{fig:B7r2_st_dphi_results}. Since all three models yield nearly identical radial profiles of $s$ and $t$, we only present the radial profile for the deviated toroidal alignment model with per-annulus $\Delta\phi$ for simplicity, along with the $\Delta\phi$ profile (Figure~\ref{fig:B7r2_st_dphi_profile}). 
Formally, the global model yields positive $\Delta\phi = 2.86\pm0.44^\circ$, and the per-annulus $\Delta\phi$ model yields a predominantly positive $\Delta\phi$ profile (Figure~\ref{fig:B7r2_st_dphi_profile}(b)). However, as shown in Section~\ref{sec:B7_limitations}, the Band~7 polarization angle is only weakly sensitive to the grain alignment direction, and the data contain a pronounced angular structure that the model cannot reproduce, whose systematic contamination of the measured angular deviation is hard to quantify. Thus, the fitted $\Delta\phi$ cannot be interpreted as quantifying the alignment deviation. 

\begin{figure}[h] 
    \centering
    \begin{subfigure}{\linewidth} 
        \centering
        \includegraphics[width=1.0\linewidth]{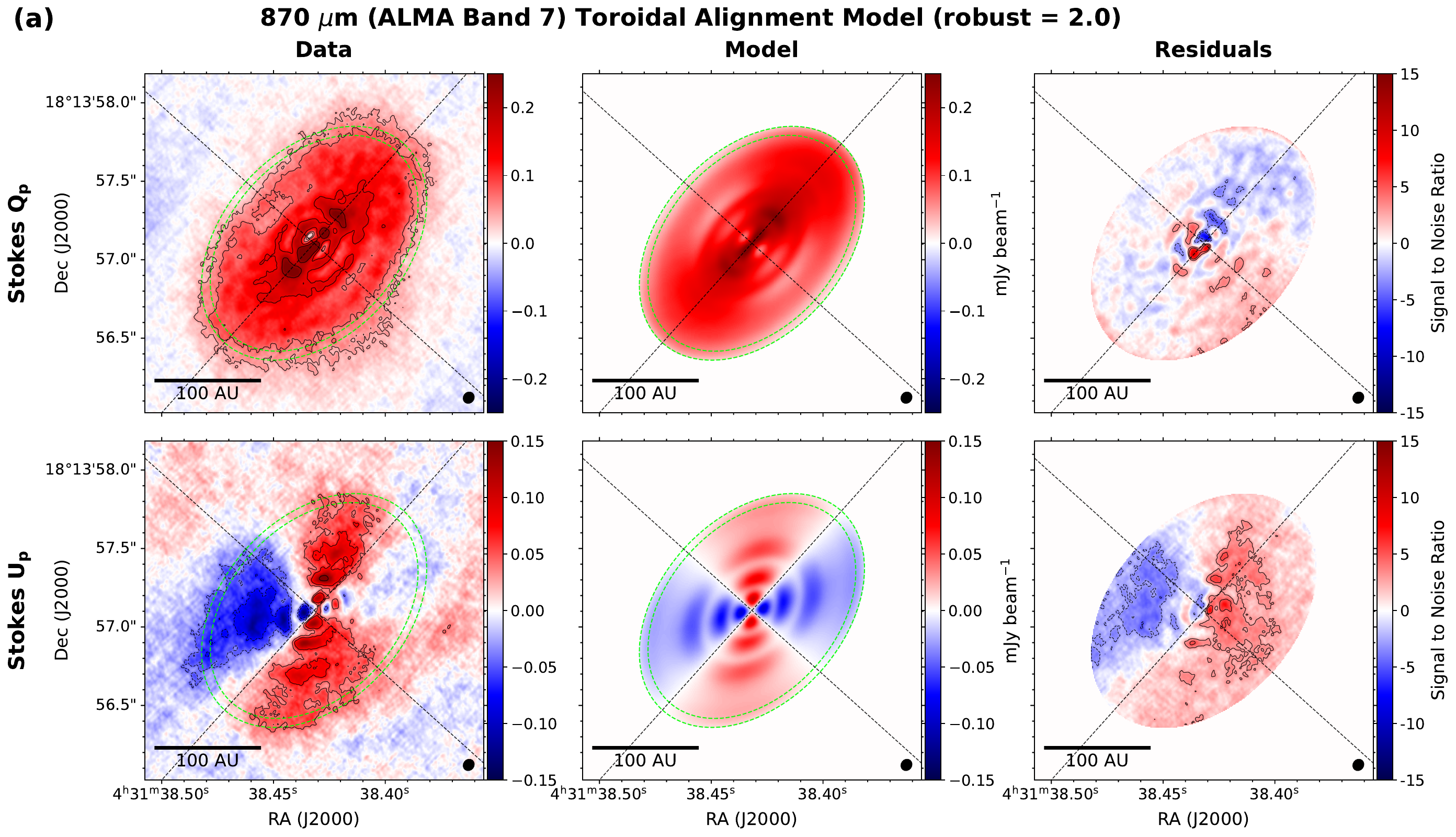}
    \end{subfigure}
    \begin{subfigure}{\linewidth}
        \centering
        \includegraphics[width=1.0\linewidth]{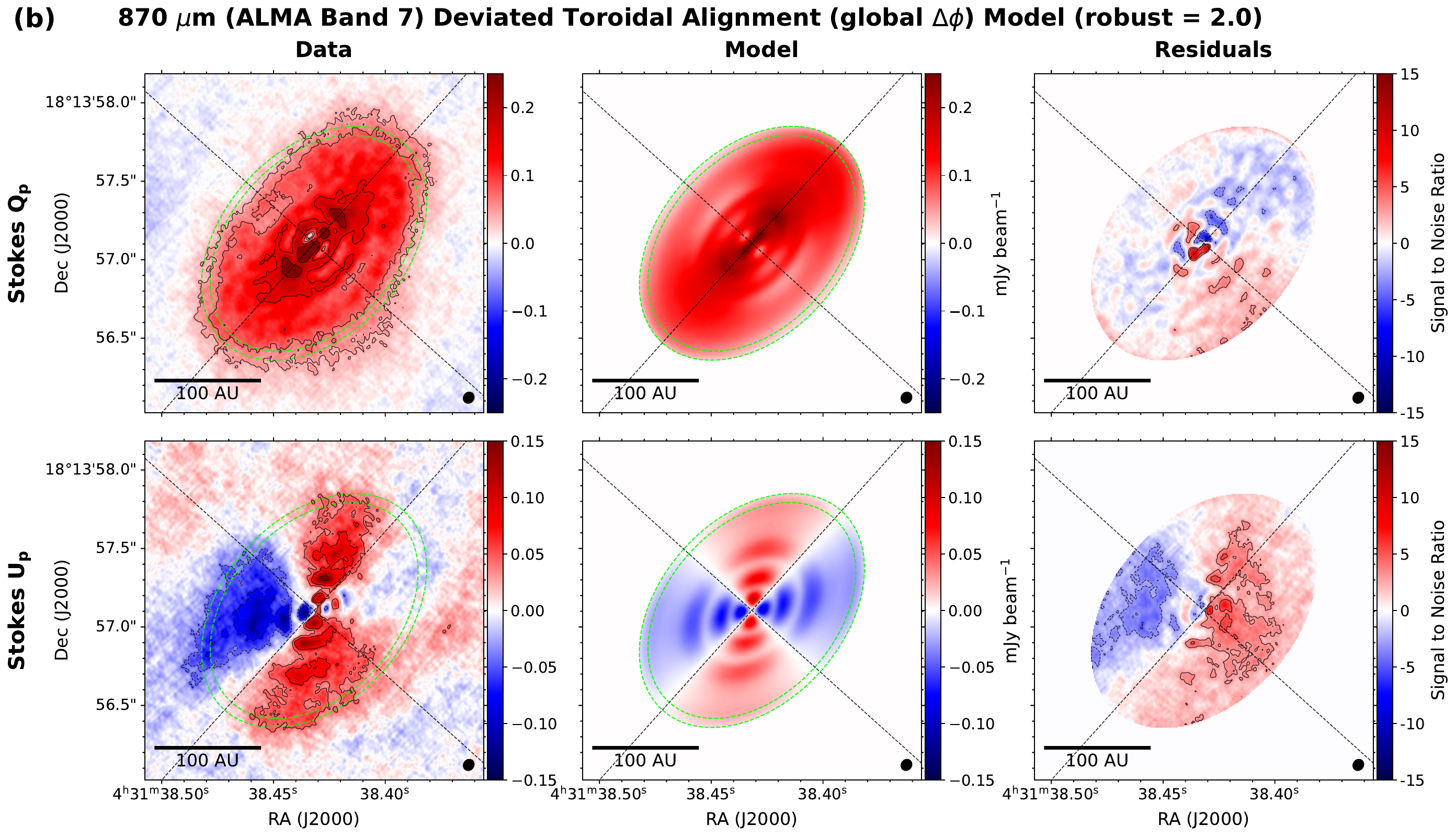}
    \end{subfigure}
    \caption{Results of the toroidal alignment model (a), deviated toroidal alignment model with global $\Delta\phi$ (b) and per-annulus $\Delta\phi$ (c) fitting for Band~7 robust=2.0. The layout and plotting conventions are the same as Figure~\ref{fig:B4st_results}.}
    \label{fig:B7r2_st_dphi_results}
\end{figure}

\begin{figure}[h]
    \ContinuedFloat
    \centering
    \begin{subfigure}{\linewidth}
        \centering
        \includegraphics[width=1.0\linewidth]{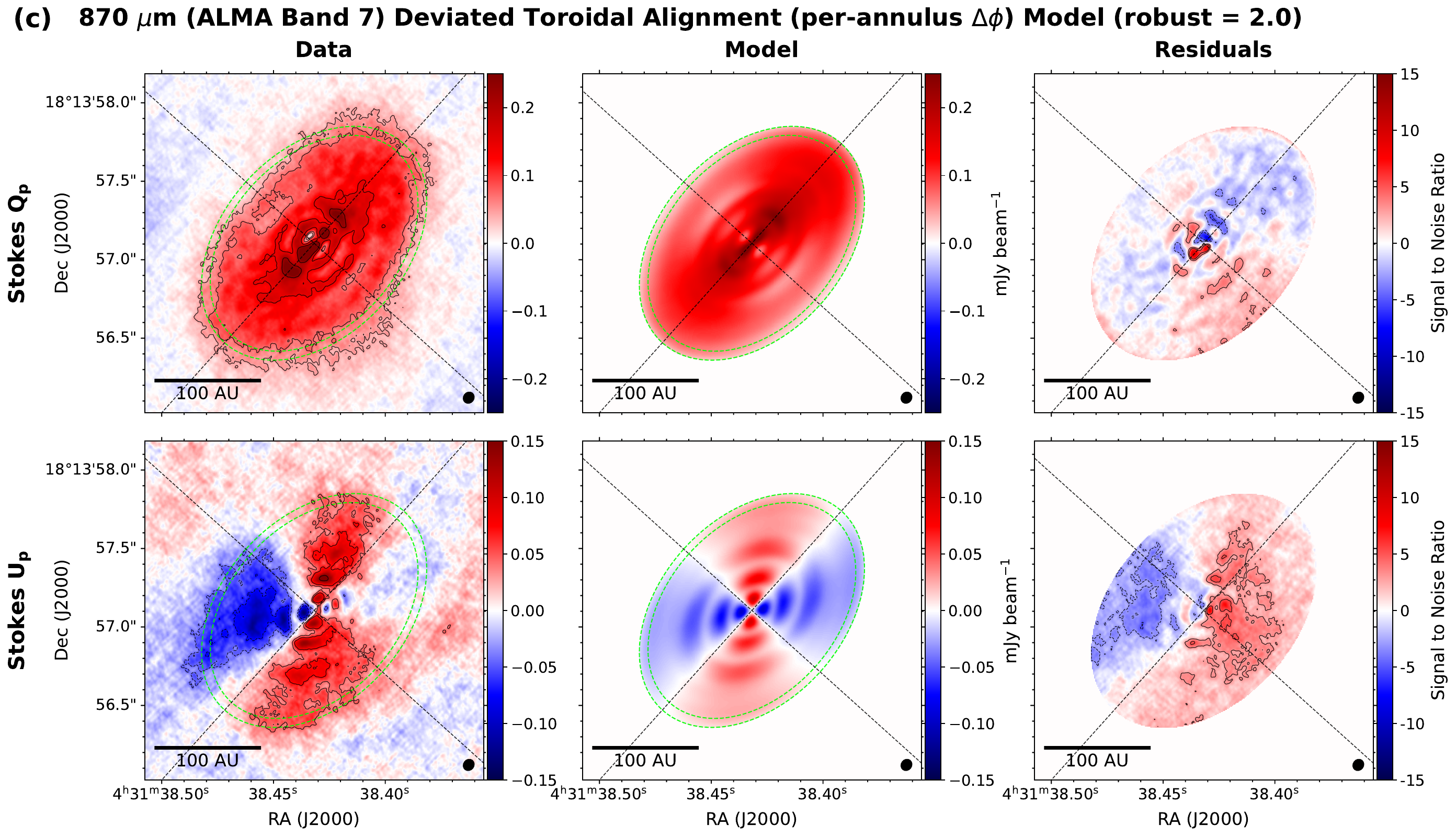}
    \end{subfigure}
    \caption[]{(continued)}
\end{figure}

\begin{figure}[h]
    \centering
        \centering
        \includegraphics[width=0.5\linewidth]{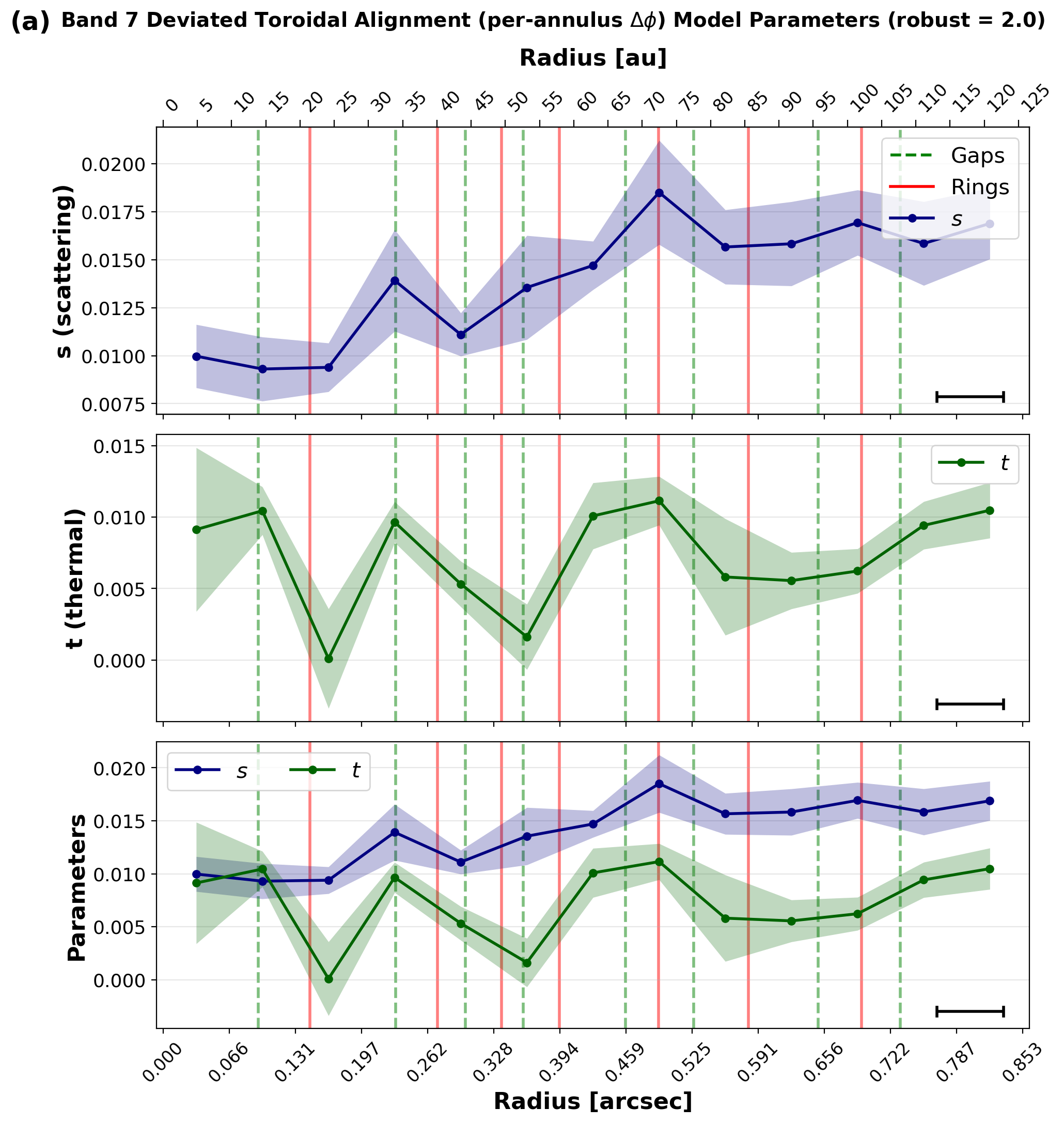}\hfill
        \includegraphics[width=0.5\linewidth]{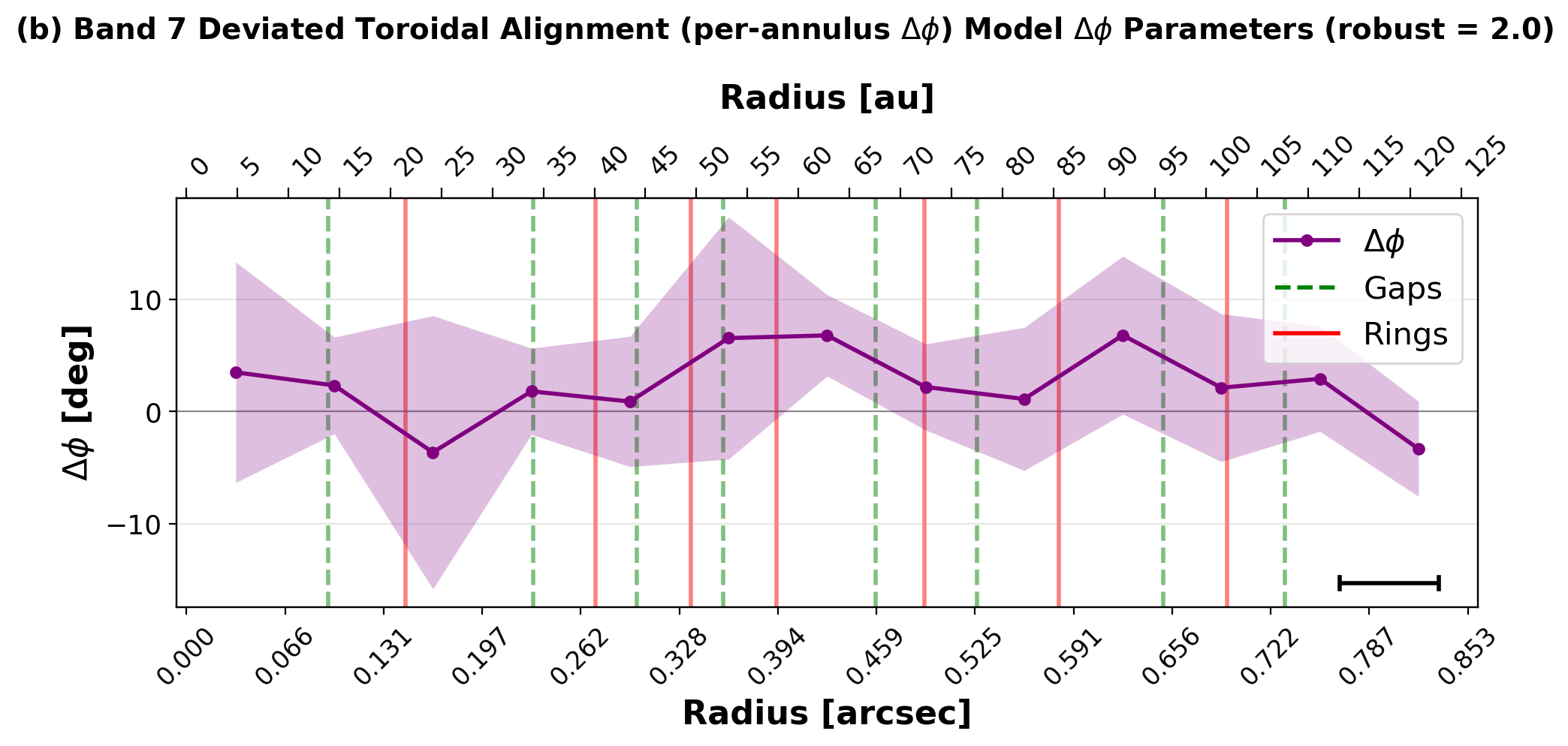}
    
    \caption{Radial profile of the parameters of the deviated toroidal alignment model with per-annulus $\Delta\phi$ fitting for Band~7 robust=2.0, (a) the profile of $s$ and $t$, (b) the profile of $\Delta\phi$. The plotting convention is the same as Figure~\ref{fig:B4_st_params}.}    
    \label{fig:B7r2_st_dphi_profile}
\end{figure}

We note that the Band~7 Stokes $U_p$ data exhibit a pronounced east-west asymmetry, with significantly suppressed signal on the western side of the disk. Additionally, an anomalous counterclockwise deviation is observed in Stokes $U_p$ on the southwestern (minor-axis near-side) portion of the disk. These asymmetries are present in both robust = 0.5 and robust = 2.0 datasets, and become more prominent with the robust = 2.0 sensitivity. These features lead to structured residuals in the model fits. The overall residual analysis and the implications of these asymmetries for the reliability of the fitted parameters will be discussed in Section~\ref{sec:B7_limitations}.

\section{DISCUSSION} \label{sec:discussion}
In this section, we discuss the implications of our modeling results. We examine the geometric properties of the toroidal alignment model and the origin of the structured residuals in Section~\ref{sec:B4_st_discussion}. The deviated toroidal alignment model, the implied trailing spiral streamlines of alignment and the possible physical origins are discussed in Section~\ref{sec:B4_st_dphi_discussion}. We discuss Band~7 modeling in comparison to Band~4 results in Section~\ref{sec:B7_limitations}, and discuss the radial variation of polarization mechanisms across both wavelengths in Section~\ref{sec:P_mechanism_discussion}. Finally, in Section~\ref{sec:stokes_number}, we test the birdie-like alignment interpretation quantitatively by estimating the Stokes number it implies for the measured $\Delta\phi$, and discuss the resulting tension with the Stokes numbers expected for the grains that dominate the millimeter emission.

\subsection{Toroidal Alignment Model} \label{sec:B4_st_discussion}
The structured residuals of the toroidal alignment model in the robust = 0.5 data (Section~\ref{sec:B4_st_result}) can be understood from a geometric property of the model. The model is axisymmetric, meaning the dust alignment configuration over the disk is invariant under arbitrary rotation about the disk rotation axis. For an inclined disk, the model produces normalized Stokes $q_p$ and $u_p$ maps that are centrosymmetric, i.e., invariant under $180^\circ$ rotation about the disk center. Because the Band~4 Stokes $I$ distribution is nearly axisymmetric, the models of $Q_p$ and $U_p$ inherit this centrosymmetry of $q_p$ and $u_p$. While weak asymmetries do exist in Stokes $I$, they are much smaller than, and inconsistent with, those observed in the polarization data. Consequently, the opposite-quadrant amplitude asymmetries in the observed residuals cannot be reproduced by the model and must originate in the polarization itself rather than in the total intensity. As noted by \cite{Ian_Nature_2023}, in Band~7, asymmetries revealed in polarization observations are not evident in non-polarimetric Stokes $I$ data, and our results confirm this finding in Band~4. 

\begin{figure}
    \centering
    \includegraphics[width=1.0\linewidth]{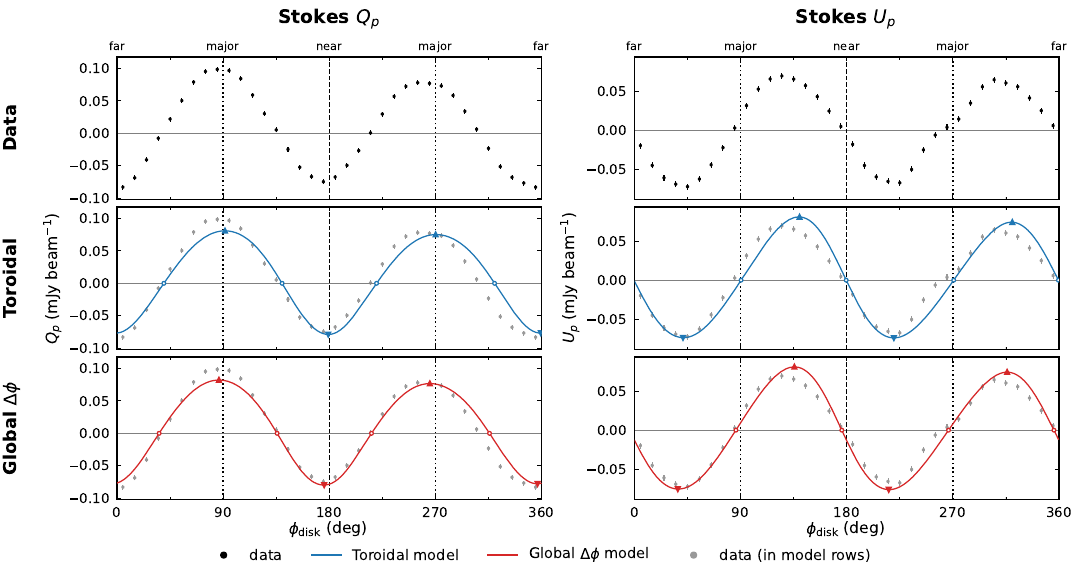}
    \caption{Azimuthal profiles of Stokes $Q_p$ (left) and $U_p$ (right) at the deprojected radius $r = 68$~au for the Band~4 robust = 0.5 data (top), the toroidal alignment model (middle), and the deviated global $\Delta\phi$ model (bottom), as a function of the disk-plane azimuth $\phi_{\rm{disk}}$, measured counterclockwise from the far-side minor axis. Dashed and dotted vertical lines mark the minor ($\phi_{\rm disk} = 0^\circ, 180^\circ$) and major ($90^\circ, 270^\circ$) axes. A clockwise rotation on the sky corresponds to a shift toward smaller $\phi_{\rm disk}$. The data are sampled every $10^\circ$ with uncertainties equal to the rms noise. Model profiles are extracted from the beam-convolved model maps. Open circles and triangles mark their zero-crossings and extrema.}
    \label{fig:B4r0.5_azimuthal_profile_68au}
\end{figure}

The geometric properties of the toroidal alignment model are also reflected in the azimuthal positions of the zero-crossings and extrema of the modeled Stokes $Q_p$ and $U_p$ (Figure~\ref{fig:B4r0.5_azimuthal_profile_68au}). In the disk plane, $q_p$ has extrema on the principal axes, at $\phi_{\rm disk} = 0^\circ, 180^\circ$ (minor axis) and $90^\circ, 270^\circ$ (major axis), where $\phi_{\rm disk}$ is the azimuthal angle in the disk plane (Equation~\ref{eq:phi_ell}). The scattering component $s$ adds a constant offset to $q_p$, which moves its zero-crossings while leaving the extrema and the centrosymmetry unchanged. Without scattering, the zero-crossings would lie at $\phi_{\rm disk} = \arctan(1/\cos i) \approx 55.6^\circ$ and the corresponding symmetric angles. The positive $s$ in our Band~4 fits moves them toward the minor axis, to $\phi_{\rm disk} \approx 39.7^\circ$ at $r = 68$~au for the toroidal alignment model. 
The modeled $u_p$ is complementary, with extrema at the intermediate azimuths $\phi_{\rm disk} = 45^\circ, 135^\circ, 225^\circ, 315^\circ$ and zero-crossings on the principal axes. Because $s$ does not contribute to $u_p$, these zero-crossings are fixed by the alignment geometry alone, which makes them the most direct tracer of the alignment direction. Convolution with the nearly circular synthesized beam slightly modifies the apparent positions of these features but does not break the centrosymmetry.

However, the observed Stokes $Q_p$ and $U_p$ show systematic deviations from these geometric predictions. In the robust = 0.5 data, the zero-crossings of $U_p$ are displaced from the principal axes toward smaller $\phi_{\rm disk}$ by a few degrees (Figure~\ref{fig:B4r0.5_azimuthal_profile_68au}), which corresponds to a clockwise rotation of the pattern on the sky (Figure~\ref{fig:B4r0.5st_result}). The extrema of $U_p$ at the intermediate azimuths are displaced in the same sense. The global $\Delta\phi$ model reproduces these displacements, whereas the toroidal model places the $U_p$ zero-crossings on the principal axes by construction. The corresponding offsets in $Q_p$ are less direct to identify, because its extrema are broad and its zero-crossings also depend on $s$. The amplitude differences between the data and both models reflect the asymmetric residual structure discussed above and do not affect the positions of the zero-crossings. These features indicate a coherent rotation of the polarization geometry, consistent with the approximately centrosymmetric residual patterns identified above, which cannot be reproduced by purely azimuthally aligned grains.

\subsection{Deviated Toroidal Alignment Model} \label{sec:B4_st_dphi_discussion}
In the deviated toroidal alignment model, the introduction of $\Delta\phi$ relaxes the geometric constraint imposed by the toroidal alignment model, allowing the grain alignment direction to deviate from purely azimuthal directions while preserving the local nature of the alignment prescription. For the robust $=-0.5$ data, the toroidal model already fits close to the noise level, and the deviated models yield almost no measurable improvement in the fit quality. They nevertheless capture the systematic angular deviation present in the observations, which is evident in the model geometry (Figure~\ref{fig:B4rm0.5st_dphi_const_result}, \ref{fig:B4rm0.5st_dphi_result}), returning consistent, positive $\Delta\phi$ in both implementations (Figure~\ref{fig:B4rm0.5_dphi_profile}). We therefore focus on the robust = 0.5 results, where the improvements are clearly revealed in both reduced $\chi^2$ and residuals (Figure~\ref{fig:B4r0.5st_dphi_const_result}, \ref{fig:B4r0.5st_dphi_result}).

Two residual structures remain in the global $\Delta\phi$ model result. First, the north--south amplitude asymmetry in Stokes $Q_p$ persists. The $q_p$ and $u_p$ maps remain centrosymmetric for any $\Delta\phi$ and therefore cannot reproduce opposite-quadrant amplitude differences. This asymmetry plausibly reflects a near/far-side asymmetry of the scattering contribution that is not modeled in this work. Second, the data exhibit radial and subtle azimuthal variations of the deviation angle itself, visible near the zero-crossings of the observed maps. The per-annulus model captures part of the radial variation with modest improvement in reduced $\chi^2$ from $\sim5.10$ (global) to $\sim4.83$ (per-annulus), while the residuals remain similar, indicating that such variation is secondary to the overall deviation captured by the global $\Delta\phi$. The remaining diffuse residuals, which retain a subtle centrosymmetric character in Stokes $U_p$, likely reflect finer azimuthal variations of the deviation that neither implementation represents, which are amplified under the sensitivity of the robust = 0.5 data.

Importantly, $\Delta\phi$ improves the fit quality without re-balancing the polarization mechanism amplitudes $s$ and $t$. The radial profiles of $s$ and $t$ remain largely unchanged from those of the toroidal alignment model (Section~\ref{sec:B4_st_dphi_result}). This indicates that the improved fitting quality arises primarily from the alignment geometry, which is the behavior expected when $\Delta\phi$ plays its design role and captures a genuine angular deviation of the grain alignment, rather than acting as an additional degree of freedom that improves the fit by adjusting the relative amplitude contributions.

\subsubsection{Orientation, Radial Profile, and Origin of the Deviated Toroidal Alignment} \label{sec:spiral_mechanism_discussion}
In the presence of deviated toroidal alignment, the grain alignment directions no longer form closed azimuthal loops, but instead form spiral-like streamlines that extend radially outward across the disk. Equivalently, this spiral geometry manifests observationally as a coherent rotation of the pattern in Stokes $Q_p$ and $U_p$ maps, where the rotation direction of the pattern is consistent with the azimuthal direction of spiral streamlines extending from inner to outer radii.

The spiral direction can be classified as either trailing or leading with respect to the disk rotation: when following the spiral curves from smaller to larger radii, if the azimuthal progression direction is opposite to the disk's rotation direction, the spiral is classified as trailing; conversely, if the azimuthal progression direction is in the same direction as the disk rotation, the spiral is classified as leading. Under this definition, a trailing spiral forms when each grain alignment direction deviates from purely azimuthal in the same direction as the disk rotation (which corresponds to an opposite rotation of patterns in Stokes $Q_p$ and $U_p$), and vice versa for a leading spiral (see Figure~\ref{fig:spiral_orientation}). 

\begin{figure} [h]
    \centering
    \includegraphics[width=0.5\linewidth]{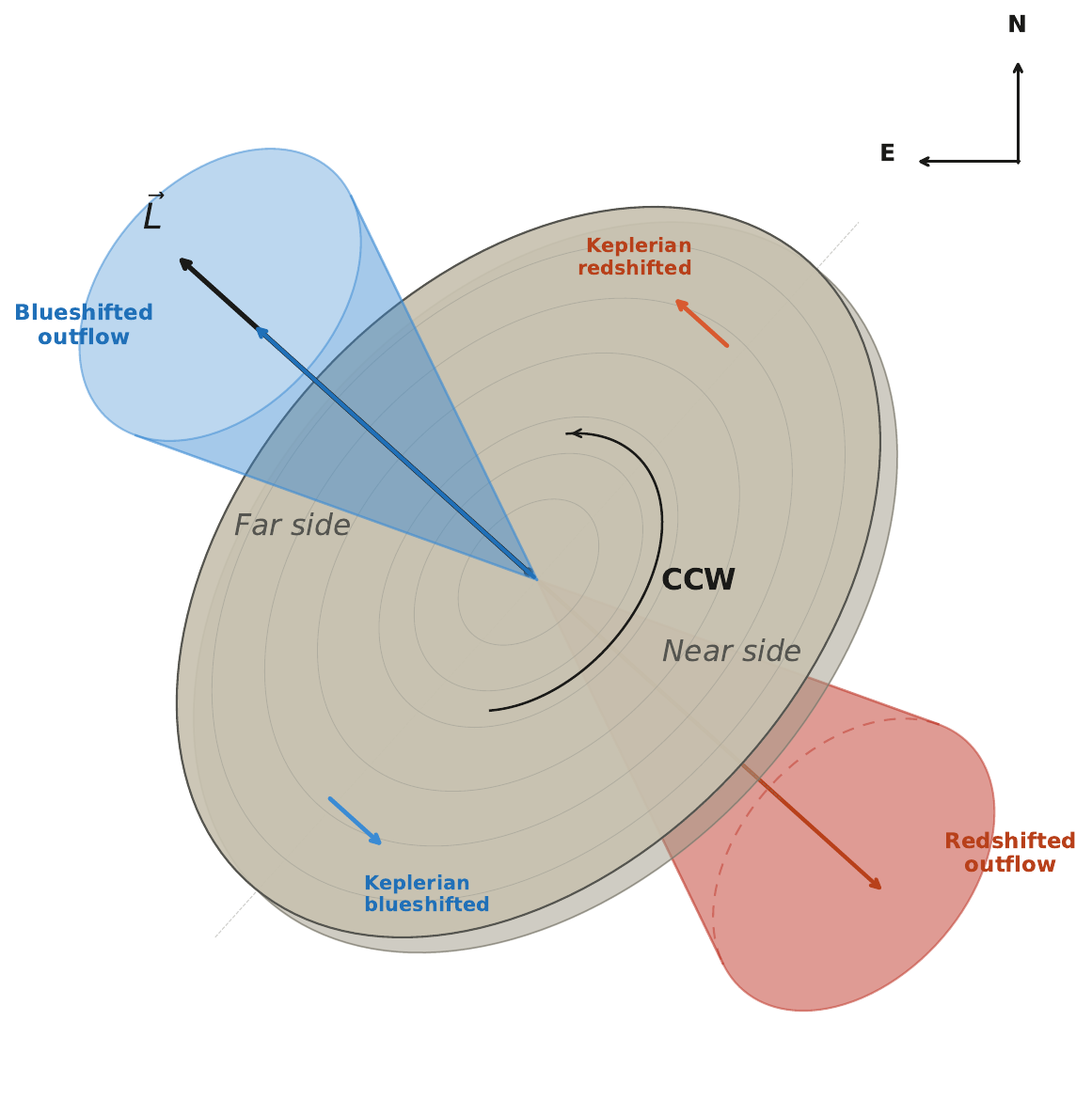}
    \caption{Illustration of the HL~Tau disk orientation and rotation geometry. The cones show the outflow directions. The blueshifted outflow is projected to the northeast, which determines the far-side and near-side of the disk. The Keplerian velocity field on the disk, with redshifted emission to the northwest and blueshifted emission to the southeast, combined with the outflow-determined disk orientation, establishes the angular momentum vector $\vec{L}$ and the corresponding counterclockwise rotation of the disk as projected on the sky.}
    \label{fig:HLTau_disk_orientation}
\end{figure}

\begin{figure}[h]
    \centering
    \includegraphics[width=1.0\linewidth]{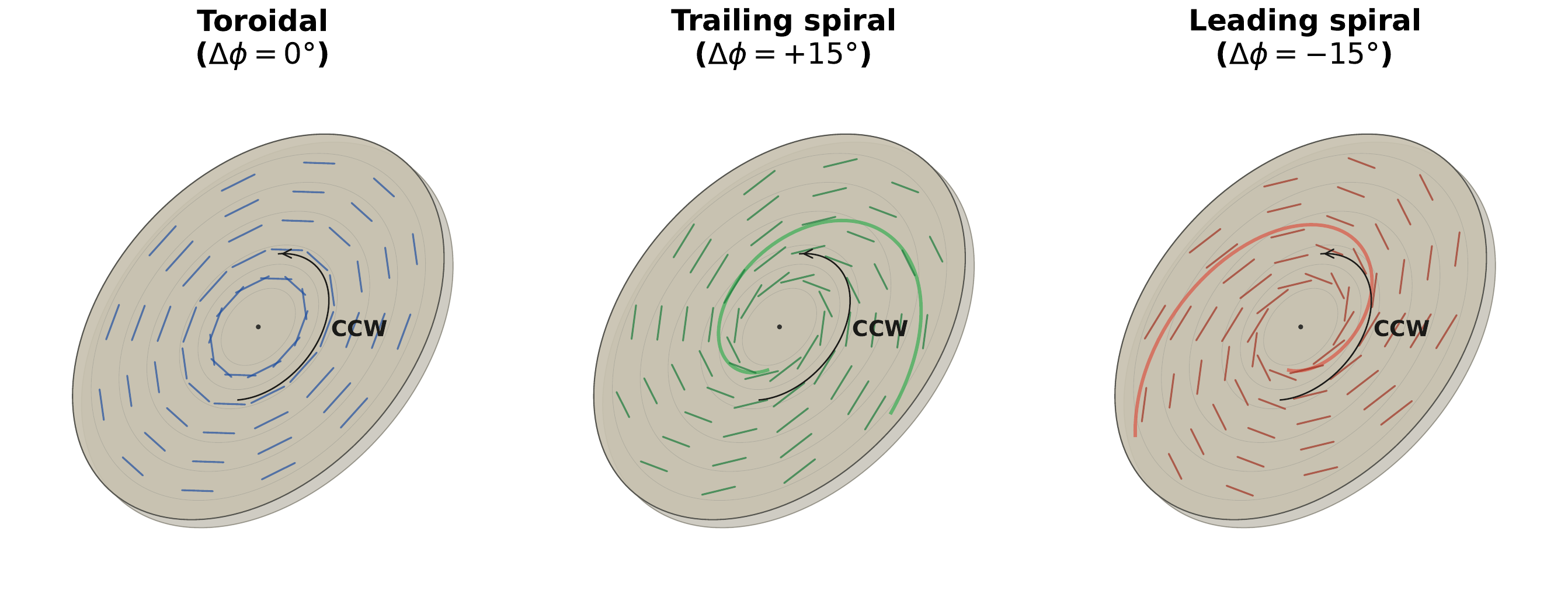}
    \caption{Illustration of grain alignment geometries. The disk has the same orientation and rotation direction (counterclockwise, shown by the black curved arrows) as HL~Tau. The short line segments represent the grain alignment directions. \textit{Left:} Toroidal alignment. \textit{Middle:} Deviated toroidal alignment with positive $\Delta\phi$, where each alignment direction is rotated counterclockwise (same as disk rotation direction) from purely azimuthal, forming trailing spiral streamlines that extend clockwise (opposite to the disk rotation direction). \textit{Right:} Negative $\Delta\phi$, where the alignment deviation is opposite to the rotation, forming leading spiral streamlines. The solid spiral curve in the middle and right panels illustrates a representative streamline of the alignment direction field. The $15^\circ$ deviation angle is exaggerated from actual data for visual clarity.}
    \label{fig:spiral_orientation}
\end{figure}

Based on the $\rm{HCO^+}$ (3-2) kinematic analysis of HL~Tau presented by \cite{Yen_2019}, the northwestern side of the disk is redshifted, while the southeastern side is blueshifted, presenting a velocity profile consistent with Keplerian rotation. Furthermore, \cite{ALMA_Partner_2015} and \cite{Yen_2017_outflow} have shown that the blueshifted outflow direction is in the northeastern side of the disk, which means the northeast is the far-side of the disk. Therefore, the angular momentum vector of the disk is projected along the northeast direction and towards the observer, and HL~Tau exhibits a counterclockwise rotation as projected on the sky (see Figure~\ref{fig:HLTau_disk_orientation}). 

We recall that our disk plane azimuthal angle $\phi_{\rm{disk}}$ is defined to increase counterclockwise (Equation~\ref{eq:phi_ell}). In our deviated toroidal alignment model (Equation~\ref{eq:spiral_alignment_qp}, \ref{eq:spiral_alignment_up}), the grain alignment direction at each location is modified by an angular offset $\Delta\phi$ such that the alignment direction at azimuthal angle $\phi_{\rm{disk}}$ is replaced by $(\phi_{\rm{disk}} + \Delta\phi)$. A positive $\Delta\phi$ in our model therefore corresponds to a systematic rotation of the grain alignment directions (and the corresponding polarization vector) toward increasing $\phi_{\rm{disk}}$ (counterclockwise), resulting in spiral streamlines that azimuthally expand from inner to outer radii toward decreasing $\phi_{\rm{disk}}$ (clockwise). Equivalently, this corresponds to an apparent rotation of the observed pattern in Stokes $Q_p$ and $U_p$ maps toward decreasing $\phi_{\rm{disk}}$, which is clockwise. Given that HL~Tau's rotation direction is counterclockwise as projected on the sky, a positive $\Delta\phi$ therefore indicates trailing spiral streamlines, as shown in Figure~\ref{fig:spiral_orientation}.

As shown in Section~\ref{sec:B4_st_dphi_global_result}, the global $\Delta\phi$ provides a direct, well-constrained measurement of the overall alignment deviation in Band~4. Both robust weightings yield positive global $\Delta\phi$ values: $\Delta\phi = 2.83^\circ \pm 0.33^\circ$ (robust $=-0.5$) and $\Delta\phi = 4.10^\circ \pm 0.09^\circ$ (robust = 0.5), indicating a trailing spiral orientation of the grain alignment streamlines. 

The per-annulus $\Delta\phi$ profile in Section~\ref{sec:B4_st_dphi_per-annulus_result} further supports this conclusion and provides additional radial information. We note that the innermost annuli (within one beam of the center of the disk) do not provide reliable constraints on $\Delta\phi$. At small radii, each annulus subtends a much shorter arc length per unit azimuthal angle, making azimuthal deviations intrinsically more difficult to capture. Additionally, the fixed synthesized beam size leads to strong azimuthal averaging in the inner disk, where a single beam can span a substantial azimuthal fraction of an annulus. These effects suppress the sensitivity to azimuthal structures. Thus, the $\Delta\phi$ values fitted in the innermost annuli are unreliable, and should not be used to determine the spiral orientation. Focusing on the intermediate radii that cover all ring and gap substructures across the disk, the $\Delta\phi$ profile is predominantly positive, consistent with the trailing orientation inferred from the global model. 
However, as shown in Section~\ref{sec:B4_dphi_robustness}, the per-annulus $\Delta\phi$ model is not statistically preferred over the global model, and the radial variation of $\Delta\phi$ should be regarded as exploratory.

\begin{figure}
    \centering
    \includegraphics[width=1.0\linewidth]{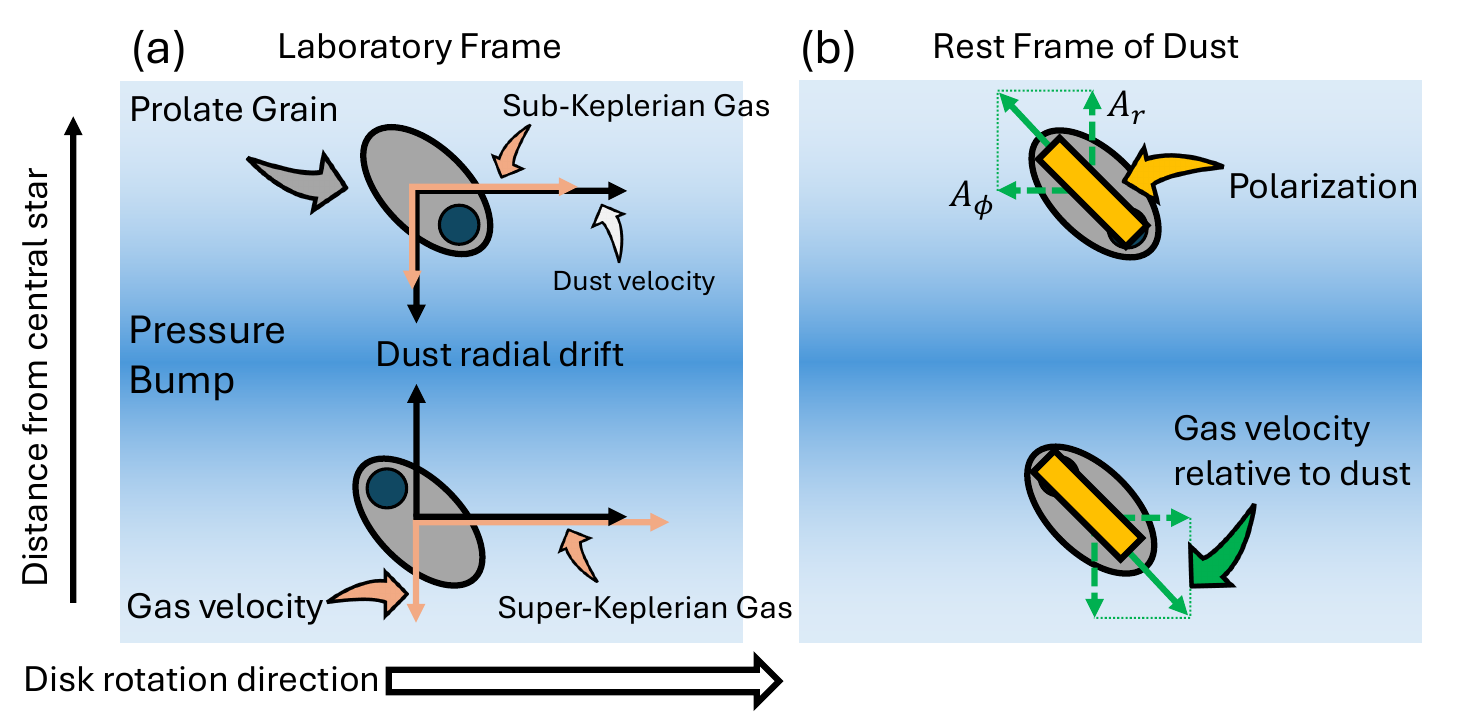}
    \caption{Birdie-like alignment of prolate grains in a pressure bump with gas-dust radial drift. Panel (a) shows the perspective in the laboratory frame, where a pressure bump causes the interior and exterior gas flows to become super-Keplerian and sub-Keplerian, respectively, while the dust grains are near Keplerian. Gas drag causes the dust grains to radially drift towards the pressure bump from both sides and accumulate. Panel (b) shows the perspective in the rest frame of dust grains. $A_\phi$ and $A_r$ are the relative gas-dust flow component in azimuthal and radial direction. Prolate, birdie-like grains align their long axes parallel to the local gas flow. The deep blue spot indicates the displaced center of mass. The resulting polarization produces trailing spirals. The schematic follows from \cite{Mori_2019_AS209}, but modified to account for the birdie-like alignment of grains.}
    \label{fig:birdie-like_pressure_bump}
\end{figure}

Similar trailing spiral polarization patterns have previously been observed in AS~209 \citep{Mori_2019_AS209} and GG~Tau~A \citep{Tang_2023_GGTau}, providing important empirical precedents for interpreting the deviated toroidal alignment observed in HL~Tau. In AS~209, \cite{Mori_2019_AS209} proposed a physical scenario in which local gas pressure maxima located near a ring produce a characteristic gas velocity structure: gas orbits at super-Keplerian velocity interior to the pressure bump due to a positive pressure gradient, and at sub-Keplerian velocity exterior to it due to a negative pressure gradient, while dust grains orbit at approximately Keplerian velocity (Figure~\ref{fig:birdie-like_pressure_bump}a). This velocity structure adopts the usual equilibrium solution where dust experiences drag as the gas orbits at a different velocity due to a pressure gradient \citep[e.g.,][]{NAKAGAWA1986375, Takeuchi_Lin_2002, Birnstiel_2024}. In the rest frame of dust grains (Figure~\ref{fig:birdie-like_pressure_bump}b), the relative gas flow interior to the pressure bump corresponds to a tailwind along the direction of disk rotation, leading to a gain in angular momentum of dust grains, causing them to radially drift outward (which is radially toward the pressure bump). Meanwhile, the gas flow relative to the dust grains exterior to the pressure bump corresponds to a headwind opposite to the direction of disk rotation, leading to a loss in angular momentum of dust grains, causing them to radially drift inward (also toward the pressure bump). Thus, dust grains symmetrically drift radially toward the pressure bump, leading to grain accumulation. Following such radial drift, the relative gas velocity experienced by the dust grains acquires an additional effective radial component in the dust rest frame: the relative gas flow appears to point radially away from the pressure bump on both sides. Combined with the azimuthal headwind or tailwind described above, the resultant gas flows experienced by the grains are: interior to the pressure bump, dust grains experience a relative gas flow azimuthally aligned with the disk rotation and radially inward, while exterior, the relative gas flow is azimuthally opposite to the disk rotation and radially outward. As a result, although the relative gas flow vectors on the two sides point in opposite directions, they are approximately collinear in orientation and would thus impose a common spiral orientation regardless of whether the grain long axis aligns parallel or perpendicular to the gas flow. The relative magnitudes of the radial and azimuthal velocity components depend on the Stokes number of the grains, which gives the efficiency of aerodynamic coupling between gas and grains \citep{Kataoka_2019_helicity}.

This configuration disfavors the helicity-induced alignment proposed by \cite{Kataoka_2019_helicity}, in which the polarization vectors are expected to be perpendicular to the gas velocity relative to the grain, producing a leading spiral, in contradiction to our measured trailing orientation. In contrast, the birdie-like model proposed by \cite{badminton_birdie-like} considers prolate grains whose center of mass is offset from the geometric center. Under aerodynamic torques, such prolate grains can align with their long axes parallel to the gas flow direction under subsonic drift. Given the relative gas flow geometry described above, this model yields polarization orientations consistent with a trailing spiral, making the birdie-like model a plausible explanation for the angular deviation observed in HL~Tau under the pressure bump scenario. We note, however, that this scenario would require a fairly large Stokes number (quantified and discussed in Section~\ref{sec:stokes_number}), which remains an open issue for the assumed velocity structure \citep{badminton_birdie-like}. 

Back to previous observational examples, we note that AS~209 and GG~Tau~A are very different sources. AS~209 is a Class II protoplanetary disk, where the spiral polarization deviation is detected near a prominent dust ring at 120 au \citep{Mori_2019_AS209}. GG~Tau~A is a hierarchical triple stellar system surrounded by a massive circumternary Keplerian ring \citep{Di_Folco_2014_GGTau}, and the observed spiral polarization pattern is also in this ring structure \citep{Tang_2023_GGTau}. Despite the substantially different astrophysical environments between these two systems, trailing spiral polarization is observed in both cases in the vicinity of ring-like dust concentrations. The proposed pressure-bump-driven radial drift scenario leads naturally to dust accumulation and potentially to the formation of rings. HL~Tau hosts a large number of concentric ring and gap substructures distributed across the disk plane, while coherent trailing spiral deviations are detected across a wide radial range including all those ring substructures. In this context, it is plausible that HL~Tau harbors multiple, small-scale pressure bumps across the disk, driving widespread accumulation of prolate grains into dust rings and causing birdie-like alignment. Such a scenario could simultaneously account for the abundance of rings, and the disk-wide presence of trailing spiral deviation in polarization. The pressure bumps do not need to completely halt inward radial drift, while a ``traffic jam'' picture, in which rapid radial drift is slowed down is sufficient to produce both dust accumulation into rings and the trailing spiral alignment pattern. Such a picture is also consistent with the optically thick inner disk interior to D14 gap, which requires a sustained inward flow of dust. 

We note that alternative alignment mechanisms under specific scenarios may also be capable of producing the observed azimuthal polarization patterns. \cite{Hoang_2009_wrong_alignment} showed that under conditions of slow internal relaxation and thermal rate rotation, grains can undergo ``wrong internal alignment'', where the grain long axes become parallel rather than perpendicular to the magnetic field. \cite{Thang_2024_MRAT} applied this concept to HL~Tau and demonstrated that large grains containing iron inclusions could undergo wrong internal alignment, and reproduce the observed azimuthal polarization patterns with a toroidal magnetic field. Separately, \cite{Guillet_2020_Mie} showed that when grain sizes approach the observation wavelength (the Mie regime), the polarization direction of thermal emission by aligned grains can reverse, which could similarly produce azimuthal polarization pattern in a toroidal magnetic field. If such a magnetic field possessed a non-toroidal component, these scenarios could in principle also produce spiral-like deviations in the alignment field. Both the birdie-like and magnetic-field scenarios can therefore reproduce the azimuthal pattern, and both can produce a spiral deviation under appropriate conditions; distinguishing them will require more detailed theoretical modeling of grain alignment.

We note that the $\Delta\phi$ profile measured from the Band~4 data does not exhibit a one-to-one correspondence between local extrema in $\Delta\phi$ and individual ring locations. This may indicate more complex dynamical processes of the radial drift, but may also be partly due to the limited angular resolution of the Band~4 observations, as several substructures resolved in Band~7 are not resolved in Band~4. 

\subsection{Band 7 Analysis, Comparison, and Limitations} \label{sec:B7_limitations}
As stated in Section~\ref{sec:B7_Modeling_Results}, we use models to fit Band~7 polarization data imaged with robust = 0.5 \citep[the same robust weighting as][]{Ian_Nature_2023} and robust = 2.0, and adopt a similar division of roles between the two robust weightings as used for Band~4: the higher resolution robust = 0.5 data are used to examine the radial behavior of the polarization mechanisms through $s$ and $t$, while robust = 2.0 data are used to examine the larger-scale polarization morphologies and the presence of deviated toroidal alignment structures.

\cite{Ian_Nature_2023} showed that the Band~7 polarization in HL~Tau is dominated by dust scattering, and the radial variation of the polarization mechanisms is already well studied with the toroidal alignment model. Stokes $Q_p$ in Band~7 is dominated by the scattering-induced polarization signal due to the alignment between the chosen reference axis and the expected scattering polarization direction, and has overall positive values.
Another notable feature in Band~7 polarization data is a pronounced asymmetry in Stokes $U_p$. The negative Stokes $U_p$ signal on the western side of the disk exhibits substantially lower signal-to-noise compared to other azimuthal regions, resulting in a strongly suppressed amplitude in that quadrant. In addition, on the southwestern side (minor-axis near-side) of the disk, the zero-crossing region of Stokes $U_p$ deviates counterclockwise relative to the expectation for a simple toroidal alignment model. The positive Stokes $U_p$ region associated with the southern side extends azimuthally in counterclockwise direction, encroaching into the area that would originally be expected to correspond to the negative quadrant. Such a behavior viewed in isolation, as we discussed above in Section~\ref{sec:spiral_mechanism_discussion}, would resemble a leading rather than trailing deviation. Both the signal suppression in the western negative region and the anomalous counterclockwise rotation are directly visible in Band~7 Stokes $U_p$ data imaged in both robust weightings (Figure~\ref{fig:B7_r0.5_st_result} and Figure~\ref{fig:B7r2_st_dphi_results}), though these features are amplified and more obvious in robust = 2.0 data. Neither of these behaviors is reflected in Stokes $Q_p$ and Stokes $I$. This suggests that the observed asymmetry and deviation originate primarily from the polarization contributed by aligned grains' thermal emission, which dominates Stokes $U_p$, while being overwhelmed by scattering in Stokes $Q_p$ and absent in Stokes $I$.

The asymmetry and deviation described above cannot be captured by the toroidal alignment model even with the angular deviation, and therefore inevitably affect the model fitting results of Band~7. However, the impact differs between the inference of polarization mechanisms amplitudes ($s$ and $t$) and the angular offset $\Delta\phi$.

The radial profiles of polarization mechanism amplitudes $s$ and $t$ remain reliable. Although the Stokes $U_p$ signal on the western side is strongly suppressed and becomes difficult to trace radially at larger radii, the toroidal alignment residuals do not exhibit arc-like or ring-like structured features in that region. Instead, the residuals appear continuous in radius, and largely uniform in color, indicating that the asymmetry manifests primarily as amplitude suppression rather than a systematic deviation in the fundamental radial structure. This amplitude asymmetry does affect the overall fitted amplitude of $t$. To achieve an optimal fit across all azimuthal regions, the model compromises by biasing the fitted $t$ toward lower values to reduce the large residuals on the suppressed western side, at the cost of introducing systematic negative residuals on the eastern side (as well as positive residuals on the northern and southern positive quadrants). Nevertheless, this compensation does not modify the radial dependence of $t$, and does not affect the model's capability to capture structures in radial profile. In addition, \cite{Ian_Nature_2023} has demonstrated that the radial behaviors of $s$ and $t$ are tightly coupled to the disk substructures traced by Stokes $I$, which shows no corresponding strong asymmetry. Thus, while quantitative comparison of the absolute $t$ amplitudes between Band~7 and Band~4 should be interpreted with caution, the radial profile of fitted $t$ retains structural fidelity and is suitable for comparison with Band~4 results. 

In contrast, the angular offset $\Delta\phi$ cannot be reliably measured at Band~7 for two independent reasons. First, the sensitivity of the observed polarization angle to the grain alignment direction is intrinsically low at this wavelength. Because scattering dominates at Band~7, the polarization angle responds weakly to a rotation of the alignment direction. For the fitted Band~7 $s$ and $t$ profiles, an alignment deviation of $\Delta\phi \approx 4^\circ$ (the Band~4 value) changes the beam-averaged polarization angle by $\lesssim1^\circ$, as measured by evaluating the fitted model with and without the deviation. The same deviation at Band~4, where thermal emission dominates, rotates the observed polarization angle by nearly the full $4^\circ$.

This limitation is compounded by the simplified treatment of the scattering component in our linear decomposition model (Section~\ref{sec:Linear_Polarization_Modeling}). Radiation anisotropy and the full three-dimensional disk geometry are not included, and the linear addition is expected to hold only at moderate optical depth \citep{Daniel_model_2022, Ian_Nature_2023}. Real scattering polarization carries azimuthal structure beyond this prescription, most notably the near/far-side asymmetries predicted for inclined, optically thick disks \citep{Haifeng_2017_s_od} and for strong forward scattering by large, unsettled irregular grains \citep{Daniel_2023_irregular}, and detected in HL~Tau \citep{Ian_Nature_2023, Daniel_Multiband_2024}. These effects enter the observed polarization angle in proportion to the scattering fraction. At Band~4, they are strongly suppressed by the dominance of the thermal term, whereas at Band~7, the observed angle inherits them at full weight. The decomposition therefore remains robust for the radial amplitude profiles at Band~7, but not for fine angular structure such as $\Delta\phi$.

Second, the dominant angular structure observed at Band~7 is not of a form that any alignment deviation can produce. The deviation of the measured polarization angles from the toroidal model reaches $\sim\pm10^\circ$ with opposite signs at the two ends of the minor axis. An azimuthal offset of the alignment direction of an axisymmetric model cannot generate such a pattern for any magnitude, sign, or radial profile of $\Delta\phi$. In the half-disk jackknife test refitting the global $\Delta\phi$ model, the near and far halves of the Band~7 data produce $\Delta\phi$ of opposite signs. Thus, fitting the current $\Delta\phi$ model to Band~7 data forces a single azimuthal offset onto a near/far-antisymmetric angular pattern, so that neither the full-disk global $\Delta\phi$ value, nor the half-disk values, should be interpreted as measurements of the alignment deviation in Band~7. This pattern indicates an additional structure contributing to the Band~7 polarization, beyond the axisymmetric scattering plus (possibly deviated) toroidal alignment framework, rather than evidence for or against any particular $\Delta\phi$. It cannot yet be directly attributed to any existing scattering model, and exploring the origin is beyond the scope of this paper. We leave it as an open question. 

Given these limitations, we conclude that the Band~7 results neither support nor contradict the trailing spiral streamline orientation measured at Band~4. The formally positive $\Delta\phi$ values in Section~\ref{sec:B7_Modeling_Results} average over a structure that the model cannot produce, and we do not use them as evidence. Conversely, the unexplained Band~7 structure does not weaken the Band~4 measurement of $\Delta\phi$, for which thermal emission dominates and the angle response to the alignment direction is direct, and the robustness and significance have been tested in Section~\ref{sec:B4_dphi_robustness}.

\subsection{Polarization Mechanisms: Radial Variation and Multi-band Comparison} \label{sec:P_mechanism_discussion}
Section~\ref{sec:B4_st_result} has presented a detailed description of the radial profiles of the polarization mechanism amplitudes $s$ and $t$ in Band~4, based on the robust $=-0.5$ toroidal alignment model result (Figure~\ref{fig:B4rm0.5_st_st_profile}). Overall, the Band~4 thermal component $t$ shows major peaks associated with the region around prominent gaps, but with radial offsets of $\sim$ 2 to 3 au relative to the gap centers (partly because the annuli centers are not designed to coincide with the gap and ring locations). The $t$ component exhibits troughs across bright belts. The scattering component $s$ displays an approximately complementary profile, which peaks in parts of the bright-belt regions, while showing troughs across the prominent dark belts, except in the outer disk around D96, where both $s$ and $t$ exhibit peaks. Notably, the scattering amplitude in Band~4 is weaker than the thermal component at most of the radii outside the very inner disk, indicating that Band~4 polarization is dominated by aligned-grain thermal emission rather than dust scattering. Given this dominance, the radial fluctuations observed in the scattering component $s$ are small in amplitude when viewed on the same scale as the variations in $t$. Furthermore, because both $s$ and $t$ contribute to $q_p$, their fitted values are partially degenerate, and adjacent annuli are additionally coupled through the beam convolution. Some apparent radial features in the $s$ profile may therefore arise from correlated adjustments of the fitted parameters rather than from physically distinct structures. As a result, the Band~4 polarization morphology is primarily governed by the radial behavior of $t$, while the detailed radial variation in $s$ should be interpreted with caution.  

We now turn to Band~7 to examine how the radial behavior of the polarization mechanisms changes at shorter wavelengths. \cite{Ian_Nature_2023} showed that the Band~7 polarization in HL~Tau is dominated by dust scattering, with higher polarization fractions and stronger azimuthal ($t$) components in gap regions compared to rings. Using the same modeling framework, we explicitly present the radial variations of the scattering and thermal emission amplitudes $s$ and $t$ in Section~\ref{sec:B7_Modeling_Results}, which were not individually emphasized in \cite{Ian_Nature_2023}. Overall, both the thermal emission and scattering components in Band~7 are enhanced in gap regions relative to rings, but their radial behaviors have substantial differences. The thermal component exhibits large-amplitude, sharply localized peaks associated with the most prominent gaps, while remaining relatively smooth across sequences of weaker substructures that are close to each other. In contrast, the scattering component responds more sensitively to fine-scale ring and gap alternations. The subtle enhancements at gaps and suppressions at rings are resolved even within narrow radial intervals, although with much smaller amplitude variations. We also notice that at radii larger than 73 au, the profiles of $s$ and $t$ started to exhibit slightly complementary behavior, similar to the complementary behavior observed in Band~4. The scattering dominance in Band~7 polarization is reflected in the fact that the mean scattering amplitude across the disk is comparable to the local peak values of the thermal component. Nevertheless, when viewed on the same scale, fluctuations in $s$ remain significantly weaker than those in $t$, and appear as modest fluctuations on a relatively smooth radial profile. 

We next compare the inferred polarization mechanism profile between Band~4 and Band~7. In the thermal component $t$, both bands show pronounced enhancements associated with the major gaps D34 and D68. At other radii, Band~4 shows a trough at D14 which differs from the sharp peak measured in Band~7, and Band~4 presents enhancements across the D96+D108 region corresponding to the gaps, whereas Band~7 shows no rises. We note that in Band~4, slight offsets are observed between measured $t$ peaks and the nominal gap centers, which should be attributed primarily to the larger beam size in Band~4 and the correspondingly wider annulus sampling adopted in the fitting, which smooths and shifts localized extrema compared to the higher-resolution Band~7 profiles.

The scattering component $s$ exhibits a stronger band-dependent contrast in its response to substructures. In Band~7, $s$ shows a more consistent and sensitive response, with subtle enhancements at gaps and suppression at rings. By contrast, $s$ in Band~4 has fluctuations less tied to individual substructures, and appears to show a reversed response to rings and gaps that is anti-correlated with the prominent $t$ variations across the intermediate disk regions. Similar reversed, complementary behavior is also observed in Band~7 at radii after 73 au. 

In terms of overall amplitudes, Band~4 displays a higher thermal component and a lower scattering component than Band~7 across most radii, which is consistent with the trend shown by \cite{Daniel_Multiband_2024}. We note that our results exhibit higher average values and stronger peaks compared to those in \citet[][their Figure 8]{Daniel_Multiband_2024}. This is expected, as our results present the intrinsic (pre-convolved) values, whereas \cite{Daniel_Multiband_2024} performed model fitting directly on convolved data. It is also consistent with the expected wavelength dependence of optical depth, where optical depth decreases with longer wavelength \citep{Carrasco-González_2019_wl_od}, such that the Band~4 emission (2.1 mm) is expected to be more optically thin than Band~7 emission (870 $\rm{\mu m}$). For observed polarized thermal emission from aligned grains, increased optical depth attenuates the polarization \citep{Hildebrand_2000}. The larger overall $t$ measured in Band~4 relative to Band~7, together with the enhancement of $t$ in prominent gap regions, is consistent with this expectation. 

The behavior of scattering-induced polarization $s$ with wavelength has a non-monotonic dependence on optical depth. As shown by \cite{Haifeng_2017_s_od}, for intermediate inclinations, the scattering-induced polarization peaks at optical depth $\tau$ of order unity, and rapidly decreases to zero with lower optical depth in the $\tau < 1$ regime. For $\tau > 1$, the scattering-induced polarization first decreases quickly from its peak, and then transitions into an asymptotically flat regime at $\tau \gg 1$, approaching a value of order half of its peak value. \cite{Pinte_2016_HLTau_od} provides a multi-band analysis of HL~Tau's radial profile of optical depth. This analysis shows that at 870 $\rm{\mu m}$ the disk is largely optically thick, with ring regions typically reaching $\tau \gg 1$ and gap regions having optical depths slightly above or close to unity ($\tau \gtrsim 1$). Such a configuration would have gaps showing maximal scattering polarization at $\tau \sim 1$, whereas the optically thicker rings lie further into the asymptotic regime where scattering polarization is reduced but still obvious. This is consistent with the measured Band~7 $s$ profile, which exhibits systematically higher values in gaps compared to rings, and whose overall amplitudes remain relatively high and stable across the disk.

Although \cite{Pinte_2016_HLTau_od} did not explicitly analyze the optical depth profile at Band~4 wavelengths, we can still reasonably predict its behavior given the wavelength dependence of optical depth. Ring regions in Band~4 may have optical depths ranging from $\tau > 1$ to $\tau \sim 1$, depending on the radius; while the gap regions are likely to fall into the optically thin regime with $\tau < 1$ or $\tau \ll 1$. Once gap optical depths drop below unity, scattering-induced polarization is expected to decline rapidly, such that the $s$ correspondence with rings and gaps could be inverted to the optically thick case. These expectations naturally explain both the overall reduction of the scattering amplitudes in Band~4 and its irregular correspondence with rings and gaps substructures. A direct modeling of the Band~4 optical depth radial profile would be required to quantitatively test this interpretation.

\subsection{Stokes Number Implied by the Birdie-like Alignment Scenario} \label{sec:stokes_number}
To estimate the Stokes number $\mathrm{St}$ inferred from $\Delta\phi$ in the birdie-like alignment scenario, we adopt the formulation from \citet[][their Section 4.1]{badminton_birdie-like}, which assumes a steady-state, axisymmetric solution to dust experiencing gas drag in the presence of a pressure gradient \citep{NAKAGAWA1986375}. The radial and azimuthal components of the gas flow experienced by the dust $A$ are:

\begin{align}
    A_R &\equiv v_{g,R} - v_{d,R} = -\frac{\alpha \mathrm{St}+\beta}{\mathrm{St} + \mathrm{St}^{-1}}\left(\frac{H}{R}\right)^2v_K \\
    A_\Phi &\equiv v_{g,\Phi} - v_{d,\Phi} = \frac{1}{2}\frac{-\alpha +\beta\mathrm{St}}{\mathrm{St} + \mathrm{St}^{-1}}\left(\frac{H}{R}\right)^2v_K
\end{align}
with $A_R>0$ pointing radially outward and $A_{\Phi}>0$ along the disk rotation direction. Here, $v_g$ and $v_d$ are the gas and dust velocities, with the subscripts $R$ and $\Phi$ denoting their radial and azimuthal components; $\alpha$ and $\beta$ parametrize the radial gas inflow and the midplane pressure gradient (Equations~\ref{eq:alpha} and \ref{eq:beta}); and $H$ and $v_K$ are the gas scale height and the Keplerian velocity, defined below Equation~\ref{eq:alpha}. Since the prolate grain long axis aligns with this flow, the alignment direction deviates from the azimuthal direction by an angle $\Delta\Phi_A$, which we define as: 
\begin{equation} \label{eq:dPhiA}
    \tan (\Delta\Phi_A) = -\frac{A_R}{A_{\Phi}} = \frac{2(\alpha\mathrm{St}+\beta)}{\beta\mathrm{St}-\alpha}
\end{equation}
where $\Delta\Phi_A$ is taken in $(-90^\circ,90^\circ]$. The common factor $(H/R)^2v_K/(\mathrm{St}+\mathrm{St}^{-1})$ cancels in the ratio, so that $\Delta\Phi_A$ depends only on $\mathrm{St}$, $\alpha$, and $\beta$.

A positive $\Delta\Phi_A$ corresponds to trailing spiral alignment streamlines, and a negative $\Delta\Phi_A$ corresponds to leading spiral alignment streamlines. Note that this correspondence holds independently of the disk's orientation on the sky (unlike the $\Delta\phi$ defined previously), as the definition of $\Delta\Phi_A$ has already adopted the rotation direction of the disk. For the rotation sense and orientation of HL~Tau (Section~\ref{sec:spiral_mechanism_discussion}), a positive observed $\Delta\phi$ corresponds to a positive $\Delta\Phi_A$ in this article, and we hereafter identify $\Delta\Phi_A = \Delta\phi$. Then, we can solve for $\mathrm{St}$ through estimates of $\alpha$ and $\beta$.

Here, $\alpha$ parametrizes the radial velocity of accreting gas in the midplane:
\begin{equation} \label{eq:alpha}
    v_{g,R} = -\alpha\left(\frac{H}{R}\right)^2v_K
\end{equation}
$v_K\equiv\Omega_KR$ is the Keplerian velocity in the midplane, $\Omega_K\equiv \sqrt{GM_\star/R^3}$ is the Keplerian frequency, $H \equiv c_s/\Omega_K$ is the scale height, and $c_s \equiv \sqrt{kT/(\mu m_p)}$ is the isothermal sound speed.

If viscosity is the origin of the inward flow, $\alpha$ is 1.5 times the Shakura--Sunyaev viscosity parameter \citep{badminton_birdie-like}. Attributing the stellar accretion rate of $\dot{M} = 8.7\times10^{-8} M_{\sun}~\mathrm{yr}^{-1}$ \citep{Beck_2010} entirely to such a flow gives $\alpha_{\mathrm{SS}} \approx 10^{-2}$ at 100 au \citep{Pinte_2016_HLTau_od}, which we take as an upper bound. Alternatively, applying a standard dust settling model to the thin midplane layer of millimeter-sized grains gives $\alpha_{\mathrm{SS}} \approx 3\times10^{-4}$ \citep{Pinte_2016_HLTau_od}, which we take as a lower bound. The two estimates differ by more than an order of magnitude, as the former measures the radial transport of angular momentum, while the latter measures the vertical diffusion of dust. For $\alpha = 1.5\,\alpha_{\mathrm{SS}}$, we therefore adopt $4.5\times10^{-4} \lesssim \alpha \lesssim 1.5\times10^{-2}$.  

The midplane pressure gradient $\beta$ is defined as: 
\begin{equation} \label{eq:beta}
    \beta\equiv \frac{\partial \ln{P_g}}{\partial \ln{R}}
\end{equation}
$P_g$ is the gas pressure in the midplane:
\begin{equation}
    P_g = \rho_g c_s^2 = \frac{\Sigma_g}{\sqrt{2\pi}H} c_s^2
\end{equation}
We have:
\begin{align}
    \ln{P_g} &= \ln{\Sigma_g} + \ln{T} - \ln{H} + \mathrm{const} \\
    \beta &= \frac{\partial\ln{\Sigma_g}}{\partial\ln{R}} + \frac{\partial\ln{T}}{\partial \ln{R}} - \frac{\partial\ln{H}}{\partial \ln{R}}    
\end{align}

According to \cite{Booth_Ilee_2020}, the gas surface density of HL Tau can be expressed as:
\begin{align}
    \Sigma_g(r) &= \Sigma_0\left(\frac{r}{r_c}\right)^{-\gamma} \exp\left[-\left(\frac{r}{r_c}\right)^{2-\gamma}\right] \\
    \Sigma_0 &= (2-\gamma)\frac{M_{\mathrm{gas}}}{2\pi r_c^2}\exp{\left[\left(\frac{r_{in}}{r_c}\right)^{2-\gamma}\right]}
\end{align}

We adopt $M_{\mathrm{gas}} = 0.20M_{\sun}$ \citep{Booth_Ilee_2020}, $r_c = 80~\mathrm{au}$, $r_{in} = 8.78~\mathrm{au}$, $\gamma = -0.20$ \citep{Kwon_2015}, and $T=55\left(\frac{r}{30~\mathrm{au}}\right)^{-0.51}~\mathrm{K}$ \citep{ueda_2025}. This gives $\beta\approx-5.2$ at 100 au. The value of $\beta$ varies from $\approx-2$ to $\approx-7$ between 30 and 120 au, but as shown below, the inferred $\mathrm{St}$ is insensitive to $\beta$ in this regime.

Solving from the definition of $\Delta\Phi_A$, we have:
\begin{equation} \label{eq:St}
    \mathrm{St} = \frac{2\beta +\alpha \tan(\Delta\phi)}{\beta \tan(\Delta\phi) - 2\alpha}
\end{equation}

For the global $\Delta\phi = 4.10^\circ$ measured with robust $=0.5$, the adopted ranges of $\alpha$ and $\beta$ give $23.1 \lesssim\mathrm{St}\lesssim27.9$. The global $\Delta\phi = 2.83^\circ$ measured with robust $=-0.5$ gives correspondingly $31.0 \lesssim\mathrm{St}\lesssim40.3$. In the limit $\alpha \to 0$, Equation~\ref{eq:St} reduces to $\mathrm{St} = 2/\tan(\Delta\phi)$, independent of $\beta$, which gives the upper ends of these ranges. Because $\alpha\,\mathrm{St} \lesssim 0.6$ remains small compared with $|\beta| \gtrsim 2$ over the adopted ranges, $\mathrm{St}$ falls below this limit by at most $\sim25\%$, so the inferred $\mathrm{St}$ is set mainly by the measured $\Delta\phi$ rather than by the adopted values of $\alpha$ and $\beta$.

These values are in strong tension with the Stokes number expected for grains dominating the millimeter emission. Adopting the same gas surface density as above, multi-wavelength continuum modeling of HL~Tau yields $\mathrm{St}\lesssim10^{-2}$ beyond 60~au, and at most $\sim10^{-1}$ in the inner disk \citep{ueda_2025}, which are two to three orders of magnitude below our inferred values. Reaching $\mathrm{St} \approx 28$ at 100~au would require either a gas surface density of $\sim10^{-2} ~\mathrm{g~cm^{-2}}$, three orders of magnitude below the $\mathrm{^{13}C^{17}O}$-based measurement \citep{Booth_Ilee_2020}, or compact grains with sizes of order a meter. A similarly large Stokes number was already noted by \cite{badminton_birdie-like} for azimuthal alignment in smooth gas disks. We therefore regard this tension as a significant limitation of the birdie-like alignment interpretation under the assumed axisymmetric velocity structure. It reproduces the observed trailing orientation, but requires grains far more decoupled from the gas than the continuum observations indicate.

\section{CONCLUSIONS} \label{sec:conclusions}
We present new high-resolution ALMA Band~4 (2.1 mm) polarization observations of HL~Tau. Using a linear decomposition framework, we separate the polarization into contributions from dust scattering and thermal emission from aligned grains, and model the polarization using several grain‑alignment geometries. We apply the same modeling approach to ALMA Band~7 (870 $\rm{\mu m}$) observation to enable a multi-band comparison. Our main findings are as follows:

(1) \textit{Observational morphology}: Band~4 polarization vectors show predominantly azimuthal orientation in an elliptical pattern. Unlike Band 7, the polarization fraction and polarized intensity distributions are more diffuse, without sharp contrasts between individual rings and gaps.

(2) \textit{Polarization mechanisms and radial variation}: Band~4 polarization is dominated by the contribution from thermal emission of aligned grains, in contrast to Band~7's scattering dominance. In Band~4, the thermal component $t$ exhibits major peaks around several prominent gaps, and shows troughs across bright belts. The scattering component $s$ displays an approximately complementary radial behavior across the inner and most of the intermediate disk regions. This contrasts with Band~7, where both $s$ and $t$ are mainly enhanced in gap regions relative to rings. This wavelength dependence is consistent with optical depth evolution and the non-monotonic optical-depth dependence of scattering-induced polarization.

(3) \textit{Deviated toroidal alignment}: In Band~4, we detected a coherent deviation in the grain alignment direction from purely azimuthal. 
The global $\Delta\phi$ model yields angular deviations of $\Delta\phi = 2.83^\circ \pm 0.33^\circ$ (robust $=-0.5$) and $\Delta\phi = 4.10^\circ \pm 0.09^\circ$ (robust = 0.5), considering statistical uncertainties only. This detection reaches a Gaussian significance of $\approx7\sigma$ under conservative uncertainty estimates. The per-annulus $\Delta\phi$ profile is predominantly positive across the intermediate disk radii (14 to 108 au), while its apparent radial variation is suggestive but not required by the data. At Band~7, where scattering dominates, the polarization angle is only weakly sensitive to the alignment direction, and the data exhibit an additional near/far-side antisymmetric angular structure that no azimuthal alignment offset can produce. Therefore, the Band~7 results neither support nor contradict this detection.

(4) \textit{Trailing spiral alignment streamline and its interpretation}:
Streamlines of deviated grain alignment directions measured in HL~Tau Band~4 form spirals that progress azimuthally in the clockwise direction with increasing radius, opposite to the disk's counterclockwise rotation, and are thus characterized as trailing spirals. Such morphology favors the birdie-like alignment where prolate dust grains experience radial gas-dust drift near pressure bumps \citep{badminton_birdie-like}. The disk-wide presence of a trailing spiral across HL~Tau's extensive ring-gap structures suggests the scenario where multiple small-scale pressure bumps slow down radial drift throughout the disk, driving accumulation into rings and grain alignment, while still allowing dust to flow inward to sustain the optically thick inner disk. 

However, under the assumed velocity structure, the observed deviation angle implies Stokes numbers of $\mathrm{St} \approx 23$--$40$, two to three orders of magnitude larger than those expected for the grains dominating the millimeter emission. Thus, the birdie-like model with an axisymmetric disk velocity structure requires grains far more decoupled from the gas than the continuum observations indicate, which is the main limitation of this interpretation.

\begin{acknowledgments}
This paper makes use of the following ALMA data: 
ADS/JAO.ALMA $\#$2016.1.00115.S, ADS/JAO.ALMA $\#$2019.1.00134.S, ADS/JAO.ALMA $\#$2019.1.01051.S, and ADS/JAO.ALMA $\#$2023.1.00515.S.
ALMA is a partnership of ESO (representing its member states), NSF (USA) and NINS (Japan), together with NRC (Canada), NSTC and ASIAA (Taiwan), and KASI (Republic of Korea), in cooperation with the Republic of Chile. The Joint ALMA Observatory is operated by ESO, AUI/NRAO and NAOJ. The National Radio Astronomy Observatory and Green Bank Observatory are facilities of the U.S. National Science Foundation operated under cooperative agreement by Associated Universities, Inc.

L.W.L. and Y.J.X. acknowledge support from NSF AST-2307844 and NSF AST-2108794. 

Z.-Y.D.L. acknowledges support from U.S. National Science Foundation Astronomy and Astrophysics Research grant \#2307612.

M.F.L. thanks the Instituto Argentino de Radioastronom\'{i}a (CONICET, Argentina) and the Facultad de Ciencias Astron\'omicas y Geof\'{i}sicas (Universidad Nacional de La Plata, Argentina) for their continuous support and warmth of their people.

Z.Y.L. is supported in part by NSF AST-2307199.

R.T. was supported by JSPS KAKENHI Grant Number JP25K07351.

C.C.-G. acknowledges support from UNAM DGAPA-PAPIIT grant IG101224.

The figures, tables, scripts used to make them, and FITS images are available in the Illinois Data Bank \citep{illinoisdatabankIDB-6766628}. The raw ALMA data and the calibration scripts delivered by ALMA are available from the ALMA archive.

\end{acknowledgments}





%
\facilities{ALMA}

\software{Astropy \citep{astropy:2013, astropy:2018, astropy:2022}, \texttt{emcee} \citep{emcee_2013}, CASA \citep{The_CASA_Team_2022}}



\bibliography{Band4}{}
\bibliographystyle{aasjournalv7}



\end{document}